\documentclass[11pt,a4paper]{article}
\pdfoutput=1

\usepackage{jheppub}
\usepackage{dsfont}
\usepackage{ulem}
\usepackage{natbib}
\usepackage{amsmath,amssymb,amsthm,enumitem}
\usepackage{mathrsfs}
\usepackage{bbm}
\usepackage{hyperref}
\usepackage{braket}
\usepackage{xcolor}
\usepackage{slashed}
\usepackage[hang,flushmargin,bottom]{footmisc}

\usepackage{amsfonts,latexsym,euscript,mathalfa}
\usepackage{graphicx,epsfig,xcolor}
\usepackage{sectsty}
\usepackage{cancel}

\makeatletter\renewcommand{\@biblabel}[1]{#1.}\makeatother

\newcommand{\im}{{\rm i} \, }
\newcommand{\id}{\mathbbm 1}

\newcommand{\nn}{\nonumber}
\newcommand{\p}[1]{\left(#1\right)}

\newcommand{\f}[2]{\frac{#1}{#2}}
\newcommand{\be}[1]{\begin{equation}#1\end{equation}}
\newcommand{\ale}[1]{\begin{align}#1\end{align}}
\newcommand{\comm}[1]{\left[#1\right]}

\newcommand{\Q}{{\mathcal Q\,}}
\newcommand{\Tr}{\mathrm{Tr}}

\newcommand{\A}{\mathcal{A}}
\newcommand{\F}{\mathcal{F}}

\newcommand{\D}{\mathcal{D}}

\def\A{{\mathcal{A}}}
\def\ep{{\epsilon}}
\def\L{{\mathcal{L}}}

\def\T{{\mathbb{T}}}

\def\CN{{\mathcal{N}}}
\def\Di{{\mathbb{D}}}
\def\d{{\rm d}}
\def\Im{{\rm Im}}
\def\Re{{\rm Re}}

\def\i{{\rm i}}
\def\e{{\rm e}}

\def\brakett #1#2{{ \langle #1,#2 \rangle}}

\preprint{
{\small{\textsf{DESY 19-214 ~~ UUITP-19/19}}}}

\title{Localization of 4d $\mathcal{N}=1$ theories  on $\mathbb{D}^2\times \mathbb{T}^2$
}

\author[a]{Pietro Longhi}
\author[b]{Fabrizio Nieri}
\author[c]{Antonio Pittelli}

\affiliation[a]{
Institute for Theoretical Physics, ETH Zurich, 8093, Zurich, Switzerland
}
 \affiliation[b]{
DESY theory, Notkestrasse 85, 20607 Hamburg, Germany
 }
\affiliation[c]{Department of Physics and Astronomy, Uppsala University,\\
Box 516, SE-75120 Uppsala, Sweden.}

\emailAdd{longhip@phys.ethz.ch}
\emailAdd{fb.nieri@gmail.com}
\emailAdd{apittelli88@gmail.com}

\abstract{
We consider 4d ${\mathcal N}=1$ gauge theories with R-symmetry on a hemisphere times a torus. We apply localization techniques to evaluate the exact partition function through a cohomological reformulation of the supersymmetry transformations. 
Our results represent the natural elliptic lifts of the lower dimensional analogs as well as a field theoretic derivation of the conjectured 4d holomorphic blocks, from which partition functions of compact spaces with diverse topology can be recovered through gluing.
We also analyze the different boundary conditions which can naturally be imposed on the chiral multiplets, which turn out to be either Dirichlet or Robin-like. We show that different boundary conditions are related to each other by coupling the bulk to 3d ${\mathcal N}=1$ degrees of freedom on the boundary three-torus, for which we derive explicit 1-loop determinants.  
}

\keywords{Supersymmetry, localization, boundary conditions.}
\begin{document}

\maketitle
\newpage

%%%%%%%%%%  Main text     %%%%%%%%%%%
\section{Introduction}

Since the pioneering work of Pestun \cite{Pestun:2007rz}, which builds on previous results on exact non-perturbative effects in 4d $\mathcal{N}=2$ supersymmetric gauge theories (8 flat-space supercharges) \cite{Lossev:1997bz,Moore:1998et,Moore:1997dj,Losev:1997tp,Nekrasov:2002qd,Nekrasov:2003rj}, localization techniques have been successfully applied to computing protected observables in theories with varying amounts of supersymmetry and in diverse dimensions and backgrounds (for recent comprehensive reviews on techniques and results we refer to \cite{Teschner:2016yzf,Pestun:2016zxk}). For theories with $\mathcal{N}=1$ supersymmetry in four dimensions (4 flat-space supercharges), most of the studies have focused on closed compact spaces with $\mathbb{S}^n\times\mathbb{T}^m$ topology (or quotient thereof), while only few results for manifolds with boundaries are available, mostly focused on theories in two and three dimensions \cite{Shadchin:2006yz,Fujimori:2015zaa,Hori:2013ika,Honda:2013uca,Sugishita:2013jca,Yoshida:2014ssa,Cabo-Bizet:2016ars,Assel:2016pgi,David:2016onq,David:2018pex,Kimura:2018axa, Pittelli:2018rpl}. Loosely speaking, we will sometimes refer to a space with a boundary (possibly asymptotic) as non-compact, in sharp contrast to the radically different case of a closed space without boundary. The study of theories on non-compact spaces is interesting for many reasons as the boundary provides a kind of refinement of the compact setups. The scarcity of literature on the subject represents the main motivation for this paper, which is devoted to studying 4d $\mathcal{N}=1$ theories with R-symmetry on curved manifolds with boundaries. Following the rigid supergravity framework developed in \cite{Festuccia:2011ws,Dumitrescu:2012ha,Klare:2012gn,Closset:2013vra,Closset:2014uda} and demanding the existence of two Killing spinors of opposite R-charges in the bulk, the allowed backgrounds are complex manifolds diffeomorphic to torus fibrations over a Riemann surface. We consider the simplest choice, namely the (twisted) product of a torus $(\mathbb{T}^2$) times a disk ($\mathbb{D}^2$). The main goal of the paper is to compute the exact partition function ($Z$) through supersymmetric localization, recovering previous results arising from factorization arguments \cite{Nieri:2015yia} as well as studying supersymmetry preserving  boundary conditions. This observable can also be interpreted as a flavored Witten index (up to contact terms)
\be{\label{eq:index}
Z[\mathbb{D}^2\times\mathbb{T}^2]=\Tr_{\mathcal{H}}(-1)^\text{F}\e^{-2\pi \text{H}}~,\nn
}
where $\text{F}$ is the fermion number and $\text{H}$ is a certain element of the bosonic subalgebra commuting with the localizing supercharge. The trace is over the Hilbert space of states on $\mathbb{D}^2\times\mathbb{S}^1$, whose definition includes a choice of boundary conditions. This four dimensional background is particularly interesting also because most of the lower dimensional results can be recovered by dimensional reduction. 

\textbf{Factorization}. The formulation of supersymmetric theories on manifolds with boundaries can be thought to be more elementary to a large extent. Indeed, the majority of the compact space results can be (non-trivially) obtained from the non-compact ones through certain sewing procedures, reflecting how a compact background can be decomposed into basic geometries while respecting supersymmetry (for an alternative approach using compact backgrounds with defect operators we refer to \cite{Closset:2017zgf,Closset:2018ghr,Closset:2017bse}). The prototypical example of this picture is provided by 3d $\mathcal{N}=2$ gauge theories, where the topology is sufficiently rich and the theories sufficiently simple to manifest these phenomena in a controllable way. In fact, in a large number of interesting examples it was explicitly shown that the partition functions of Yang-Mills-Chern-Simons-Matter theories on $\mathbb{S}^3$,  $\mathbb{S}^2\times\mathbb{S}^1$ or lens spaces $\mathbb{L}(p,1)$ \cite{Kapustin:2009kz,Gang:2009wy,Hama:2010av,Hama:2011ea,Kapustin:2011jm,Imamura:2011su,Imamura:2011wg,Imamura:2012rq,Alday:2012au,Alday:2013lba,Benini:2015noa,Closset:2016arn} can be assembled from two copies of the partition functions on the half-space $\mathbb{D}^2\times\mathbb{S}^1$, a.k.a. 3d holomorphic blocks \cite{Beem:2012mb}. This can intuitively be understood as a manifestation of the genus one Heegaard decomposition/gluing of the compact spaces into/from a pair of solid tori $\mathbb{D}^2\times\mathbb{S}^1$ and the quasi-topological nature of the partition functions \cite{Closset:2012ru,Closset:2013vra}. Together, these observations lead to the expected (schematic) result
\be{
Z[(\mathbb{D}^2\times\mathbb{S}^1)\cup_g (\mathbb{D}^2\times\mathbb{S}^1)]=\sum_{\gamma}\, Z_\gamma[\mathbb{D}^2\times\mathbb{S}^1]Z_\gamma[\mathbb{D}^2\times\mathbb{S}^1]^{(g)}~,\nn
}
where $g\in\text{SL}(2,\mathbb{Z})$ is the group element associated with the homeomorphism implementing the sewing along the boundaries $\partial(\mathbb{D}^2\times\mathbb{S}^1)\simeq \mathbb{T}^2$ of the solid tori into the compact space (also acting on the disk partition function), while $\gamma$ represents a label for the IR boundary conditions (Higgs vacua) to be summed over. This intriguing type of factorization, very reminiscent of the $tt^*$ geometries \cite{Cecotti:1991me,Cecotti:2013mba}, was first observed by studying the functional structure of the partition functions on different compact spaces in several examples \cite{Pasquetti:2011fj,Dimofte:2011py,Imamura:2013qxa,Taki:2013opa,Nieri:2013yra,Dimofte:2014zga,Nieri:2015yia}, whereas a derivation of the 3d holomorphic blocks was obtained through supersymmetric localization on $\mathbb{D}^2\times\mathbb{S}^1$ \cite{Yoshida:2014ssa}, lifting the results for 2d $\mathcal{N}=(2,2)$ theories on $\mathbb{D}^2$  \cite{Hori:2013ika,Honda:2013uca}. Similarly, a factorized structure for the partition functions of 4d $\mathcal{N}=1$ theories on $\mathbb{S}^3\times\mathbb{S}^1$,  $\mathbb{S}^2\times\mathbb{T}^2$ and more generally $\mathbb{L}(p,1)\times\mathbb{S}^1$ \cite{Benini:2011nc,Closset:2013sxa,Yamazaki2013,Razamat:2013opa,Assel:2014paa,Nishioka:2014zpa,Gadde:2015wta,Honda:2015yha,Benini:2016hjo} was also observed \cite{Peelaers:2014ima,Chen:2014rca,Yoshida:2014qwa,Nieri:2015yia}, where the solid tori are naturally replaced by $\mathbb{D}^2\times\mathbb{T}^2$ patches
with the sewing along the boundaries $\partial(\mathbb{D}^2\times\mathbb{T}^2)\simeq \mathbb{T}^3$ implemented by an element $g\in\text{SL}(3,\mathbb{Z})$. One of the main motivations behind this work is to provide a derivation of the 4d holomorphic blocks proposed in \cite{Nieri:2015yia} from an independent computation though supersymmetric localization on $\mathbb{D}^2\times\mathbb{T}^2$, thus completing the 2d-3d-4d or rational-trigonometric-elliptic hierarchy \cite{Nekrasov:2009uh,Nekrasov:2009ui} of the exact effective twisted superpotential of gauge theories with 4 supercharges. 

\textbf{Boundary conditions}. The formulation of supersymmetric theories on manifolds with boundaries is also necessary in order to study interesting aspects which would be lost otherwise, such as the physics of boundary conditions, interfaces and bulk/boundary coupled systems. Once again, 3d $\mathcal{N}=2$ theories have provided a very useful laboratory so far \cite{Gadde:2013wq,Gadde:2013sca,Okazaki:2013kaa,Aprile:2016gvn}, and the lift to 4d $\mathcal{N}=1$ theories provides another strong motivation for this paper (for a recent general analysis we refer to \cite{DiPietro:2015zia}). An interesting and localization-friendly approach has been recently put forward in \cite{Dimofte:2017tpi} for 3d $\mathcal{N}=2$ theories and a class of dual boundary conditions preserving 2d $\mathcal{N}=(0,2)$ supersymmetry on the boundary, including the familiar Dirichlet and Neumann conditions. In this paper, we begin to develop the four dimensional lift of that approach, focusing on a subset of boundary conditions which naturally arises from our localization framework. Indeed, it turns out that we are able to perform localization upon imposing either Dirichlet or Robin-like conditions on the chiral multiplets of the theory, which preserve some supersymmetry at the boundary. We argue that the two boundary conditions can be flipped by coupling additional degrees of freedom supported on the boundary, and we provide a non-trivial check by computing the $\mathbb{D}^2\times\mathbb{T}^2$ partition function for both boundary conditions and showing that their ratio does indeed reproduce the partition function of a 3d boundary theory on $\partial(\mathbb{D}^2\times\mathbb{T}^2)\simeq \mathbb{T}^3$, which we derive by cohomological localization. This feature is reminiscent of dualities of boundary conditions in the context of 3d $\mathcal{N}=2$ theories \cite{Dimofte:2017tpi}, and it also provides a physical interpretation of well-known shift properties of the elliptic Gamma function featuring in our computations. However, a direct uplift would correspond to 4d $\CN=1$ supersymmetry in the bulk and 3d $\CN=1$ on the boundary, whereas our background only preserves a certain subalgebra in both cases due to curvature and bulk-boundary couplings.

\textbf{Localization}. On the technical side, our approach to computing exact partition functions is based on the cohomological reformulation of the supersymmetry transformations, closely following the 3d $\mathcal{N}=2$ framework of \cite{Kallen:2011ny,Ohta:2012ev}. In a nutshell, the central observation is that the supercharge which we use for localization can be identified with an equivariant differential acting on the supermanifold of quantum fields, therefore inducing the structure of an equivariant cohomological complex. Upon specifying its structure and the pairing between the different fields, the fluctuations around the localization locus can be integrated out to obtain the 1-loop determinants for both vector and chiral multiplets. However, we encounter several subtleties along the way, such as the presence of fermionic zero modes, singularities of the path integral measure and the related question of identifying the correct integration contour(s). Many of these issues are already familiar from studies of localization on compact manifolds \cite{Benini:2013nda,Benini:2013xpa,Benini:2015noa,Closset:2015rna} and can be addressed by existing arguments, whereas other subtleties are strictly tied to our choice of background. Amongst the main differences, we can mention the global constraints imposed by the boundary. A key property of the selected supercharge is that its action squares to a (twisted) Lie derivative along a Killing vector parallel to the boundary, hence ensuring the cancellation of several boundary terms and leaving some freedom in the choice of boundary conditions. However, some of the boundary terms must be killed by a specific choice of boundary conditions, which turns out to be either Dirichlet or Robin-like for the chiral multiplets, while we consider Neumann for the vector to preserve some gauge symmetry on the boundary. These restrictions discard some of the fluctuations which would otherwise contribute to the 1-loop determinants, hence yielding different results w.r.t. the compact backgrounds (essentially, half of those). Finally, we discuss how 4d $\mathcal{N}=1$ multiplets can be decomposed into 3d $\mathcal{N}=1$ multiplets when restricted to the boundary, which is a convenient step before considering bulk-boundary coupled systems. The literature on 3d supersymmetric theories with minimal supersymmetry is rather limited (we refer to \cite{Gates:1983nr} for a general analysis), but this very interesting subject has lately gained great attention \cite{Drukker:2017xrb,Drukker:2017dgn,Braun:2018joh,Eckhard:2018raj,Benini:2018bhk,Benini:2018umh}. Our setup naturally allows us to consider such theories on (twisted) $\mathbb{T}^3$, and opens up a new perspective for computing their partition functions through cohomological localization in conjunction with the bulk theory. 

\textbf{Summary of the main results}. Our main results are concrete expressions for the 1-loop determinants of 4d $\mathcal{N}=1$ vector and chiral multiplets on $\mathbb{D}^2\times\mathbb{T}^2$, which constitute the building blocks of integral expressions of gauge theory partition functions. For a vector multiplet of a gauge group $\text{G}$ with Cartan subalgebra $\mathfrak{h}$ we find
\be{\nn
\mathcal{Z}_\text{1-loop}^{\rm vec}(\Phi_0)
=
\left[\frac{\e^{-\frac{\i\pi}{3} P_{3}(0)}}{\text{Res}_{u=0}\Gamma(u;\tau,\sigma)}\right]^{\text{rk}(\text{G})}
{\det_\text{ad}}'\left[\frac{\e^{-\frac{\i\pi}{3} P_{3}(\Phi_{(0)})}}{\Gamma(\Phi_{(0)};\tau,\sigma)}\right]~,
}
where $\Phi_{(0)}\in\mathfrak{h}_\mathbb{C}$ is the zero mode of the gauge connection along the (anti-holomorphic) Killing vector field on the torus, $P_3(u)$ is a cubic polynomial arising from regularization, $\Gamma(u;\tau,\sigma)$ is the elliptic Gamma function with parameters $\tau,\sigma$ associated with the torus modulus and disk equivariant parameter respectively (that is the moduli of the complex structure), while the prime stands for excluding the zero roots (corresponding to Cartan generators). Similarly, for a chiral multiplet of R-charge $r$ in a representation $\mathcal{R}$ of the gauge group we find
\be{\nn
\mathcal{Z}^{\textrm{chi}(\text{D})}_{\textrm{1-loop}}(\Phi_{(0)})
 = \det_{\mathcal{R}}\left[
\dfrac{
	\e^{-\frac{\i\pi}{3} P_3(\sigma(1-r/2)-\Phi_{(0)})}
}{\Gamma(\sigma(1-r/2)-\Phi_{(0)};\tau,\sigma)
}
\right]
}
for Dirichlet conditions, and 
\be{\nn
\mathcal{Z}^{\textrm{chi}(\text{R})}_{\textrm{1-loop}}(\Phi_{(0)})
 = \det_{\mathcal{R}}\left[
	\e^{\frac{\i\pi}{3} P_3(\sigma  r/2 +\Phi_{(0)})} \,\Gamma(\sigma r/2+\Phi_{(0)};\tau,\sigma)
	\right]
	}
for Robin-like conditions. One can notice that the ratio of these results is a Jacobi Theta function (up to the exponential of a quadratic polynomial)
\be{\nn
\frac{\mathcal{Z}^{\textrm{chi}(\text{D})}_{\textrm{1-loop}}(\Phi_{(0)})}{\mathcal{Z}^{\textrm{chi}(\text{R})}_{\textrm{1-loop}}(\Phi_{(0)})}= \det_{\mathcal{R}}\left[\e^{-\i\pi P_2(\sigma r/2+\Phi_{(0)})}\Theta(\sigma r/2+\Phi_{(0)};\sigma)\right]~,
}
which can naturally be interpreted as the contribution of boundary degrees of freedom flipping the boundary conditions. This is an interesting prediction, which we are able to support by explicit computation of the 1-loop determinants of a pair of 3d $\mathcal{N}=1$ real multiplets on $\partial(\mathbb{D}^2\times\mathbb{T}^2)\simeq \mathbb{T}^3$, for which we find the r.h.s. of the ratio above.

\textbf{Organization of the paper}. The paper is organized as follows. In section~\ref{sec:geometry}, we collect relevant definitions and properties of the background geometry that we consider.
In section~\ref{sec:bulk-susy}, we provide a self-contained discussion of the supersymmetric multiplets, their actions and boundary terms.
In section~\ref{sec:localization}, we discuss the cohomological localization for vector and chiral multiplets, introducing the relevant complexes and evaluating the 1-loop determinants subject to boundary conditions. 
In section~\ref{sec:boundary}, we describe how bulk degrees of freedom split up into boundary multiplets and summarize how the surviving curved space supersymmetry algebra acts on the latter.
In section~\ref{sec:bc-flips}, we review how bulk fields can be decomposed into multiplets of the minimal supersymmetry preserved by the boundary, laying out the foundation for the analysis of the interplay between boundary matter and boundary conditions for bulk fields. In section~\ref{sec:applications}, we discuss the inclusion of some observables and classical terms in the theory which are consistent with our localization framework, and we also test our results for the gauge theory partition functions against Seiberg duality for SQCD. In section~\ref{sec:conclusions}, we conclude with a discussion of the main open questions that arose from our analysis and suggestions for future works. The paper is accompanied by several appendices, where we summarize our conventions and notations used in the main text as well as few side technical aspects of our analysis.

\section{Background geometry}\label{sec:geometry}
Let us start by reviewing the geometry of the class of supergravity backgrounds we are interested in, closely following \cite{Dumitrescu:2012ha,Closset:2013sxa,Assel:2014paa,Nishioka:2014zpa}. 
The purpose of this section is to provide a detailed description of the geometry from various angles.
We begin by expressing various geometric quantities such as the complex structure or the metric in terms of Killing spinor bilinears. This will be useful in the construction of supersymmetric actions and in the description of the cohomological complexes adopted for localization.
The choice of adapted complex coordinates will be useful for explicit computations of 1-loop determinants. An alternative choice of real coordinates will also be useful for dealing with the boundary. Our conventions are collected in appendix \ref{app:conventions} together with some useful identities.

\subsection{General aspects}
We consider Riemannian 4-manifolds with metric $g_{\mu\nu}$ admitting solutions to the following Killing spinor equations 
\be{
\begin{split}
&(\nabla_\mu-\i A_\mu+\i V_\mu+\i \sigma_{\mu\nu}V^\nu)\zeta=0~,\\
&(\nabla_\mu+\i A_\mu-\i V_\mu-\i \tilde\sigma_{\mu\nu}V^\nu)\tilde\zeta=0~,
\end{split}
}
where $A_\mu$ is a background Abelian connection for the R-symmetry line bundle $\mathscr{R}$, while $V_\mu$ is a background field satisfying $\nabla_\mu V^\mu=0$. The existence of solutions implies that the manifold has to be Hermitian, and we denote the Hermitian metric by $g_{\mu\nu}$. We focus on the case where two solutions $\zeta$ and $\tilde\zeta$ of R-charges $\pm 1$ and opposite chiralities exist. In this case, we can define the following fundamental vectorial Killing spinor bilinears\footnote{In our conventions, the non-zero contractions of the vectors evaluate to $2$ instead of $1/2$ as \cite{Closset:2013sxa}.}
\begin{align}
K^\mu &\equiv \tilde{\zeta} \tilde{\sigma}^\mu \zeta, \quad
\bar{K}^\mu \equiv \frac{1}{|\zeta|^2|\tilde{\zeta}|^2} \tilde{\zeta}^\dagger \tilde{\sigma}^\mu \zeta^\dagger~,\quad Y^\mu\equiv\frac{1}{|\tilde\zeta|^{2}}\tilde\zeta^\dagger\tilde\sigma^\mu\zeta~,\quad \bar{Y}^\mu \equiv -\frac{1}{|\zeta|^2} \tilde{\zeta} \tilde{\sigma}^\mu \zeta^\dagger~.
\end{align}
The vectors $Y^\mu$, $\bar Y^\mu$ have R-charge $\pm 2$, while $K^\mu$ is a complex Killing vector and we assume that $K^{\mu}$ commutes with its conjugate.\footnote{Note that, in general, $\bar K\neq K^\dagger$.} When there is no confusion, we will omit all the indexes to avoid cluttering. In particular, the dual 1-forms obtained by lowering the indexes with the metric will be denoted by the same symbols. Next, we define the following 2-form Killing spinor bilinears\footnote{We follow \cite{Closset:2013sxa} for the definition of $J,\tilde{J}$, which is minus that of \cite{Assel:2014paa}.}   
\be{
 J_{\mu\nu}\equiv -\frac{2\i}{|\zeta|^2}\zeta^\dagger\sigma_{\mu\nu}\zeta~,\quad  P_{\mu\nu} \equiv \zeta\sigma_{\mu\nu}\zeta~,\quad \tilde J_{\mu\nu}\equiv -\frac{2\i}{|\tilde\zeta|^2}\tilde\zeta^\dagger\tilde\sigma_{\mu\nu}\tilde\zeta~,\quad  \tilde P_{\mu\nu} \equiv \tilde\zeta\tilde\sigma_{\mu\nu}\tilde\zeta~.
}
Notice that the tensor $P$ is self-dual and carries R-charge $2$, while $\tilde P$ is anti-self-dual and carries R-charge $-2$. The tensor ${J^\mu}_{\nu}$ squares to $-1$, it is integrable and provides a complex structure.\footnote{A similar property applies to $\tilde J$.} In fact, the 2-form $J_{\mu\nu}$ is the K{\"a}hler form associated with the Hermitian metric. Notice that the vectors $K,Y$ are anti-holomorphic while $\bar K,\bar Y$ are holomorphic. In terms of the 1-form bilinears, we also have the expressions
\begin{align}\label{JsPs}
J&= -\frac{\i}{2} \left(K\wedge\bar{K} + Y\wedge\bar{Y}\right)~, & P &= \frac{1}{2} K \wedge Y~,\nn\\
\tilde{J} &= -\frac{\i}{2}\left(K\wedge\bar{K} - Y\wedge\bar{Y} \right)~, & \tilde{P} &= -\frac{1}{2} K \wedge \bar{Y}~.
\end{align} 
The 1-form Killing spinor bilinears provide an orthonormal frame, and the metric reads
\be{
g\equiv \frac{1}{2}\p{K\otimes \bar K+\bar K\otimes  K+Y\otimes \bar Y+\bar Y\otimes Y}~.
}
The volume form is taken to be
\be{
\textrm{vol}_4\equiv \frac{1}{2}J\wedge J=-\frac{1}{4}K\wedge  \bar K\wedge Y \wedge\bar Y~.
}

The Killing spinor equations also impose some integrability condition on the background fields, in particular 
\be{
R-6 V^\mu V_\mu=2J^{\mu\nu}F_{\mu\nu}=-2\tilde J^{\mu\nu}F_{\mu\nu}~,
}
where $R$ is the Ricci curvature (negative for the round sphere) and $F_{\mu\nu}$ is the R-symmetry field strength. Also, the existence of the Killing spinors, and hence a choice of the above geometric structures, can be used to determine the background fields $A_\mu$ and $V_\mu$, in particular
\be{
V_\mu=\frac{1}{2}\nabla_\nu {J^\nu}_\mu+ \kappa K_\mu~,
}
where  $\kappa$ is an arbitrary $K$-invariant function, namely $K^\mu\partial_\mu \kappa=0$. The R-symmetry connection can be similarly written in terms of the Chern connection, the complex structure and the Killing vector, but we will consider its explicit form later on in specific coordinates. 
Finally, given a connection 1-form $\A$ for an additional vector bundle, we denote the total covariant derivative by
\be{
\mathcal{D}_\mu\equiv\nabla_\mu-\i q_{\rm R}A_\mu-\i \A_\mu.~,
}
where the dot denotes the action in the appropriate representation and $q_{\rm R}$ is the R-charge. In fact, $\A$ will be the gauge connection for a gauge group $\text{G}$ with Lie algebra $\mathfrak{g}$ and choice of Cartan subalgebra $\mathfrak{h}$. For a given vector $v$, we define the (total) covariant derivative along $v$ by  
\be{
\mathcal{L}_v\equiv v^\mu\mathcal{D}_\mu~.
}
In particular, the Killing spinor equations imply the conservation laws   
\be{\label{cons}
\mathcal D_\mu K^\mu = 0, \quad 
\mathcal D_\mu \bar{K}^\mu = 0, \quad\mathcal D_\mu Y^\mu=0,\quad \mathcal D_\mu \bar Y^\mu=0~.
}

\subsection{Complex coordinates}\label{globalHol}

Under the  above assumptions, a complex 4-manifold admitting such geometric structures is a $\mathbb{T}^2$ fibration over a Riemann surface. Introducing local complex coordinates $(w,\bar w)$ on the torus and $(z,\bar z)$ on the base, the Hermitian metric can be locally written as 
\be{
\d\text{s}^2\equiv\Omega^2\left(|\d w+h\d z|^2+c^2\d z\d\bar z\right)~,
}
where $\Omega=\Omega(z,\bar z)$ and $c=c(z,\bar z)$ are nowhere vanishing real functions while $h=h(z,\bar z)$ is generically complex. This suggest to work in the holomorphic frame
\be{
\theta^1+\i\theta^2\equiv \Omega(\d w+h\d z)~,\quad \theta^3+\i\theta^4 \equiv \Omega c\, \d z~,
}
where $\theta^a$ form a real orthonormal frame and $a=1,2,3,4$ is a flat space index. In this frame, the Killing spinors we consider can be explicitly given as
\be{
\zeta_{\alpha}=\sqrt{\frac{s}{2}}\ \delta^+_{\alpha}~,\quad \tilde\zeta^{\dot\alpha}=-\frac{\Omega}{\sqrt{2 s}}\ \delta^{\dot\alpha}_{\dot +}~,
}
where $s$ is a nowhere vanishing global section of the canonical line bundle tensored with the square of the R-symmetry line bundle. In fact, since $P$ is a section of the R-symmetry line bundle $\mathscr{R}$ with R-charge $2$ and a section of the canonical bundle $\mathscr{K}$ of self-dual $(2,0)$-forms, it can be defined by 
\be{\label{eq:Pform}
P \equiv \frac{1}{2}s\sqrt[4]{g}\ \d w \wedge \d z ~,
}
with the square root of the determinant of the metric being $\sqrt{g}=c^2\Omega^4/4$. In particular, this shows that $s$ transforms by  phases under  holomorphic coordinate changes. Local R-symmetry transformations can then be used to make $s$, and hence the Killing spinors, scalar w.r.t. changes of adapted coordinates. The other Killing spinor bilinears  read 
\begin{equation}
K = \partial_{\bar{w}}, \qquad \bar{K} = \frac{4}{\Omega^2} \partial_w, \qquad
Y = \frac{2s}{\Omega^2 c}(\partial_{\bar{z}}- \bar{h}\partial_{\bar{w}}), \qquad
\bar{Y} = \frac{2s^{-1}}{c}(\partial_z - h \partial_w)~,
\end{equation}
and the dual 1-forms are 
\begin{equation}
K = \frac{1}{2} \Omega^2(\d w + h \d z), \quad
\bar{K} = 2 (\d\bar{w} + \bar{h} \d\bar{z}), \quad
Y = c s \, \d z, \qquad
\bar{Y} = \Omega^2 cs^{-1} \, \d \bar{z}~,
\end{equation}
while the volume form becomes 
\be{
\textrm{vol}_4=\textrm{vol}_{\mathbb{T}^2}\wedge\textrm{vol}_{\mathbb{D}^2}~,\quad \textrm{vol}_{\mathbb{T}^2}\equiv \i\frac{\Omega^2}{2}\d w\wedge\d \bar w~,\quad \textrm{vol}_{\mathbb{D}^2}\equiv \i\frac{\Omega^2c^2}{2}\d z \wedge\d \bar z~.
}
Finally, the R-symmetry connection in the holomorphic frame reads
\be{\label{ARhol}
A_\mu=-\frac{1}{2}{J_{\mu}}^{\nu}\partial_\nu\ln\sqrt[4]{g}-\frac{\i}{2}\partial_\mu\ln s+\frac{1}{4}\p{g_{\mu\nu}+\i J_{\mu\nu}}\nabla_\alpha J^{\alpha\nu}+\frac{3}{2}\kappa K_\mu~.
}

After the previous general and local analysis, let us next discuss the global properties of the background we are interested in. 
We consider the product of a torus $\T^2$ with a disk $\mathbb{D}^2$. 
The modular parameter $\tau\in\mathbb{H}^+$ of the torus determines the periodicities of the complex coordinate $w$  to be $w\sim w+2\pi\sim w+2\pi\tau$. 
The boundary of the disk is defined to be a circle located at $|z|=|z|_\partial$, corresponding to the actual boundary $\partial(\mathbb{D}^2\times\mathbb{T}^2)\simeq \mathbb{T}^3$. In later sections, we will make a few simplifying choices w.r.t the most general setup that has been discussed so far. 
In particular, we will consider the standard flat metric on the torus and the standard K{\"a}hler metric on the disk
\be{\label{eq:Kmetric}
\d \text{s}^2=\d w \d \bar w +\frac{4\, \d z \d\bar z}{(1+\epsilon|z|^2)^2}~,\qquad \Omega^2=1~,\quad c^2=\frac{4}{(1+\epsilon|z|^2)^2}~,\quad h=\bar h=0~,
}
where $\epsilon=\pm 1,0$ for the spherical, hyperbolic or flat space. In the following, we will adapt our discussion to the spherical metric, in which case we can consider a boundary at finite distance. Note that this metric posses an additional Killing vector corresponding to rotations parallel to the boundary of the disk generated by $-\i(z\partial_z-\bar z\partial_{\bar z})$. It will be convenient to consider the unit normalized real vector 
\be{\label{eq:tcoord}
 T\equiv -\frac{\i}{c|z|}(z\partial_z-\bar z\partial_{\bar z})~,
}
and introduce the normal direction to the boundary as its orthogonal (unit normalized) complement on the disk, namely 
\be{\label{eq:Nvec}
N\equiv J(T)= \frac{1}{c|z|}\, (z\partial_z+\bar z\partial_{\bar z})~.
}
Even though these are ill-defined at the origin of the disk, we only need them close to the boundary, and we will often use the symbol $\perp$ to denote the normal direction. As far as the background fields are concerned, the K{\"a}hler condition $\d J=0$ implies $V=\kappa K$, we may further specialize to $\kappa=0$, in which case the R-symmetry connection reduces to
\be{
A=-\frac{\i}{2(1+|z|^2)}\, (\bar z \d z-z\d \bar z)-\frac{\i}{2}\d \ln s~.
}
The R-symmetry connection is real provided that $|s|$ is constant, which is what we assume. From $\d A=\text{vol}_{\mathbb{D}^2}/2$, we see that this configuration gives a unit flux on the full sphere. We also consider a refinement of the background consisting of the following global identifications of the holomorphic coordinates \cite{Closset:2013sxa}
\be{\label{twist}
(w,z)\sim (w+2\pi,\e^{2\pi\i\alpha}z)\sim (w+2\pi\tau,\e^{2\pi\i\beta}z)~,
} with $\alpha,\beta\in\mathbb{R}$, $\alpha\sim \alpha+1$, $\beta\sim\beta+1$. The parameters $\tau$ and $\sigma\equiv \alpha\tau-\beta$ can be interpreted as complex structure moduli. 
Note that the 2-form (\ref{eq:Pform}) must be well-defined under the quotient (\ref{twist}), implying that $s$ must also be subject to the identifications  
\be{\label{eq:stransf}
s\sim \e^{-2\pi\i\alpha}\, s~,\qquad s\sim \e^{-2\pi\i\beta}\, s~,
}
where in each equation the two sides are evaluated at different points precisely as in (\ref{twist}).
As we have already mentioned, since $s$ is a nowhere-vanishing global section of $\mathscr{R}^2\otimes\mathscr{K}$ transforming by phases w.r.t. holomorphic coordinate changes, we can always offset these identifications by a suitable R-transformation so that $s$ (and hence the Killing spinors) behaves as a scalar w.r.t. changes of adapted coordinates. In other words, one can identify $\mathscr{R} \simeq \mathscr{K}^{-\frac{1}{2}}$ (up to a trivial line bundle). After that, since we restrict to constant $|s|$ in order for the R-symmetry connection to be real, we could treat $s$ as a constant as well, however it will be more convenient to retain it for bookkeeping purposes. Any other field $X$ of the theory carrying R-charge $q_\text{R}$ also acquires {\it twisted periodicities} under the identifications (\ref{twist}), namely
\be{\label{twistbc}
X\sim \e^{\i \pi q_\text{R} \alpha}\, X~,\qquad X\sim \e^{\i \pi q_\text{R} \beta}\, X~~.
}
For our background, $\mathscr{K}$ is topologically trivial, hence the R-charges do not need to be quantized. This is in contrast with the case where the disk is replaced by a compact sphere.

\subsection{Real coordinates}\label{globalReal}

It is also useful to have a description of the background in terms of real coordinates trivializing the identifications (\ref{twist}). This corresponds to another presentation of the background, instead of a disk fibered over a torus, we will here present it as a torus fibered over a disk.
For the torus we employ real coordinates $(x,y)$ subject to periodic identifications 
\be{
x\sim x+2\pi~,\qquad y\sim y+2\pi~.
}
Similarly, we can parametrize the unit disk through a radial coordinate $ \tan\vartheta/2$, $\vartheta\in[0,\pi/2]$  and an angular coordinate $\varphi\sim \varphi+2\pi$. Here we are adapting the notation to the hemisphere with polar angle $\vartheta$, and thus it is natural to place the boundary at $\vartheta=\pi/2$, namely at $|z|=1$ in complex coordinates. The change from real to complex coordinates is as follows
\begin{align}\label{coord1}
w &= x + \tau y~,& \bar w &= x+\bar \tau y~,\nn\\
z &= \tan \frac{\vartheta}{2} \, \e^{\i(\varphi + \alpha x + \beta y)} ~,& \bar z &= \tan \frac{\vartheta}{2} \, \e^{- \i(\varphi + \alpha x + \beta y)}~,
\end{align}
which inverts to
\begin{align}\label{coord2}
x &= \frac{\i}{2 \Im(\tau)}\left(w \, \bar \tau - \bar w \, \tau\right)~,&  y &= -\frac{\i}{2 \Im(\tau)}\left(w-\bar w\right) ~,\nn\\
\vartheta &= 2\arctan \sqrt{z\bar z}~, & \varphi &= -\frac{\i}{2}\log\frac{z}{\bar z}  - \frac{\i}{2\Im(\tau)}  \left(w \, \bar \sigma - \bar w \,  \sigma\right) ~,
\end{align}
where $\sigma = \tau\alpha - \beta$.
In real coordinates, the metric reads
\be{
\d \text{s}^2=(\d x+\text{Re}(\tau) \d y)^2+\text{Im}(\tau)^2\d y^2+\d\vartheta^2+\sin^2\vartheta(\d\varphi+\alpha\d x+\beta\d y)^2~,
}
while the R-symmetry connection becomes (up to flat connections)
\be{
A=\frac{1}{2}(1-\cos\vartheta)(\d\varphi+\alpha\d x+\beta\d y)~.
}
These expressions clarify that the refinement provided by $\alpha,\beta$ is equivalent to turning on a flat connection in the $\d\varphi$ direction, namely an equivariant deformation by the chemical potential $\sigma$ for the angular momentum on the disk. 
In fact, the Killing vector is a complex combination of the three real angles in geometry
\be{
K=-\frac{\i}{2\text{Im}(\tau)}(\tau\partial_x-\partial_y-\sigma\partial_\varphi)~,
}
and the twisted periodicities on the fields can be interpreted as shifting the effective angular momentum by the R-charge due to the presence of a magnetic field.\footnote{In real coordinates, it is more convenient to work with periodic fields along the torus and turning on a corresponding flat connection. 
While we have repeated all computations also in the real frame, here we only present them in the holomorphic frame, which is more elegant.} Finally, the normal and tangent vectors to the boundary introduced in (\ref{eq:tcoord}), (\ref{eq:Nvec}) are simply identified with 
\be{
N=\partial_\vartheta~,\qquad T=\frac{1}{\sin\vartheta}\partial_\varphi~.
}

\textbf{Remark}. When there is a boundary, an important piece of data is the transition between the bulk and boundary frames, and this is particularly delicate for spinorial objects. We refer to \cite{Hori:2013ika} for an exhaustive discussion on the disk in the context of supersymmetric localization. We discuss the geometry of the boundary and its relation to the bulk in more detail in section \ref{sec:boundary}.

\section{Bulk supersymmetry}\label{sec:bulk-susy}

Having discussed the background geometry in the previous section, we now turn to a description of the supersymmetry algebra preserved by this background.
We will first review how supersymmetry acts on 4d $\mathcal{N}=1$ vector and chiral multiplets in the supergravity background of the previous section.
This will set the stage for the discussion of supersymmetric actions and suitable boundary terms.

\subsection{Supersymmetry multiplets}

In Euclidean flat space, the 4d $\mathcal{N}=1$ supersymmetry algebra is generated by left and right handed supercharges $Q_\alpha$, $\tilde Q_{\dot\alpha}$ with R-charges $\mp 1$ satisfying
\be{
\{Q_\alpha,\tilde Q_{\dot\alpha}\}=2 \sigma^a_{\alpha\dot\alpha}\text{P}_a~,
}
where $\text{P}_a$ is the (covariant) momentum operator and $a=1,2,3,4$ a flat index.  In the curved background we are considering, we will denote by $\delta_\zeta$, $\delta_{\tilde\zeta}$ the odd supersymmetry transformations generated by the commuting Killing spinors $\zeta$, $\tilde\zeta$ respectively, and the supersymmetry algebra is broken down to the subalgebra
\begin{align}\label{deltaK}
\{\delta_\zeta,\delta_{\tilde\zeta}\}=& \ 2\i\delta_K~,\qquad \delta_K\equiv L_K-\i K^\mu A_\mu+G_{\Phi}~,\nn\\
\{\delta_\zeta,\delta_\zeta\}=& \ 
\{\delta_{\tilde\zeta},\delta_{\tilde\zeta}\}=[\delta_K,\delta_\zeta]=[\delta_K,\delta_{\tilde\zeta}]=0~,
\end{align}
where $L_K$ is the Lie derivative along $K$ and $G_{\Phi}$ represents an infinitesimal gauge transformation with parameter $\Phi$. Gauge transformations act on the gauge connection $\A$ and any other field $X$ as follows
\be{\label{gaugeT}
G_{\Phi}\mathcal{A}_\mu\equiv\mathcal{D}_\mu \Phi~,\quad G_{\Phi}X\equiv\i \Phi . X~,
}
where $\Phi.$ denotes the action in the relevant linear representation. It turns out that 
\be{
\Phi\equiv-\iota_K \A\equiv -K^\mu \A_\mu
} 
is a component of the gauge field itself. We also introduce the neutral combination 
\be{\label{eq:4dSUSYalggg}
\Q\equiv\delta_\zeta+\delta_{\tilde\zeta}~,\qquad \Q^2=2\i\delta_K~,
}
which we are going to use for localization. In particular, using the real coordinates for the simplified K{\"a}hler background described in section \ref{globalHol}, we can readily identify the Hamiltonian operator $\text{H}$ that appeared in the introduction of the paper with
\be{
\delta_K=0\quad \Rightarrow \quad \i\text{H}\equiv \text{P}_y=\tau \text{P}_x-\sigma \text{P}_\varphi+u~,
}
where $u$ can be an element in the Cartan of a global symmetry for which a background flat connection can be turned on. The almost perfect symmetry amongst the three boundary translation operators already hints at interesting modular properties of the partition functions.\footnote{From this viewpoint, it might be more useful to think about our index as a kind of elliptic genus for a theory quantized on $\mathbb{T}^2$. This can be made more explicit by restoring the term $\exp[-2\pi \text{Im}(\tau)\Q^2]$ under the trace defining the index. For a similar perspective in the compact case, we refer to \cite{Closset:2013sxa,Honda:2015yha}.} 

We can now move on to discuss how supersymmetry is realized on elementary fields. Even though for actual computations we are eventually interested in the simplified geometry where $\d J=V=0$, in this section we are going to discuss the general setup, unless otherwise stated.

\subsubsection{Vector multiplet}
A vector multiplet  $({\cal A},\lambda,\tilde\lambda,D)$ includes a gauge field, gauginos, and an auxiliary scalar  $D$ whose R-charges are $(0,1,-1,0)$. In Euclidean signature, $\lambda$ and $\tilde \lambda$ are independent Weyl spinors of opposite chirality. All fields are valued in the Lie algebra $\mathfrak{g}$ of the gauge group $\text{G}$. The field strength $\F\equiv \d_\A \A=\d\A-\i \A\wedge\A$ is
\be{
\cal F_{\mu\nu}= \partial_\mu \cal A_\nu-\partial_\nu \cal A_\mu-\i[\cal A_\mu,\cal A_\nu]~,
}
with Hodge dual $\tilde\F\equiv\star\F$,\footnote{We use the $\varepsilon$ tensor $\varepsilon_{w z \bar w \bar z}=\sqrt{|g|}\times1$.} that is 
\be{
\tilde\F_{\mu\nu}=\frac{1}{2}\varepsilon_{\mu\nu\rho\sigma}\F^{\rho\sigma}~.
}

\textbf{SUSY transformations}. The off-shell supersymmetry transformations are
\ale{\label{eq:vecsusytrans}
& & \Q \A_\mu &= \i\zeta\sigma_\mu\tilde \lambda+\i\tilde\zeta\tilde\sigma_\mu\lambda~,& &\nn\\
\Q\lambda&=\F_{\mu\nu}\sigma^{\mu\nu}\zeta+\i D\zeta~,& & &\Q\tilde\lambda&=\F_{\mu\nu}\tilde\sigma^{\mu\nu}\tilde\zeta-\i D\tilde\zeta~,\nn\\
& & \Q D &=-\zeta\sigma^\mu\mathcal{\hat D}_\mu\tilde\lambda+\tilde\zeta\tilde\sigma^\mu\mathcal{\hat D}_\mu\lambda~,& &
}
where 
\be{
\mathcal{\hat D_\mu}\lambda\equiv\mathcal{ D_\mu}\lambda+\i\frac {3}{2}V_\mu\lambda~,\qquad \mathcal{\hat D_\mu}\tilde\lambda\equiv\mathcal{ D_\mu}\tilde\lambda-\i\frac{3}{2}V_\mu\tilde\lambda~.
}

\subsubsection{Zero mode multiplet}

For later purposes, it is useful to separately study the multiplet of zero modes, which we now define.  This is motivated by the BPS localization locus that we will encounter later on. Since we are eventually interested in flat connections, it is useful to define the constant variables 
\be{
\Phi_{(0)}\equiv -\iota_K\A^{(0)}=- \A^{(0)}_{\bar w}~,\qquad \bar \Phi_{(0)}\equiv -\iota_{\bar K}\A^{(0)}= -\frac{4}{\Omega^2}\A^{(0)}_{w}~,
}
which coincide with the torus zero modes of the connection and  encode the commuting holonomies around its cycles. 
In fact, using the gauge freedom, we can restrict to constant configurations along the torus, therefore $\Phi_{(0)}$ and $\bar \Phi_{(0)}$ are proportional to the periods
\be{
\oint_y\A^{(0)}-\tau\oint_x\A^{(0)}~,\qquad
\oint_y\A^{(0)}-\bar\tau\oint_x\A^{(0)}~.
}
We can further assume that the holonomies can be simultaneously conjugated to the same Cartan torus,\footnote{For simplicity, we restrict to simply connected or unitary Lie groups.} hence effectively defining $\text{rk(G)}$ independent complex coordinates on the flat connection moduli space. Also, since the periods shift by $n_y-n_x\tau$ and $n_y-n_x\bar\tau$ respectively under the large gauge transformation $\A_{x,y}\to \A_{x,y}+2\pi n_{x,y}$,\footnote{With an abuse of notation, we denote by $n_x,n_y$ elements of the co-root system.} they are defined on a $\text{rk(G)}$-dimensional torus.  The K{\"a}hler background we are eventually interested in admits non-trivial fermionic zero modes coming from the Cartan gaugini 
\be{\label{cartangaugini}
\lambda=\lambda_{(0)}\in\mathfrak{h}_\mathbb{C}~,\quad \tilde\lambda=\tilde\lambda_{(0)}\in\mathfrak{h}_\mathbb{C}~,
}
which feel only the spin and R-symmetry connections. Therefore, they must be proportional to the Killing spinors. This is particularly simple to see in the case $V=0$, when the Killing spinors are annihilated by the Dirac operator. Altogether, the zero modes form a complete supermultiplet, and by defining the following constant fermionic scalar zero modes 
\be{
\Lambda^0_{(0)}\equiv \frac{\zeta^\dagger}{|\zeta|^2}\lambda_{(0)}~,\quad \tilde \Lambda^0_{(0)}\equiv \frac{{\tilde\zeta}^\dagger}{|\tilde\zeta|^2}\tilde\lambda_{(0)}~,
}
or equivalently
\be{
\Xi^0_{(0)}\equiv 2\i \left(\Lambda^0_{(0)}+\tilde\Lambda^0_{(0)}\right)~,\quad \Psi_{(0)}\equiv 2\i\left(\tilde\Lambda^0_{(0)}-\Lambda^0_{(0)}\right)~,
}
then the supersymmetry transformations are
\ale{\label{eq:vecsusytransZero}
\Q \Phi_{(0)} &= 0~, &\Q \bar \Phi_{(0)} &= -\Xi^0_{(0)}~,\nn\\
\Q\Psi_{(0)}&= \Delta_{(0)}~, & \Q \Delta_{(0)} &=0~,
}
where we have defined the constant (generically complex) combination 
\be{\label{eq:Delta0}
\frac{\Delta_{(0)}}{4}\equiv D_{(0)}-\frac{\i}{2}Y^\mu \bar Y^\nu \F^{(0)}_{\mu\nu}\in\mathfrak{h}_\mathbb{C}~.
}
Note that on the zero mode supermultiplet we have $\Q^2=2\i\delta_K=0$, and that the BPS configurations necessarily have $\Delta_{(0)}=0$, namely $\F_{(0)}=D_{(0)}=0$ upon imposing the usual reality conditions $\F_{(0)}\in\mathfrak{h}_\mathbb{R}$, $ D_{(0)}\in\i\mathfrak{h}_\mathbb{R}$.

\subsubsection{Chiral multiplet}
The 4d $\CN=1$ chiral multiplet $(\phi,\psi,F)$ contains a complex scalar, a Weyl spinor, and an auxiliary complex scalar, all transforming in the same representation of the gauge group~$\text{G}$.  
We will say that the multiplet has R-charge charge $r$ if its component fields have charges $(r,r-1,r-2)$. 
In Euclidean signature, each chiral multiplet is accompanied by an independent anti-chiral multiplet, with component fields denoted by $(\tilde\phi,\tilde\psi,\tilde F)$. These have opposite R-charges, and transform in the conjugate representation of $\text{G}$.

\textbf{SUSY transformations}. The off-shell supersymmetry transformations for a chiral multiplet coupled to a vector multiplet are
\ale{\label{eq:chisusytrans}
\Q \phi &=\sqrt2\,\zeta\psi~,\quad & \Q\tilde\phi &=\sqrt2\,\tilde\zeta\tilde\psi~, \nn\\
\Q\psi &=\sqrt2\,F\zeta+\i\sqrt2(\sigma^\mu\tilde\zeta)\mathcal D_\mu\phi~,\quad & \Q\tilde\psi &=\sqrt2\,\tilde F\tilde\zeta+\i\sqrt2(\tilde\sigma^\mu\zeta)\mathcal D_\mu\tilde\phi~,\nn\\
\Q F &= \i\sqrt2\,\tilde\zeta\tilde\sigma^\mu\mathcal{\hat D_\mu}\psi-2\i\tilde\zeta\tilde\lambda\phi~,\quad &
\Q \tilde F &= \i\sqrt2\,\zeta\sigma^\mu\mathcal{\hat D}_\mu\tilde\psi+2\i\,\tilde\phi \zeta\lambda~,
}
where\footnote{Using the Killing spinor equations we can also write $\tilde\zeta\tilde\sigma^\mu\mathcal{\hat D_\mu}\psi =\mathcal{D_\mu}(\tilde\zeta\tilde\sigma^\mu\psi)$, $\zeta\sigma^\mu\mathcal{\hat D}_\mu\tilde\psi=\mathcal{D_\mu}(\zeta\sigma^\mu\tilde\psi)$.}
\be{
\mathcal{\hat D_\mu}\psi\equiv\mathcal{ D_\mu}\psi-\frac{\i}{2}V_\mu\psi,\qquad \mathcal{\hat D_\mu}\tilde\psi\equiv\mathcal{ D_\mu}\tilde\psi+\frac{\i} {2}V_\mu\tilde\psi.
}

\subsection{Supersymmetric actions}

Having reviewed how the basic 4d $\CN=1$ multiplets transform under the supercharges preserved by our background, we now turn to supersymmetric actions, without superpotential terms. 
The inclusion of a superpotential will be discussed in section \ref{sec:potential} since it is most conveniently carried out in terms of twisted fields, to be introduced later on. Also, in order to avoid cluttering, for non-Abelian theories the  $\Tr$ in all the actions is left implicit. 

\subsubsection{Vector multiplet}
The usual Lagrangian for the vector multiplet is
\be{
\mathscr{L}_\text{vec}\equiv \f14\F_{\mu\nu}\F^{\mu\nu}-\f{D^2}2+ \f \i 2\lambda\sigma^\mu\mathcal{\hat D}_\mu\tilde\lambda+\f \i 2\tilde\lambda\tilde\sigma^\mu\mathcal{\hat D}_\mu\lambda~.
}
However, in the presence of a boundary the action
\be{
\mathscr{S}_\text{vec}\equiv \int\d^4 x\sqrt{g}\; \mathscr{L}_\text{vec}~
}
is generically neither $\Q$-exact nor supersymmetric due to boundary terms, namely
\be{
\mathscr L_\text{vec}=\Q\left(\cdots\right)-\nabla_\mu\left(\cdots\right)^\mu~,\qquad \Q\mathscr L_\text{vec}=\nabla_\mu\left(K^\mu\cdots\right)-\nabla_\mu\left(\Q\; \cdots\right)^\mu~.
}
Let us compute the boundary terms. We define an involution ${}^\vee$ acting  as $^{\dagger}$ on Killing spinors, background geometric quantities and $\mathbb{C}$-numbers, while we will specify its action on dynamical fields momentarily. We define the following fermionic functionals 
\be{\label{eq:V-vec}
\mathscr{V}_\text{vec}\equiv\frac{1}{4|\zeta|^2}(\Q\lambda)^\vee\lambda~,\qquad \tilde{\mathscr{V}}_\text{vec}\equiv\frac{1}{4|\tilde\zeta|^2}(\Q\tilde\lambda)^\vee\tilde\lambda~,
}
and compute the variations
\begin{align}
\Q\lambda&=\F_{\mu\nu}\p{\zeta\frac{\i}{2}J^{\mu\nu}-\frac{\zeta^\dagger}{|\zeta|^2}P^{\mu\nu}}+\i D\zeta~,\nn\\
(\Q\lambda)^\vee&=-\i\p{\frac{1}{2}J^{\mu\nu}\F_{\mu\nu}^\vee+D^\vee}\zeta^\dagger+|\zeta|^2 \bar K^\mu \bar Y^\nu \F_{\mu\nu}^\vee \zeta~,\nn\\
\Q\tilde\lambda&=\F_{\mu\nu}\p{\tilde\zeta\frac{\i}{2}\tilde J^{\mu\nu}-\frac{\tilde\zeta^\dagger}{|\tilde\zeta|^2}\tilde P^{\mu\nu}}-\i D\tilde\zeta~,\nn\\
(\Q\tilde\lambda)^\vee&=-\i\p{\frac{1}{2}\tilde J^{\mu\nu}\F^\vee_{\mu\nu}-D^\vee}\tilde\zeta^\dagger-|\tilde\zeta|^2\bar K^\mu Y^\nu \F_{\mu\nu}^\vee\tilde\zeta~,
\end{align}
where we used the  identities $\zeta^\dagger \sigma^{\mu\nu}\zeta^\dagger=|\zeta|^4 \bar K^{[\mu}\bar Y^{\nu]}$, $\tilde\zeta^\dagger \tilde\sigma^{\mu\nu}\tilde\zeta^\dagger=|\tilde\zeta|^4 \bar K^{[\mu} Y^{\nu]}$. Then the bosonic parts of the Lagrangians $\Q \mathscr{V}_\text{vec}$, $\Q \tilde{\mathscr{V}}_\text{vec}$ read
\begin{align}\label{eq:vec-action-bos}
\Q \mathscr{V}_\text{vec}\Big|_\text{B}=\frac{1}{8}\F^\vee_{\mu\nu}(\F_{\mu\nu}+\tilde \F_{\mu\nu})+\frac{1}{4}D^\vee D+\frac{1}{8}J^{\mu\nu}(\F^\vee_{\mu\nu}D+\F_{\mu\nu}D^\vee)+\i {J^\mu}_{\rho} \F^\vee_{\mu\nu}\F^{\nu\rho}~,\nn\\
\Q \tilde{\mathscr{V}}_\text{vec}\Big|_\text{B}=\frac{1}{8}\F^\vee_{\mu\nu}(\F_{\mu\nu}-\tilde \F_{\mu\nu})+\frac{1}{4}D^\vee D+\frac{1}{8}\tilde J^{\mu\nu}(\F^\vee_{\mu\nu}D+\F_{\mu\nu}D^\vee)+\i \tilde{J}^\mu_{\,~\rho} \F^\vee_{\mu\nu}\F^{\nu\rho}~.
\end{align}
If we choose the involution to act as $(\A,D)^\vee = (\A,-D)$, then in the summation of the two contributions most of the terms will cancel out leaving us simply with
\be{
\Q (\mathscr{V}_\text{vec}+ \tilde{\mathscr{V}}_\text{vec})\Big|_\text{B}=\mathscr{L}_\text{vec}\Big|_\text{B}~.
}
Similarly, for the fermionic parts after this choice we obtain
\be{\label{eq:vec-action-fermi}
\Q \mathscr{V}_\text{vec}\Big|_\text{F}=\f \i 2 \lambda \sigma^\mu\mathcal{\hat D}_\mu\tilde\lambda  - \f \i 4 \bar Y^\mu\mathcal D_\mu\p{\lambda\lambda}~,\qquad \Q \tilde{\mathscr{V}}_\text{vec}\Big|_\text{F}=\f \i 4 Y^\mu\mathcal D_\mu(\tilde\lambda\tilde\lambda)+\f \i 2\tilde\lambda\tilde\sigma^\mu\mathcal{\hat D}_\mu\lambda~,
}
and using the conservation (\ref{cons}) we get
\be{
\Q(\mathscr{V}_\text{vec}+\tilde{\mathscr{V}}_\text{vec})=\mathscr{L}_\text{vec}+\nabla_\mu \mathscr{B}^\mu_\text{vec}~,\qquad \mathscr{B}^\mu_\text{vec}\equiv \frac{\i}{4}\left(\bar Y^\mu \lambda\lambda+Y^\mu \tilde\lambda\tilde\lambda\right)~.
}
Note that the other $\Q$-exact combination is a boundary Lagrangian
\be{
\Q (\mathscr{V}_\text{vec}-\tilde{\mathscr{V}}_\text{vec})=\f18\epsilon^{\mu\nu\rho\sigma}\F_{\mu\nu}\F_{\rho\sigma} + \nabla_\mu\p{\frac{\i}{2}\lambda\sigma^\mu\tilde\lambda}+\mathcal L_{\tilde Y}\p{\frac{\i}{4}\lambda\lambda}-\mathcal L_{ Y}\p{\frac{\i}{4} \tilde\lambda\tilde\lambda}~,
}
which reduces to a 3d $\mathcal{N}=1$ Chern-Simons term on the boundary.

\textbf{Remark}. In order to properly define the theory, we need to discuss the allowed boundary conditions. 
We will do that when computing the partition function of the theory through localization. The Lagrangian that we will use differs from the usual one by boundary terms, 
and to apply our framework to the latter one would need to additionally restrict to those boundary conditions that also kill the normal component of $\mathscr{B}_\text{vec}^\mu$ (and its ${\cal Q}$-variation if some supersymmetry has to be preserved on the boundary). Since our Lagrangian is ${\cal Q}$-exact and  the supercharge squares to an isometry generated by a Killing vector parallel to the boundary, supersymmetry alone does not enforce a choice of boundary conditions, leaving the freedom to explore different choices.

\subsubsection{Chiral multiplet}\label{sec:chiral-action}
The usual  Lagrangian for the chiral multiplet coupled to a vector is
\begin{multline}\label{Vchi}
	\mathscr L_\text{chi}
		\equiv
	\mathcal D_\mu\tilde\phi\mathcal D^\mu\phi+\tilde\phi\p{D+\frac{r}{4}R}\phi
	-\tilde FF
	+\i\tilde\psi\,\tilde\sigma^\mu\hat\D_\mu\psi\, 
	+\\
	+\i\sqrt2\p{\tilde\phi\lambda\psi - \tilde\psi\tilde\lambda\phi}+
	\i V_\mu\p{\phi\mathcal D^\mu\tilde\phi-\tilde\phi\mathcal D^\mu\phi-\i \frac{3r}{2}V^\mu\tilde\phi\phi}~.
\end{multline} 
However, in general the action
\be{
\mathscr S_\text{chi}\equiv\int\d^4 x\sqrt{g}\;\mathscr L_\text{chi}
}
is not supersymmetric nor $\Q$-exact due to boundary terms, namely
\be{
\mathscr L_\text{chi}=\Q\left(\cdots\right)-\nabla_\mu\left(\cdots\right)^\mu~,\qquad \Q\mathscr L_\text{chi}=\nabla_\mu\left(K^\mu\cdots\right)-\nabla_\mu\left(\Q\; \cdots\right)^\mu~.
}
Let us compute the boundary terms. We define the following fermionic functionals
\begin{align}\label{Vpsi}
\mathscr{V}_\text{chi}&\equiv\f1{2|\zeta|^2}\left((\delta_\zeta \psi)^\vee\psi+(\delta_\zeta\tilde\psi)^\vee\tilde\psi\right)=\sqrt{2}F^\vee \zeta^\dagger \psi +\sqrt{2}\left(\i\tilde\zeta\L_Y\tilde\phi-\i\frac{\tilde\zeta^\dagger}{|\tilde\zeta|^2}\L_K\tilde\phi\right)^\vee \tilde\psi~,\nn\\
\tilde{\mathscr{V}}_\text{chi}&\equiv\f1{2|\tilde\zeta|^2}\left((\delta_{\tilde\zeta} \tilde\psi)^\vee\tilde\psi+(\delta_{\tilde\zeta}\psi)^\vee\psi\right)=\sqrt{2}\tilde F^\vee \tilde\zeta^\dagger\tilde\psi+\sqrt{2}\left(-\i\zeta\L_{\bar Y}\phi-\i\frac{\zeta^\dagger}{|\zeta|^2}\L_K\phi\right)^\vee\psi~,\\
	\label{eq:V-lambda}
	\mathscr{V}_\lambda
	&\equiv-\i\frac{\zeta^\dagger}{|\zeta|^2}\tilde\phi\lambda\phi~,\qquad \qquad \quad ~~~~ \tilde{\mathscr{V}}_{\lambda}\equiv\i\frac{{\tilde\zeta}^\dagger}{|\tilde\zeta|^2}\tilde\phi\tilde\lambda\phi~,\nn\\
\quad \mathscr{V}_\kappa&\equiv-\f{\phi}{\sqrt2|\zeta|^2}\,\kappa K_\mu\,\zeta^\dagger\sigma^\mu\tilde\psi~,\qquad \tilde{\mathscr{V}}_\kappa \equiv\f{\tilde\phi}{\sqrt2|\tilde\zeta|^2}\,\kappa K_\mu\,\tilde\zeta^\dagger\tilde\sigma^\mu\psi~.
\end{align}
We let the involution introduced previously to act as $(\phi,F)^\vee=(\tilde \phi,-\tilde F)$ and assume a real R-symmetry connection so that $A^\vee = A$. The  variations of the fermionic functionals yield 
\ale{
\delta_\zeta\p{\mathscr{V}_\text{chi}+\mathscr{V}_\lambda+\mathscr{V}_\kappa}&=\mathscr{L}_\text{chi}+\nabla_\mu \mathscr{B}^{\mu}_\text{chi}~,\quad &\delta_{\tilde\zeta}\p{\mathscr{V}_\text{chi}+\mathscr{V}_{\lambda}+\mathscr{V}_\kappa}&=0~,\nn\\
\delta_\zeta \p{\tilde{\mathscr{V}}_\text{chi}+\tilde{\mathscr{V}}_{\lambda}+\tilde{\mathscr{V}}_\kappa}&=0~,\quad & \delta_{\tilde\zeta}\p{\tilde{\mathscr{V}}_\text{chi}+\tilde{\mathscr{V}}_{\lambda}+\tilde{\mathscr{V}}_\kappa}&=\mathscr{L}_\text{chi}
+\nabla_\mu \tilde{\mathscr{B}}^{\mu}_\text{chi}~,\label{deltaV}
}
where
\be{\label{Bmu}
\begin{split}
\mathscr{B}^{\mu}_\text{chi}&\equiv\i\tilde\phi\phi\p{2\kappa K^\mu-V^\mu}-\f \i2\tilde\psi\tilde\sigma^\mu\psi-\f12 J^{\mu\nu}\tilde\psi\tilde\sigma_\nu\psi-\i J^{\mu\nu}\tilde\phi\mathcal D_\nu\phi~,\\
\tilde{\mathscr{B}}^{\mu}_\text{chi}&\equiv-\i\tilde\phi\phi\p{2\kappa K^\mu-V^\mu}-\f \i2\tilde\psi\tilde\sigma^\mu\psi+\f12 \tilde J^{\mu\nu}\tilde\psi\tilde\sigma_\nu\psi-\i\tilde J^{\mu\nu}\phi\mathcal D_\nu\tilde\phi~.
\end{split}
}

\textbf{Remark}. In order to properly define the theory, we need to discuss the allowed boundary conditions. We will do that when computing the partition function of the theory through localization.
As for the vector multiplet, let us reiterate that the Lagrangian that we are going to use for localization differs by boundary terms from the usual one. Therefore, in order to apply our framework to the usual Lagrangian one would need to additionally check that the boundary conditions also kill the the normal components of $\mathscr{B}_\text{chi}^\mu$, $\tilde{\mathscr{B}}_\text{chi}^\mu$  (and their ${\cal Q}$-variations if some supersymmetry has to be preserved on the boundary).

\section{Localization}\label{sec:localization}

We have now collected all the necessary ingredients for the off-shell Lagrangian formulation of 4d $\CN=1$ theories on a manifold with boundary. In this section, we will employ the previous analysis to set up a computation of the path integral by localization. After some general remarks on the idea behind the cohomological approach to localization, we will discuss a preliminary reduction of the functional integral to a contour integral over the bosonic zero modes of the vector multiplet.
We will then introduce the localization Lagrangians for vector and chiral multiplets and discuss the BPS locus defined by each of them.
The next step will be to switch to twisted fields in order to make the structure of cohomological complexes manifest, both  for vector and chiral multiplets.
The identification of the complexes will then allow us to proceed with an explicit evaluation of the 1-loop determinants, taking into account the choice of boundary conditions.
We will conclude the section with a brief discussion of the peculiar modular properties of the partition functions.

\subsection{Preliminaries on the cohomological approach}
The path integral we wish to compute takes the schematic form
\be{
Z=\int[\d X]\, \e^{-\mathscr{S}[X]}~,
}
where we have generically denoted by $X$ the collection of all quantum fields. We also recall that the combination $\Q$ of supercharges  introduced in (\ref{deltaK}) represents an odd symmetry which is preserved even in the presence of the boundary. The localization principle relies on the fact that, upon very general and mild conditions,\footnote{Incidentally, we will soon see that in our case a very careful analysis of the $\Q$-exact terms is needed.}  $\mathcal{Q}$-exact deformations of the supersymmetric measure do not modify the path integral (we refer to \cite{Pestun:2016zxk} for an exhaustive review of the subject, here we simply recall the main points).  Therefore, one can either work in the semiclassical limit  if the defining action is already $\mathcal{Q}$-exact, or one can consider $\mathcal{Q}$-exact deformations 
\be{
\mathscr{S}\to \mathscr{S} +\frac{1}{e^2}\mathcal{Q}(\cdots)_\textrm{vec}+\frac{1}{\textsl{g}^2}\mathcal{Q}(\cdots)_\textrm{chi}~
}
and then take the semiclassical limit $e^2,\textsl{g}^2\to 0$, which becomes an exact approximation. The crucial requirement is that the localizing action has positive semi-definite bosonic part along the integration contour in the complexified field space. Upon such procedure, the final result can be expressed in terms of a matrix-like integral of classical and 1-loop contributions over the moduli of the localizing equations, usually given by constant field configurations, namely\footnote{In this paper, we do not have to deal with non-perturbative saddles. Also, we focus on the so-called {\it Coulomb branch} localization scheme. In principle, {\it Higgs branch} localization would also be possible and interesting to analyze, see e.g. \cite{Benini:2012ui,Doroud:2012xw,Fujitsuka:2013fga,Benini:2013yva,Peelaers:2014ima,Pan:2014bwa,Closset:2015rna,Pan:2015hza,Gomis:2016ljm,Pan:2016fbl}.}
\be{
Z=\int\d X_{(0)}\, \e^{-\mathscr{S}[X_{(0)}]}\, \mathcal{Z}_\text{1-loop}(X_{(0)})~.
}
Different techniques have been developed to compute 1-loop determinants, from brute force diagonalization of kinetic operators to sophisticated index theorems. Our approach will be based on the cohomological reformulation of the supersymmetry transformations.\footnote{For recent work in the context of 4d $\mathcal{N}=2$ theories, see e.g. \cite{Festuccia:2018rew,Festuccia:2019akm}.} The key observation is that the supercharge $\mathcal{Q}$ can be identified with an equivariant differential acting on the field supermanifold, which can be in turn neatly divided into base ($\varphi$) and fiber $(\varphi')$ coordinates forming an equivariant cohomological complex
\be{
X= (\varphi,\varphi')~,\quad \mathcal{Q}\varphi=\varphi'~,\quad \mathcal{Q}\varphi'=2\i\delta_K \varphi~.
}
After this identification, one can linearize the cohomological complex around the localization locus and 
integrate out the Gaussian fluctuations, obtaining the reduction
\be{\label{eq:cohdet}
Z=\int\d \varphi_{(0)}\, \e^{-\mathscr{S}[\varphi_{(0)}]}\, \mathcal{Z}_\text{1-loop}(\varphi_{(0)})~,\qquad
\mathcal{Z}_\text{1-loop}(\varphi_{(0)})=\sqrt{\text{sdet}_{\varphi} 2\i\delta^{(0)}_K}~,
}
which allows one to bypass the brute force diagonalization of the relevant kinetic operators. In fact, this result does not even require a choice of the $\Q$-exact deformation terms and relies only on supersymmetry, but it is a good practice to define these terms as the field space is infinite dimensional and some care may (and will) be needed. Another advantage of this approach is that, along the way, one naturally discovers the operators which pair up the bosonic and fermionic modes, which may be used to simplify the  problem even further. 

The main goal of this section is to compute the determinant (\ref{eq:cohdet}) for the vector and chiral multiplets. However, we should stress that there are several subtleties in the blind application of the localization principle to our setup, such as the presence of fermionic zero modes, singularities and divergences in the quantum measure at special loci as well as the identification of the correct integration cycle(s) in field space. These difficulties have already appeared, and have been extensively studied, in the literature \cite{Benini:2013nda,Benini:2013xpa,Benini:2015noa,Benini:2016hjo,Closset:2015rna}. 
However, most works focus on lower-dimensional theories, or on spaces without boundaries. In the following, we will adapt those results to our setup, emphasizing the additional working assumptions, similarities and differences, while we refer to the original references for a full account.

\subsection{Reduction to a contour integral}\label{sec:contours}

Before embarking on the proper localization of the path integral and the computation of the 1-loop determinants, we would like to discuss two main sources of subtleties: 
\begin{enumerate}
	\item The presence of fermionic zero modes ($\Xi^0_{(0)},\Psi_{(0)}$), making the direct application of the cohomological localization approach non-obvious.
	\item The presence of singularities at special loci ($\Phi_{(0)}=\Phi^\text{sing.}_{(0)}$), making the analysis of the Gaussian integration around the localization locus and the choice of integration contour very subtle. 
\end{enumerate}
Here we argue that these issues are related and that can be cured together to obtain a concrete answer for the partition function, along the lines of \cite{Benini:2015noa,Benini:2016hjo,Closset:2015rna}. In fact, the original analysis of \cite{Benini:2013nda,Benini:2013xpa} applies essentially unchanged to our background too, with the  modification of taking into account the disk modes. 
However, for the sake of completeness 
let us briefly recall how one can arrive at the expression  (\ref{eq:cohdet}), focusing on the Abelian case which is much simpler and eventually generalizes to the non-Abelian case too.

Even before any application  of the localization machinery, the partition function of the whole theory can schematically be represented as an integral over the zero mode supermultiplet, and we can thus write
\be{
Z=\int  \d \Phi_{(0)}\, \d \bar \Phi_{(0)}\,\d \Delta_{(0)}\, \d \Xi^0_{(0)}\, \d \Psi_{(0)}\, \mathcal{Z}(\Phi_{(0)},\bar \Phi_{(0)},\Delta_{(0)},\Xi^0_{(0)},\Psi_{(0)}) ~,
}
where $\mathcal{Z}$ is the result of path integration over all the field configurations but the zero modes. The integration over the fermionic modes can be performed using a shortcut exploiting supersymmetry. Using the property $\Q \Phi_{(0)}=0$, one can deduce that 
\be{
\Q \mathcal{Z}=0\qquad \Rightarrow \qquad  \Delta_{(0)}\frac{\partial^2}{\partial\Xi^0_{(0)}\partial \Psi_{(0)}}\mathcal{Z}\Big|_{\Xi^0_{(0)}=\Psi_{(0)}=0}=\frac{\partial}{\partial \bar \Phi_{(0)}}\mathcal{Z}\Big|_{\Xi^0_{(0)}=\Psi_{(0)}=0}~,
}
and hence 
\be{
Z=\int\d \Phi_{(0)}\, \d \bar \Phi_{(0)}\,  \frac{\partial}{\partial \bar \Phi_{(0)}}\int\frac{\d \Delta_{(0)}}{\Delta_{(0)}}\, \mathcal{Z}(\Phi_{(0)},\bar \Phi_{(0)},\Delta_{(0)}) ~,
}
where the specialization $\Xi^0_{(0)}=\Psi_{(0)}=0$ is to be understood.  Now  the integrand on the r.h.s. can be computed exactly  by localization, namely in the limit $e^2,\textsl{g}^2\to 0$. While the $\textsl{g}^2\to 0$ limit does not pose serious problems, the limit $e^2\to 0$ must be taken with some care because the integrand may develop singularities at certain points in the $\Phi_{(0)},\bar \Phi_{(0)}$ plane where chiral multiplets develop scalar massless modes.

\textbf{Remark}. Intuitively, the dangerous loci come from the scalar zero modes of the kinetic operator on which $\Q^2$ acts diagonally. In fact, for the modes $|\mathcal{L}_{\bar Y}\phi|^2=0$ the only damping term is the mass $|\Phi_{(0)}-\Phi^\text{sing.}_{(0)}|^2|\phi|^2$, which would vanish at $\Phi_{(0)}=\Phi^\text{sing.}_{(0)}$. As expected, these points are exactly the poles (mod $\mathbb{Z}\tau-\mathbb{Z}$) of the meromorphic 1-loop determinant we will compute later on. 

As a partial fix, one can make sure that there is the usual D-term coupling between the chirals and the vector, which is easy to fulfill by including in the localization action the $\Q$-exact term $\Q(\mathscr{V}_\lambda+\tilde{\mathscr{V}}_\lambda)$ from (\ref{Vpsi}). This will generate a quartic potential providing a damping even at the dangerous locus. Unfortunately, this is true only as long as $e^2\neq 0$. The recipe to control this bad behavior is to keep $\Delta_{(0)}$ as a regulator, and remove a small tubular neighborhood of the singular loci, of size $\varepsilon$ shrinking faster than any power of $e^2$. With this double scaling limit in mind, one is left with the contour integral
\be{
Z=\oint\d \Phi_{(0)}\,\oint\frac{\d \Delta_{(0)}}{\Delta_{(0)}}\, \mathcal{Z}(\Phi_{(0)},\bar \Phi_{(0)},\Delta_{(0)}) ~,
}
which still needs to be properly defined.\footnote{We are already using that after the $\Delta_{(0)}$ integral, the integrand has to be meromorphic in $\Phi_{(0)}$.} Now the disk kinetic action includes a mass term $(|\Phi_{(0)}-\Phi^\text{sing.}_{(0)}|^2+\Delta_{(0)}/2)|\phi|^2$, so the (towers of) singularities in the complex \mbox{$\Delta_{(0)}$-plane} become quite involved and one should specify the integration contour such that the scalar modes do not develop a tachyonic potential. The prescription that does the job is as follows: $i$) the contour of $-\i \Delta_{(0)}$ must asymptote the real axis and passes either slightly above or below the obvious singularity at $\Delta_{(0)}=0$; $ii$) assuming the chiral multiplets have (effective) R-charges all of the same sign, one splits the singularities to lie on either side of the contour according to the sign of their gauge charges. In the double scaling limit $\varepsilon\ll e^2\to 0$, the poles in the $\Delta_{(0)}$-plane will start moving around, towards the real axis and the origin. However, as long as the contour can be shifted to avoid possible collisions, the integral is well defined and in fact trivial because of the smoothness of the integrand and the Gaussian damping. However, a pole which moves towards the origin from the same side as the contour will inevitably pinch it against the pole at the origin. 
The contour must then cross the origin, picking up the residue at the unique BPS pole $\Delta_{(0)}=0$. This is the net contribution to the path integral, yielding the final expression
\be{
Z=\oint_\text{J.K} \d \Phi_{(0)}\, \mathcal{Z}_\text{BPS}(\Phi_{(0)}) ~,
}
where we have set $\mathcal{Z}(\Phi_{(0)},\bar \Phi_{(0)},\Delta_{(0)})|_{\Delta_{(0)}=0}\equiv \mathcal{Z}_\text{BPS}(\Phi_{(0)})$ since this is the meromorphic 1-loop determinant around the BPS saddles, and the subscript simply means that this pole prescription is equivalent to using the Jeffrey-Kirwan cycle/residue. The non-Abelian generalization is much more involved but the final result is the same, with the J.K. contour being a specific middle-dimensional cycle in the complex bosonic moduli space. 

\textbf{Remark}. We expect that on the disk there should be multiple (in fact, a basis of) integration cycles (roughly speaking, corresponding to different vacua as described in \cite{Beem:2012mb}). While the J.K. cycle arises from the above arguments, there are more choices that one may take. In particular, in the case of diverse chiral multiplets supporting a flavor symmetry, one can first resolve the degeneracies by turning on distinct background flat connections, and then one may consider contours going around the singularities associated with a selected multiplet $\gamma$, namely
\be{\label{eq:contours}
Z\to Z_\gamma=\oint_\gamma \d \Phi_{(0)}\, \mathcal{Z}_\text{BPS}(\Phi_{(0)}) ~.
} 
The basis of integration contours can be determined after the computation of the integrand.

\subsection{BPS localization locus}\label{sec:BPSlocus}

Having laid out some of the subtleties involved in the computation, we now turn to the characterization of the BPS locus.
In the cohomological localization approach, the computation of $\mathcal{Z}_\text{BPS}(\Phi_{(0)})$ amounts to the evaluation of two quantities: $i)$ the classical action on the ``trivial" BPS localization locus, in which $\Psi_{(0)}=\Xi^0_{(0)}=\Delta_{(0)}=0$; $ii)$ the 1-loop determinant of Gaussian fluctuations around this background. The BPS locus is determined by a suitable choice of $\Q$-exact localization actions, which we now construct.

\textbf{Vector multiplet}. As a localizing action  we choose the $\Q$-exact term
\be{\label{eq:Svecloc}
\mathscr{S}^{\rm loc}_{\rm vec}\equiv \Q\int\d^4 x\,\sqrt{g}\;\mathscr{V}^{\rm loc}_{\rm vec}~,\qquad \mathscr{V}^{\rm loc}_{\rm vec}\equiv \mathscr{V}_{\rm vec}+\tilde{\mathscr{V}}_{\rm vec}~,
}
with $\mathscr{V}_{\rm vec}$ and $\tilde{\mathscr{V}}_{\rm vec}$ defined in (\ref{eq:V-vec}).
Thanks to $\Q^2=2\i L_K$ on a neutral scalar, this action is automatically supersymmetric even in the presence of a boundary and regardless the boundary conditions if the Killing vector is tangent to the boundary, which is indeed our case. Notice that the bosonic part is also manifestly positive semi-definite on the {\it real contour} $(\A,D)^\dagger=(\A,-D)=(\A,D)^\vee$, i.e. along $D\in\i\mathbb{R}$ and $\A\in\mathbb{R}$.\footnote{Notice that the path integral we started with makes sense as an integral over a complexified field space along a suitable contour. In fact, the SUSY transformations do not generically preserve the reality conditions one is interested in. As usual, the action of the involution ${}^\vee$ must be taken as a definition, and it coincides with Hermitian conjugation only on the real contour. This is manifestly convergent but other choices may be possible.} The localization locus is thus given by the BPS equations
\be{
\delta_\zeta\lambda=\delta_{\tilde\zeta}\tilde\lambda=0~,
}
leading to
\be{
P^{\mu\nu}\F_{\mu\nu}= \tilde P^{\mu\nu}\F_{\mu\nu}=K^\mu\bar K^\nu\F_{\mu\nu}=0,\quad \frac{\Delta}{4}\equiv D-\frac{\i}{2}Y^\mu\bar Y^\nu\F_{\mu\nu}=0~.
}
On the real contour, these are simply equivalent to $\mathcal{F}=\Delta=0$. Therefore, the  BPS locus is given by flat connections (mod gauge transformations)
\be{
\mathcal{M}^\text{BPS}_\text{vec}\simeq \left\{ \F=\Delta=\lambda=\tilde\lambda=0\right\}/G~.
}
These are classified by the commuting holonomies  along $\mathbb{T}^2$, encoded by $\Phi_{(0)},\bar \Phi_{(0)}$, which can be assumed simultaneously conjugated to the same Cartan torus.\footnote{We remind that we restrict to simply connected or unitary Lie groups.} In fact, using the gauge freedom, the BPS locus can be reduced to the complexified Cartan torus (mod the Weyl group)
\be{\label{eq:MBPSvec}
\mathcal{M}^\text{BPS}_\text{vec}\simeq\left\{ \e^{2\pi\i \Phi_{(0)}},\e^{2\pi\i\bar \Phi_{(0)}}\in\e^{\i\mathfrak{h}_\mathbb{C}},\Delta=\lambda=\tilde\lambda=0\right\}/W~.
}

\textbf{Remark}.  Relaxing the reality condition on the contour of integration for $D$ allows for a much larger set of BPS solutions, which can be called \textit{flux configurations}. Taking a contour parallel to the imaginary axis, but not passing through the origin, the new BPS loci correspond to
\be{\label{eq:BPSflux}
0=\Re\left(\frac{\Delta}{4}\right)= \Re(D)- \frac{\i}{2}Y^\mu\bar Y^\nu\F_{\mu\nu}~,\quad 0=\Im\left(\frac{\Delta}{4}\right)= \Im(D)~.
}
The configurations with $\Re(\Delta)\neq 0$ are non-BPS, and therefore they are expected not to contribute in the limit $e^2\to 0$. However, configurations with constant $\Delta=\Delta_{(0)}$ and constant disk flux proportional to $\iota_{\bar Y}\iota_{Y}\F_{(0)}$ would deserve a separate analysis because they pass through the BPS point and can reach the real contour \mbox{$D\in\i\mathbb{R}$}. As pointed out in \cite{Benini:2015noa,Closset:2015rna}, these non-trivial non-BPS flux saddles play indeed an important role in A-twisted theories on compact spaces. In our case, since we are mainly interested in preserving same gauge symmetry at the boundary through Neumann conditions, these saddles can be discarded. However, they can play a role for Dirichlet conditions, we will briefly return to this in section \ref{subsec:Dirichletvec}.

\textbf{Chiral multiplet}. As a localizing action we choose the $\Q$-exact term
\be{
\mathscr{S}^{\rm loc}_{\rm chi}\equiv\Q\int\d^4 x\;\sqrt{g}\;\mathscr{V}^{\rm loc}_{\rm chi}~,
}
where the localizing functional is
\be{\label{V-loc-chi}
\mathscr{V}^{\rm loc}_{\rm chi}\equiv\frac{1}{2}\left(\mathscr{V}_{\rm chi}+\tilde{\mathscr{V}}_{\rm chi}+ \mathscr{V}_\lambda+\tilde{\mathscr{V}}_\lambda\right)~,
}
with $\mathscr{V}_{\rm chi}$, $\tilde{\mathscr{V}}_{\rm chi}$, ${\mathscr{V}}_{\lambda}$ and $\tilde{\mathscr{V}}_{\lambda}$, defined in (\ref{Vpsi}) and  (\ref{eq:V-lambda}).

As already mentioned for the vector, this action is automatically supersymmetric even in presence of the boundary. However, the bosonic part of the action is not manifestly positive semi-definite because of mixed terms involving $\delta_\zeta,\delta_{\tilde\zeta}$-variations and the D-term coupling. 
On the one hand, the mixed terms simply cancel out upon the standard choice $(F,\tilde F)^\vee=-(\tilde F,F)$ as shown in appendix \ref{app:chiralactions}, and the {\it real contour} where ${}^\vee$ acts as ${}^\dagger$ yields a manifestly positive semi-definite action. The convergence conditions imposed by the presence of the D-term coupling have  already been discussed in section \ref{sec:contours}. Therefore, the bosonic localization locus is given by the BPS equations
\be{
\delta_\zeta\psi=\delta_{\tilde\zeta}\psi=
\delta_{\zeta}\tilde\psi=\delta_{\tilde\zeta}\tilde\psi=0~,
}
in the (trivial) BPS vector background $\A=\A^{(0)}$. Contracting these equations with the Killing spinors, we simply get
\be{
F=\tilde F=0~,\quad \L_K^{(0)} \phi=\L_K^{(0)} \tilde\phi=0~,\quad \L_{\bar Y}^{(0)}\phi=\L_{Y}^{(0)}\tilde\phi=0~.
} 
In order to solve these equations, we need some information about the R-symmetry connection. We see from (\ref{ARhol}) that $\iota_K  A\sim K^\mu\partial_\mu(\sqrt[4]{g}s)$, implying that the general solution to $\L^{(0)}_K \phi=0$ takes the form
\be{
\phi=(\sqrt[4]{g}s)^\frac{r}{2}\, f(w,z,\bar z)\, \e^{-\i\Phi_{(0)}(\bar w- w)}~.
}
Once the twisted periodicities (\ref{twistbc}) are taken into account, one can conclude that the solution has to be $\phi=0$, and a similar argument also shows $\tilde\phi=0$. As we will explain below, the other equations can be thought as consistency conditions for $F-\i\L^{(0)}_{\bar Y}\phi=0$, $\tilde F+\i\L^{(0)}_{Y}\tilde\phi=0$.  
We can thus conclude that the BPS localization locus is trivial, namely
\be{
\mathcal{M}^\text{BPS}_\text{chi}\simeq \Big\{ \phi=\tilde\phi=F=\tilde F=\psi=\tilde\psi=0\Big\}~.
}

\subsection{Cohomological complex}
Having determined the BPS locus in the previous subsection, the next  step is to introduce the 4d $\mathcal{N}=1$ cohomological complexes. 
This will set up the coordinate system (on the space of fields) in which we are going to compute 1-loop determinants.
In practice, the cohomological complexes are obtained by converting all fields to differential forms through appropriate contractions with the Killing spinors. The selected supercharge then clearly acts as an equivariant differential on the supermanifold of cohomological field variables. 

\subsubsection{Vector multiplet}

We start by introducing twisted variables for the gauginos by using the Killing spinors. We introduce scalar fermionic variables 
\be{
\Lambda^{0}\equiv \frac{\zeta^\dagger}{|\zeta|^2}\lambda~,\qquad \tilde\Lambda^0\equiv \frac{\tilde\zeta^\dagger}{|\tilde\zeta|^2}\tilde\lambda~,\qquad 
\Lambda^+\equiv \zeta\lambda~,\qquad \tilde\Lambda^{-}\equiv \tilde\zeta\tilde\lambda~,
}
and fermionic 1-forms 
\be{
\Lambda_\mu \equiv \i\zeta \sigma_\mu \tilde\lambda ~,\qquad \tilde\Lambda_\mu \equiv \i\tilde\zeta \tilde\sigma_\mu \lambda~.
}
The superscripts of the scalar modes $(\Lambda^{0},\tilde \Lambda^{0},\Lambda^+,\tilde\Lambda^-)$ emphasize their R-charges which equal $(0,0,+2,-2)$, while the 1-forms are neutral. Note that the relations between scalars and 1-forms are
\be{
\Lambda_\mu = \i\tilde\Lambda^0 K_\mu - \i\tilde\Lambda^- Y_\mu ~,
\qquad 
\tilde \Lambda_\mu = \i \Lambda^0 K_\mu + \i \Lambda^+ \bar Y_\mu~.
}
The SUSY transformations for the twisted fields are
\begin{align}\label{eq:twistedSUSYVEC}
\delta_\zeta {\A_\mu} &= \Lambda_\mu~,\quad& \delta_{\tilde \zeta} \A_\mu &= \tilde\Lambda_\mu ~,\nn\\
\delta_\zeta {\Lambda_\mu} &=0~, \quad& \delta_{\tilde \zeta} \Lambda_\mu &= -2 \i \F^{-}_{\mu\nu} K^\nu + K_\mu D~,\\
\delta_\zeta {\tilde \Lambda_\mu} &=  -2 \i \F^{+}_{\mu\nu} K^\nu - K_\mu D~,  \quad& \delta_{\tilde \zeta} \tilde\Lambda_\mu &= 0~,\nn\\
\delta_\zeta D &= \i \hat {\cal D}_\mu \Lambda^\mu~, \quad & \delta_{\tilde \zeta} D &= -\i \hat {\cal D}_\mu \tilde\Lambda^\mu~,\nn
\end{align}
where
\be{
\hat {\cal D}_\mu \Lambda^\mu  \equiv (D_\mu - 2 \i V_\mu)\Lambda^\mu~,
\qquad \hat {\cal D}_\mu \tilde \Lambda^\mu  \equiv (D_\mu + 2 \i V_\mu)\tilde\Lambda^\mu~.
}
Since we are using the supercharge $\Q=\delta_\zeta+\delta_{\tilde\zeta}$ in order to perform localization, the SUSY transformations suggest to introduce the combinations 
\be{
\Xi_\mu\equiv \Lambda_\mu+\tilde\Lambda_\mu~,\qquad \Pi_\mu\equiv \Lambda_\mu-\tilde\Lambda_\mu~,
}
and to choose the cohomological multiplets
\be{\label{eq:vs-basis}
(\A_\mu, \Xi_\mu=\Q\A_\mu)~,\qquad (\Psi , \Delta=\Q\Psi)~,
}
where we have defined\footnote{We use that $\varepsilon_{\mu\nu\alpha\beta}J^{\alpha\beta}=2J_{\mu\nu}$, $\varepsilon_{\mu\nu\alpha\beta}\tilde J^{\alpha\beta}=-2\tilde J_{\mu\nu}$.}
\begin{align}
\Psi&\equiv \bar K^\mu \Pi_\mu =  2\i (\tilde\Lambda^0-\Lambda^0) ~,\qquad \Delta\equiv4 D +2\i \tilde \F_{\mu\nu} \bar K^\mu K^\nu =4D-2\i Y^\mu \bar Y^\nu \F_{\mu\nu}~.
\end{align}
We refer to these multiplets and $\Q$ as the SUSY complex of the vector multiplet.\footnote{In principle, one can also use scalar variables only. This might also be a convenient choice, but we decided to keep the 1-form nature of the connection manifest.} 

\textbf{Remark}. The decomposition 
\be{\label{eq:Xi-mu}
 \bar\Psi\equiv \bar K^\mu \Xi_\mu=2\i(\tilde\Lambda^0  +\Lambda^0)~,\qquad \Xi_\mu = \frac{1}{2}\bar\Psi K_\mu - \i\tilde\Lambda^- Y_\mu + \i \Lambda^+ \bar Y_\mu
}
makes it manifest that $\Xi_\mu$ and $\Psi$ capture all four fermionic (off-shell) scalar degrees of freedom and that $\iota_K\Xi=0$, implying the  property 
\be{\label{QKA}
\Q(\iota_K \A)=0 ~.
}
As a consequence, on the r.h.s. of $\Q\A$ what really appears is $\Xi=p_{\bar K}\Xi$, where we defined the following projector onto the space of horizontal 1-forms w.r.t. $\bar K$
\be{\label{eq:ProjKbar}
p_{\bar K}\equiv 1-\frac{1}{2}\bar K\iota_K~.
}
This is ultimately due to the fact that the component $\Phi=-\iota_K \A$ appears as the gauge parameter in $\Q^2=2\i\delta_K$. We will come back to this important point later on when computing the 1-loop determinant for the vector multiplet. 

\medskip

Finally, since the cohomological fields are R-symmetry neutral forms on the manifold, we can introduce the usual inner product between  $\mathfrak{g}$-valued $n$-forms on spacetime
\be{
\brakett{\omega_1}{\omega_2}\equiv \text{Tr}\int\omega_1^\dagger\wedge\star\omega_2~,\quad \omega_{1,2}\in\Omega^n~,
}
which can be used to rewrite the actions, compute the adjoint of the relevant operators as well as determine normalizability of the fields.

\subsubsection{Chiral multiplet}

We start by introducing scalar fermions by using the Killing spinors. We define 
\begin{align}\label{eq:twistCHI}
\psi&\equiv \zeta B-\frac{\zeta^\dagger}{|\zeta|^2}C~, & B&\equiv\frac{\zeta^\dagger}{|\zeta|^2}\psi~, & C&\equiv\zeta\psi~,\nn\\
\tilde \psi&\equiv\tilde\zeta \tilde B-\frac{\tilde\zeta^\dagger}{|\tilde\zeta|^2}\tilde C~, &\tilde B&\equiv\frac{\tilde\zeta^\dagger}{|\tilde\zeta|^2}\tilde\psi~, & \tilde C&\equiv \tilde\zeta\tilde\psi~.
\end{align}
Note that the R-charges of $(\phi,C)$ and $(B,F)$ and  are $r$ and $r-2$ respectively, while those of $(\tilde\phi,\tilde C)$ and $(\tilde B,\tilde F)$  are $-r$ and $2-r$ respectively. In terms of the twisted variables the SUSY transformations (\ref{eq:chisusytrans}) read 
\ale{
\delta_\zeta \phi &=\sqrt{2}C~,\quad& \delta_{\tilde\zeta} \phi &=0~,\nn\\
\delta_\zeta C &=0 ~,\quad &\delta_{\tilde \zeta} C &=\sqrt{2}\i\mathcal{L}_K\phi~,\label{eq:twistSUSYCHI1}\\
\delta_\zeta B &=\sqrt{2}F~,\quad& \delta_{\tilde\zeta} B &=-\sqrt{2}\i\mathcal{L}_{\bar Y}\phi~,\nn\\
\delta_\zeta F&=0~,\quad& \delta_{\tilde\zeta} F&=\sqrt{2}\i\left(\mathcal{L}_K B+\mathcal{L}_{\bar Y} C\right)-2\i\tilde\Lambda^-\phi~,
}
and similarly
\ale{
\delta_\zeta\tilde\phi&=0~,\quad & \delta_{\tilde\zeta}\tilde\phi &=\sqrt{2}\tilde C~, \nn\\
\delta_\zeta \tilde C &=\sqrt{2}\i\mathcal{L}_K \tilde \phi~,\quad &\delta_{\tilde\zeta}\tilde C&=0~,\label{eq:twistSUSYCHI2}\\
\delta_\zeta\tilde B &=\sqrt{2}\i\mathcal{L}_{Y}\tilde\phi~,\quad &\delta_{\tilde\zeta}\tilde B &=\sqrt{2}\tilde F~,\nn\\
\delta_\zeta \tilde F &=\sqrt{2}\i\left(\mathcal{L}_K \tilde B-\mathcal{L}_{Y} \tilde C\right)+2\i\Lambda^+\tilde\phi~,\quad& \delta_{\tilde\zeta}\tilde F&=0~.
}
Since we are using the supercharge $\Q=\delta_\zeta+\delta_{\tilde\zeta}$ to perform localization, these variations led us to choose the cohomological multiplets
\be{
(\phi,\sqrt{2}C=\Q\phi)~,\quad (B,\sqrt{2}X_F=\Q B)~,
}
and similarly 
\be{
(\tilde\phi,\sqrt{2}\tilde C=\Q\tilde\phi)~,\quad (\tilde B,\sqrt{2}\tilde X_F=\Q \tilde B)~,
}
where 
\be{
X_F\equiv F-\i\L_{\bar Y}\phi~,\qquad \tilde X_F\equiv \tilde F+\i\L_Y\tilde\phi~.
}
We refer to these multiplets and $\Q$ as the SUSY complex of the chiral multiplet. 

Finally, having rewritten all the fields in terms of scalars, we can introduce the usual inner product between fields with the same quantum numbers (in particular, the same R-charge)
\be{\label{chiinner}
\brakett{\omega_1}{\omega_2}\equiv \int \star \, \omega_1^\dagger\,\omega_2~,\qquad \omega_{1,2}\in\Omega^0_{(r)}~,\quad \Omega^0_{(r)}\equiv \mathscr{R}^{\, r}\otimes \Omega^0~,
}
which can be used to rewrite the actions, compute the adjoint of the relevant operators as well as determine normalizability of the fields.

\subsection{1-loop determinants}
Since the BPS locus and the relevant cohomological complexes have been found, we are now in a position to finally compute 1-loop superdeterminants of Gaussian fluctuations.
For concreteness and ease of computations, we focus on the real K{\"a}hler background described in section \ref{globalHol}, but the final results are expected to hold more generally.

\subsubsection{Vector multiplet}\label{subsec:vecloop}

In order to compute the 1-loop determinant of Gaussian fluctuations around the localization locus, it may be useful to recast $\mathscr{V}^{\rm loc}_{\rm vec}$ into the operatorial matrix form 
\be{\label{eq:V-loc-vec-matrix-form}
\sqrt{g}\, \mathscr{V}^{\rm loc}_{\rm vec}=\left(\begin{array}{c} \Xi \\ \Psi \end{array}\right)^\text{T}\wedge \star\left(\begin{array}{cc} D_{11}&D_{10} \\D_{01}&D_{00} \end{array}\right)\left(\begin{array}{c} \A \\ \Delta \end{array}\right)~.
}
Since in cohomological variables we have 
\be{
\mathscr{V}^{\rm loc}_{\rm vec}=\frac{\i}{16}\p{-\Psi Y^\mu\bar Y^\nu +\Xi_\alpha \bar K^\alpha \bar K^\mu K^\nu -4\Xi^\nu\bar K^\mu}\F^\vee_{\mu\nu}+\frac{1}{8}\Psi D^\vee~,
}
recalling that  we have set $\F^\vee=\F$ and  $D^\vee=-D=-\frac{\Delta}{4}-\frac{\i}{2}Y^\mu \bar Y^\nu \F_{\mu\nu}$,\footnote{We recall that the real contour is defined by ${}^\vee$ acting as Hermitian conjugation, that is $\F\in\mathbb{R}$, $D\in\i\mathbb{R}$. In general, this functional may be taken as a definition.} we can easily read off the resulting matrix
\be{\label{vecOp}
\left(\begin{array}{cc} D_{11}&D_{10} \\D_{01}&D_{00} \end{array}\right)=-\frac{\i}{8}\left(\begin{array}{cc} (1+p_{\bar K})\iota_{\bar K}\d_\A&0 \\ 2\iota_{\bar Y}\iota_Y\d_\A&-\frac{\i}{4} \end{array}\right)~,
}
which can be simply linearized around the trivial BPS locus by substituting $\A=\A^{(0)}$ in ${\d_\A}$. Given the lower-triangular nature of this matrix and the supersymmetric pairing between bosons and fermions, simple linear algebra (see e.g. \cite{Pestun:2016qko,Qiu:2016dyj} in the review \cite{Pestun:2016zxk} for a general derivation) would naively lead us to conclude that the superdeterminant of the kinetic operators in $\Q\mathscr{V}^{\rm loc}_{\rm vec}$ around the localization locus would be given by
\be{\label{naivedet}
\sqrt{\frac{\det_{\Omega^0} 2\i\delta_K^{(0)}}{\det_{\Omega^1} 2\i\delta_K^{(0)}}}~,
}
where $\delta_K^{(0)}$ denotes the operator (\ref{deltaK}) at the BPS locus, and the determinants are taken over the fermionic $\Psi\in\Omega^0$ and bosonic $\A\in\Omega^1$ coordinates . However, it is easy to understand that this cannot be quite correct for at least two reasons:
\begin{enumerate}
\item As remarked around (\ref{eq:ProjKbar}), $\Xi$ is horizontal w.r.t. $\bar K$. Therefore, the operator $D_{11}$ is dangerously ambiguous by left-multiplication by the projector $p_{\bar K}$, and the argument leading to the simplification of its determinant between bosons and fermions fails. 
\item The gauge symmetry has not been fixed yet, therefore the space of 1-forms over which we are evaluating the bosonic determinant is highly redundant. 
\end{enumerate}
As we are going to explain, these two problems are actually related, and we outline how they can be solved together to yield a sensible answer. First of all, it is natural to resolve the first  ambiguity by restricting the evaluation of all the 1-form determinants on the image of $p_{\bar K}$, where it simply restricts to the identity. This means that the determinant should  be computed on the subspace $\Omega^1_K\equiv \{\omega\in \Omega^1|\, \iota_K\omega=0\}$. One can see the necessity for such a restriction on the bosonic side too. For the gauge field this amounts to splitting 
\be{\label{eq:A-splitting}
	\A \equiv \hat\A - \frac{1}{2} \Phi\, \bar K ~,\qquad \Phi \equiv -\iota_K  \A~,
}
and restricting to $\hat\A$. In the cohomological formalism, $\Phi$ should not be treated as a coordinate because $\Q \Phi=0$ and it rather it appears as a gauge parameter in $\Q^2=2\i\delta_K$. Therefore, it is natural to exclude it from the computation of the determinant. Secondly, in order to properly fix the gauge, one should introduce the usual Faddeev-Popov ghosts $\textsl{c},\bar{\textsl{c}}$ and the Lagrange multiplier $\textsl{b}$, together with the BRST charge $\Q_\text{B}$ generating gauge transformations with parameter $\textsl{c}$, for instance $\Q_\text{B} \A=\d_\A \textsl{c}$. However, in order to preserve supersymmetry, the BRST complex must be combined with the SUSY complex, in particular $\Q \textsl{c}=-2\i \Phi$, and then localization must be performed w.r.t. the total odd charge $\Q+\Q_\text{B}$. 
This is indeed consistent with $\Phi$ being not a coordinate field but rather the differential of a fermionic variable, and as such it does not contribute to the determinant. The bottom line of these arguments is that the correct ratio of 1-loop determinants can be obtained from (\ref{naivedet}) by enlarging the space of 0-forms to include the missing ghosts and reducing the space of 1-forms, namely 
\be{\label{Z1loopvec}
\mathcal{Z}_{\textrm{1-loop}}^{\textrm{vec}}(\Phi_0) \equiv 
\sqrt{\frac{(\det_{\Omega^0} 2\i\delta_{K}^{(0)}) \ (\det'_{\Omega^0} 2\i\delta_{K}^{(0)})\  (\det'_{\Omega^0}2\i\delta_{K}^{(0)})}{\det_{\Omega^1_K}2\i\delta_{K}^{(0)} }}~,
}
where the prime means exclusion of the constant modes. Unfortunately, we are not able to provide a complete proof of this formula. A posteriori, the strongest support for our proposal comes from the consistency of the final answer with known results and comparison with dimensional reductions. Therefore, before computing the concrete expression for (\ref{Z1loopvec}), let us make a few more observations that support our result. Upon dimensional reduction, the 4d $\mathcal{N}=1$ cohomological complex we are considering should reduce to the 3d $\mathcal{N}=2$ cohomological complex studied for example in \cite{Kallen:2011ny,Ohta:2012ev}, which will be referred to as K{\"a}llen's complex. Hence, our setup is expected to be related to a lift of the 3d setup by the inclusion of an additional tower of KK modes. In order to push the similarities between the 4d and 3d setups even further, let us comment on some features of the splitting of the gauge field introduced in (\ref{eq:A-splitting}):
\begin{enumerate}
\item Under a general gauge transformation $\hat \A$ and $\Phi$ {\it do not mix}. The full gauge symmetry of the 4d theory is {\it preserved}, however $\hat \A$ and $\Phi$ transform in a funny way w.r.t. arbitrary gauge transformations. On the other hand, $\hat\A$ transforms as an ordinary gauge field and $\Phi$ is covariant w.r.t. gauge transformations which are {\it holomorphic on the torus}, thus preserving the condition $\iota_K \hat\A=0$.
\item The field $\hat \A$ really contains \textit{three} independent components due to $\iota_K \hat \A = 0$. At the same time, $\Phi$ is a gauge parameter in the SUSY transformations, hence it should not enter the cohomological complex as a bosonic coordinate.
\item In complex coordinates $\hat\A=\A_z\d z+\A_{\bar z}\d\bar z+\A_w\d w$, hence our cohomological complex is formally the same as K{\"a}llen's   but with a {\it complex direction} and a {\it partial connection}\footnote{This resembles Costello-Yamazaki-Witten theory \cite{Costello:2013zra,Witten:2016spx,Costello:2017dso,Costello:2018gyb}. It would be interesting to explore connections.} upon the identifications
\be{
\Psi^{\rm 3d} \sim \Xi~,\quad%
\alpha^{\rm 3d}\sim \Psi ~, \quad%
\A^{\rm 3d}\sim \hat \A~,  \quad%
\Phi^{\rm 3d} \sim \Phi ~, \quad%
\tilde D^{\rm 3d} \sim \Delta~.
}
These formal substitutions into K{\"a}llen's computation  \cite{Kallen:2011ny} reproduce (\ref{Z1loopvec}). 
\end{enumerate}

We can now move on and compute the actual expression of the proposed determinant. Once the localizing action is brought to the cohomological form, the usual strategy is to look for the off-diagonal operator (commuting with $\delta_K$) pairing bosons and fermions and to study the equivariant index to extract the contributing modes and weights. In our case, we have to solve two other problems: $i$) we have not fully worked out the combined SUSY-BRST complex explicitly and $ii$) we cannot easily exploit the usual index theorems because of the boundary. We propose to solve these problems as follows. From (\ref{vecOp}), we see that the off-diagonal operator written in the holomorphic frame is essentially the pullback to the disk of the de Rham differential. This suggests that the right complex to look at is eventually  the de Rham one, restricted to the disk with differential $\d_2$, as also argued in \cite{Nishioka:2014zpa}. We can use the complex structure to split $\Omega^1_K=\Omega^{1,0}_{\T^2}\oplus \Omega^{1,0}_{\Di^2}\oplus \Omega^{0,1}_{\Di^2}$ and notice that there is a natural bijection between $\Omega^{1,0}_{\T^2}$ and $\Omega^{0,0}$ given by acting on the former with the interior product $\iota_{\bar K}$.\footnote{
Note that it is also important to demand that this bijection is compatible with the boundary conditions.}
This implies that the contribution of modes from these two spaces exactly cancel each other in the 1-loop determinant.
The remaining modes can also be mapped into each other by replacing the de Rham differential with the Dolbeault operators, namely $\d_2=\partial_2+\bar\partial_2$. 
We are thus led to consider the complex 
\be{
\Omega^{0,0}\oplus\Omega^{0,0} 
\xrightarrow{\partial_2\oplus\bar\partial_2}
\Omega^{1,0}_{\Di^2}\oplus \Omega^{0,1}_{\Di^2}~,
}
and (ignoring possible phases) we can remove the square root from (\ref{Z1loopvec}) by just considering $(\bar \partial_2:\Omega^0\to\Omega^{0,1}_{\Di^2})\times (c.c.)$. Hence our problem has reduced to studying the Dolbeault complex on $\Di^2$ subject to the desired boundary conditions (of course, with two towers of KK modes). The map $\bar \partial_2:\Omega^0\to\Omega^{0,1}_{\Di^2}$ is surjective without restrictions on the domain and codomain, in which case the contributing modes come entirely from the kernel of $\bar \partial_2$, that is holomorphic ghosts on the disk. This conclusion still holds if we restrict the codomain to 1-forms whose field strengths have vanishing contractions with $Y,\bar Y$ at the boundary (Neumann conditions). Because of the quotient (\ref{twist}) and the twisted periodicities (\ref{twistbc}) we are imposing on all the fields, let us introduce  twisted Fourier modes on the torus
\be{\label{twistF}
F^{(m)}_{n_x,n_y}(w,\bar w)\equiv\exp\left[-\frac{1}{2\Im(\tau)}\Big(w(\bar \tau n_x-n_y+m\bar \sigma)-\bar w(\tau n_x-n_y+m \sigma)\Big)\right]~,
}
satisfying the property 
\begin{align}
F^{(m)}_{n_x,n_y}(w+2\pi k+2\pi n \tau ,\bar w+2\pi k+2\pi n\bar\tau)&=\e^{2\pi\i m(k\alpha+n\beta)}F^{(m)}_{n_x,n_y}(w,\bar w)~.
\end{align}
Then we can expand the ghost field in a basis of holomorphic modes on the disk and twisted Fourier modes as
\be{
\textsl{c}(z,w,\bar w)=\sum'_{n_x,n_y,m\in\mathbb{Z}}\textsl{c}^{(m)}_{n_x,n_y}\, z^m \, F^{(-m)}_{n_x,n_y}(w,\bar w)~,
}
where we recall that this field has zero R-charge and hence should not get any phase under the identifications (\ref{twist}), consistently with (\ref{twistbc}). The prime on the summation reminds us to exclude the constant mode $(n_x,n_y,m)=(0,0,0)$. Finally, we demand $m\geq 0$ for the modes to be regular at the origin. The eigenvalues of $2\i\delta^{(0)}_K$ on such modes are
\be{
\lambda_c=\frac{\i}{\Im(\tau)}\Big(\tau\, n_x-n_y-\sigma\, m +\Phi_{(0)}\Big)~,
}
leading to the result
\be{\label{eq:1-loopvecHOL}
\mathcal{Z}_\text{1-loop}^{\rm vec}(\Phi_{(0)})={\det_\text{ad}}\left[-\frac{\Im(\tau)}{\i\Phi_{(0)}}\prod_{n_x,n_y\in\mathbb{Z}}\prod_{m\geq 0}-\frac{\i}{\Im(\tau)}\Big(\tau\, n_x-n_y+\sigma\, m -\Phi_{(0)}\Big)\right]~,
}
which of course needs to be regularized. We can do that by using $\zeta$-function regularization as discussed in appendix \ref{sec:1-loop-details}, with the final result (up to an overall constant)
\be{\label{eq:Neumannvec}
\mathcal{Z}_\text{1-loop}^{\rm vec}(\Phi_0)
=
\left[\frac{\e^{-\frac{\i\pi}{3} P_{3}(0)}}{\text{Res}_{u=0}\Gamma(u;\tau,\sigma)}\right]^{\text{rk(G)}}
{\det_\text{ad}}'\left[\frac{1}{\Phi_{(0)}}\frac{\e^{-\frac{\i\pi}{3} P_{3}(\Phi_{(0)})}}{\Gamma(\Phi_{(0)};\tau,\sigma)}\right]~,
}
where $\Gamma$ is the elliptic Gamma function defined in appendix \ref{appell}, the prime denotes  exclusion of zero roots and $P_3$ is a cubic polynomial. The factor ${\det_\text{ad}}'(\Phi_{(0)})^{-1}$ coming from the constant mode can be canceled by the Vandermonde determinant in the integration measure once a change of variables from the algebra to its Cartan  is performed.

\subsubsection{Chiral multiplet}

We now move our focus to the chiral multiplet.
Ideally, in order to compute the 1-loop determinant around the BPS locus, we would like to recast $\mathscr{V}^{\rm loc}_{\rm chi}$ into the matrix form
\begin{multline}\label{MatrixChi}
\mathscr{V}^{\rm loc}_{\rm chi}=\left(\begin{array}{c}  \tilde \phi \\  \sqrt{2}\tilde X_F \end{array}\right)^\text{\!\!T}\!\! \left(\begin{array}{cc} D_{1}^{(r,r)}&D_{2}^{(r,r-2)} \\D_3^{(r-2,r)}&D_4^{(r-2,r-2)} \end{array}\right) \!\!\left(\begin{array}{c} \sqrt{2}C \\ B \end{array}\right)+\\
+\left(\begin{array}{c}  \sqrt{2}\tilde C \\  \tilde B \end{array}\right)^\text{\!\!T}\!\! \left(\begin{array}{cc} \tilde D_{1}^{(r,r)}&\tilde D_{2}^{(r,r-2)} \\ \tilde D_3^{(r-2,r)}& \tilde D_4^{(r-2,r-2)} \end{array}\right)\!\!\left(\begin{array}{c} \phi \\ \sqrt{2}X_F \end{array}\right)+\frac{1}{4}\tilde\phi\Psi\phi~,
\end{multline}
up to terms whose $\Q$-variations can be killed by choosing suitable boundary conditions. The superscripts in $D_i^{(r,r')}:\Omega^0_{(r')}\to \Omega^0_{(r)}$ remind us what is the space the operator acts on as the cohomological variables have generically non-zero R-charges. Also, note that the last term can be dropped because we are eventually interested in linearizing this functional around the trivial BPS locus in which $\Psi=\Psi_{(0)}=0$. Since we have\footnote{We are  assuming a real background and  contour. In general, this functional may be taken as a definition.}
\be{
\mathscr{V}^{\text{loc}}_{\text{chi}}=\frac{1}{2}\left(\sqrt{2}F^\vee B+\sqrt{2}\i (\L_Y \tilde\phi )B-\sqrt{2}\i(\L_{\bar K}\tilde\phi)C-\tilde B\sqrt{2}\i \L_{\bar Y}\phi-\tilde C\sqrt{2}\i \L_{\bar K}\phi+\sqrt{2}\tilde B \tilde F^\vee\right)~,
}
recalling that we have defined $(F,\tilde F)^\vee=(-\tilde F,-F)$,\footnote{Correspondingly $(X_F,\tilde X_F)^\vee=(-\tilde X_F+2\i\L_Y \tilde\phi,-X_F-2\i\L_{\bar Y}\phi)$.} we can easily read off the resulting matrices
\begin{align}
\left(\begin{array}{cc} D_{1}^{(r,r)}&D_{2}^{(r,r-2)} \\D_3^{(r-2,r)}&D_4^{(r-2,r-2)} \end{array}\right)&=
\frac{1}{2}\left(\begin{array}{ccc} \i\L_{\bar K}&&-2\sqrt{2}\i\L_Y\\0 &&-1 \end{array}\right)~,\\
\left(\begin{array}{cc} \tilde D_{1}^{(r,r)}&\tilde D_{2}^{(r,r-2)} \\ \tilde D_3^{(r-2,r)}& \tilde D_4^{(r-2,r-2)} \end{array}\right)&=\frac{1}{2}\left(\begin{array}{ccc} -\i\L_{\bar K}&&0 \\ 2\sqrt{2}\i\L_{\bar Y}&& -1 \end{array}\right)~,
\end{align}
which can be easily linearized around the BPS localization locus by substituting $\A=\A^{(0)}$ in the covariant derivatives. 

\textbf{Remark}. In order to achieve this form, we had to integrate by parts, i.e. to define the adjoints of $\L_{\bar K},\L_{Y}$ w.r.t. (\ref{chiinner}), which imposes the vanishing of the  boundary terms
\be{\label{phiBbc}
\int\d^4\, x\sqrt{g}\, \L_{\bar K}(\tilde\phi \, C)~,\qquad \int\d^4 x\, \sqrt{g}\, \L_{Y}(\tilde\phi \, B)~.
}
The first term vanishes because $\bar K$ is tangent to the boundary, while the vanishing of the second term requires suitable boundary conditions to which we will return below. 

As we have already discussed for the vector multiplet, now simple linear algebra implies that the 1-loop determinant is computed by
\be{\label{eq:chiral-ker-coker}
\frac{\det_{\Omega^0_{(r-2)}}2\i\delta_K^{(0)}}{\det_{\Omega^{0}_{(r)}}2\i\delta_K^{(0)}}~,
}
with the numerator and denominator capturing the contributions of the fermionic and bosonic coordinates $B\in \Omega^0_{(r-2)}$  and $\phi\in \Omega^0_{(r)}$ respectively. In appendix \ref{chiralKbKf}, we present a detailed derivation. This result may be further simplified  by noting that the off-diagonal operators commute with $\delta_K^{(0)}$ and provide maps between 0-forms of different R-charges
\be{
\mathcal{L}_{\bar Y}~:~\Omega^0_{(r)}\longrightarrow \Omega^0_{(r-2)} ~, \qquad \mathcal{L}_{Y}~:~\Omega^0_{(r-2)}\longrightarrow \Omega^0_{(r)}~.
}
In order for these operators to provide meaningful maps, we have to impose compatible boundary conditions which are in fact dictated by the vanishing of the boundary terms appearing in (\ref{phiBbc}), which we call:
\begin{itemize}
\item \textit{Robin-like} boundary conditions
\be{
(\text{R}):~\mathcal{L}_{\bar Y}\phi |_\partial=B|_\partial=0~,
}
\item \textit{Dirichlet} boundary conditions
\be{
(\text{D}):~\phi |_\partial=\mathcal{L}_Y B|_\partial=0~.
}
\end{itemize}
The ratio of determinants can then be computed on smaller spaces given by the kernels and cokernels of the pairing operators, namely
\be{\label{Chiraldet}
\mathcal{Z}^{\textrm{chi}\text{(R)}}_{\textrm{1-loop}}(\Phi_{(0)})\equiv \frac{\det_{\textrm{Coker}\mathcal{L}_{\bar Y}|_{\Omega^0_{(r-2)}}} 2\i\delta_K^{(0)}}{\det_{\textrm{Ker}\mathcal{L}_{\bar Y}|_{\Omega^0_{(r)}}}2\i\delta^{(0)}_K}~,\quad \mathcal{Z}^{\textrm{chi}\text{(D)}}_{\textrm{1-loop}}(\Phi_{(0)})\equiv \frac{\det_{\textrm{Ker}\mathcal{L}_{Y}|_{\Omega^0_{(r-2)}}} 2\i\delta_K^{(0)}}{\det_{\textrm{Coker}\mathcal{L}_{ Y}|_{\Omega^0_{(r)}}}2\i\delta^{(0)}_K}~.
}
In our simplified choice of K{\"a}hler background, the relevant operators read
\be{
\mathcal{L}_{\bar Y}= \frac{2s^{-1}}{c}(\partial_z-\i q_\text{R} A_z)~,\qquad \mathcal{L}_{Y}=\frac{2s}{c}(\partial_{\bar z}-\i q_\text{R} A_{\bar z}) ~,
}
and we can perform explicit computations. The kernels  are easily seen to be parametrized by (anti-)holomorphic functions on the disk, namely
\begin{align}
\textrm{Ker}\mathcal{L}_{\bar Y}\ni \phi=(s/\sqrt[4]{g})^{\frac{r}{2}}f_\phi(w,\bar w,\bar z)~,\qquad  \textrm{Ker}\mathcal{L}_{Y}\ni B=(\sqrt[4]{g}s)^{\frac{r-2}{2}}f_B(w,\bar w,z)~,
\end{align}
while the cokernels are empty, since the corresponding operators are surjective thanks to boundary conditions, as can be seen from a Laurent expansion on the disk.\footnote{It may be more convenient to use ``gauge" transformed operators to simplify various factors of $\sqrt[4]{g}$.} 

\textbf{Remark}. In general, the selection of modes depends on the analytic properties of the metric. Assuming $\sqrt[4]{g}=(1+\varepsilon z\bar z)^{-1}$ as for the standard K{\"a}hler metric on the disk, there are three cases corresponding to $\varepsilon=\pm 1,0$ (spherical, hyperbolic, flat). As we have already mentioned, our analysis is adapted to the spherical case, and the mode expansion of a scalar function assumes the form $f(z,\bar z)=\sum_{m,n\in\mathbb{Z}}f_{mn}z^m \bar z^n$. Normalizability of the individual modes on the hemisphere restricts $m+n\geq 0$.

The modes that contribute to (\ref{Chiraldet}) can be expanded in a basis of (anti-)holomorphic modes on the disk and twisted Fourier modes (\ref{twistF}) on the torus according to
\begin{align}\label{eq: disktorusholomodes}
f_\phi(w,\bar w,\bar z) =& \ \sum_{n_x,n_y,m\in\mathbb{Z}} f_{\phi;n_x,n_y}^{(m)}\bar z^{m}F^{(r/2+m)}_{n_x,n_y}(w,\bar w)~,\nn\\
f_B(w,\bar w,z) = &\ \sum_{n_x,n_y,m\in\mathbb{Z}}f_{B;n_x,n_y}^{(m)}z^{m}F^{(r/2-1-m)}_{n_x,n_y}(w,\bar w)~.
\end{align}
Assuming that not only $\d |s|=0$ (as required for a real background) but also $\d s=0$ after imposing the twisted periodicities, 
then $\iota_K  A=0$ and the spectrum of $2\i\delta_K^{(0)}$ is given by the eigenvalues
\be{
\lambda_\phi=\frac{\i}{\Im(\tau)}\Big(\tau n_x-n_y+\sigma(r/2+m)+\Phi_{(0)}\Big)~,\quad \lambda_B=\frac{\i}{\Im(\tau)}\Big(\tau n_x-n_y+\sigma(r/2-1-m)+\Phi_{(0)}\Big)~.
}
Keeping only regular modes at the origin, Dirichlet conditions yield a net contribution from $B$-modes only
\be{\label{eq:1loopB}
\mathcal{Z}^{\textrm{chi}(\text{D})}_{\textrm{1-loop}}(\Phi_{(0)})=\det_{\mathcal{R}}\Bigg[\prod_{n_x,n_y\in\mathbb{Z}}\prod_{m\geq 0}-\frac{\i}{\Im(\tau)}\Big(\tau n_x-n_y+\sigma(1-r/2+m)-\Phi_{(0)}\Big)\Bigg]~,
}
where $\mathcal{R}$ is the gauge (and/or flavor) group representation. 
Viceversa, imposing Robin-like conditions leaves only a net contribution from $\phi$-modes
\be{\label{eq:1loopphi}
\mathcal{Z}^{\textrm{chi}(\text{R})}_{\textrm{1-loop}}(\Phi_{(0)})=\det_{\mathcal{R}}\Bigg[\prod_{n_x,n_y\in\mathbb{Z}}\prod_{m\geq 0}-\i\Im(\tau)\Big(\tau n_x-n_y+\sigma(r/2+m)+\Phi_{(0)}\Big)^{-1}\Bigg]~.
}
In this case it should be noted that imposing $B|_\partial=0$ also requires $\i{\cal L}_{\bar Y} \phi|_\partial = F|_\partial$ by compatibility with supersymmetry, and we can consistently set $F|_\partial=0$. 

\textbf{Remark}. If the matter content support a global symmetry, we can turn on chemical potentials $u_\text{F}$, $\bar u_\text{F}$ for the associated background flat connection. This will simply shift $\Phi_{(0)}\to \Phi_{(0)}+u_\text{F}$ and will resolve the 1-loop singularities into simple poles. Following the discussion in section \ref{sec:contours}, we see that the contours mentioned in (\ref{eq:contours}) must be middle dimensional cycles around the singularities associated with a selection of chiral multiplets with Robin-like conditions, which are indeed the dangerous modes as expected.

The regularization of the 1-loop determinants can be performed by using Hurwitz $\zeta$-function regularization as discussed in appendix \ref{sec:1-loop-details}, with the final result (up to an overall constant)
\begin{align}
\label{eq:Z-1loop-B}
\mathcal{Z}^{\textrm{chi}(\text{D})}_{\textrm{1-loop}}(\Phi_{(0)})
& = \det_{\mathcal{R}}\left[
\dfrac{
	\e^{-\frac{\i\pi}{3} P_3(\sigma(1-r/2)-\Phi_{(0)})}
}{\Gamma(\sigma(1-r/2)-\Phi_{(0)};\tau,\sigma)
}
\right]\,,
\\
\label{eq:Z-1loop-phi}
\mathcal{Z}^{\textrm{chi}(\text{R})}_{\textrm{1-loop}}(\Phi_{(0)})
& = \det_{\mathcal{R}}\left[
	\e^{\frac{\i\pi}{3} P_3(\sigma  r/2 +\Phi_{(0)})} \,\Gamma(\sigma r/2+\Phi_{(0)};\tau,\sigma)
	\right]\,.%
\end{align}
Interestingly, using the shift property (\ref{Gammashift}), we can observe the relation 
\begin{equation}\label{eq:Gammaflip}
	\mathcal{Z}^{\textrm{chi}(\text{D})}_{\textrm{1-loop}}(\Phi_{(0)}) 
	=\mathcal{Z}^{\textrm{chi}(\text{R})}_{\textrm{1-loop}}(\Phi_{(0)}) 
	\det_\mathcal{R}\left[\e^{-\i\pi P_2(\sigma r/2+\Phi_{(0)})}\Theta(\sigma r/2+\Phi_{(0)}; \sigma) \right]
	 \,, 
\end{equation}
where $\Theta$ is the short Jacobi Theta function defined in appendix \ref{appell} and $P_2$ is a quadratic polynomial. In section  \ref{sec:dirichlet-robin}, we will provide a microscopic interpretation of this relation between different boundary conditions in terms of  boundary degrees of freedom.

\subsection{Lower dimensional limits}

The 1-loop determinants for the dimensionally reduced gauge theories on $\mathbb{D}^2\times\mathbb{S}^1$ \cite{Yoshida:2014ssa} and $\mathbb{D}^2$  \cite{Hori:2013ika,Honda:2013uca} can be obtained from the results above by taking suitable limits. These simply amount to discarding either one or two towers of KK modes. For instance, employing Hurwitz $\zeta$-function regularization and focusing on the zero modes $n_x=0$, the determinants in (\ref{eq:1loopB}) and (\ref{eq:1loopphi}) reduce to (up to exponentials of quadratic polynomials)
\be{
\mathcal{Z}^{\textrm{chi}(\text{D})}_{\textrm{1-loop}}(\Phi_{(0)}) \to \det_{\mathcal{R}}\left[
\dfrac{
	1
}{\Gamma(\sigma(1-r/2)-\Phi_{(0)};\sigma)
}
\right]~,\quad 
\mathcal{Z}^{\textrm{chi}(\text{R})}_{\textrm{1-loop}}(\Phi_{(0)})
\to \det_{\mathcal{R}}\left[
	\Gamma(\sigma r/2+\Phi_{(0)};\sigma)
	\right]\,,%
}
where $\Gamma(u;\sigma)\equiv \prod_{n\geq 0}(1-\e^{2\pi\i (u+n\sigma)})^{-1}$ is the inverse of the $q$-factorial (\ref{eq:qPocch}) and hence proportional to the $q$-Gamma function. These results coincide with those in \cite{Yoshida:2014ssa} for chiral multiplets on $\mathbb{D}^2\times\mathbb{S}^1$ with Dirichlet or Neumann boundary conditions respectively. The two determinants are still related by (\ref{eq:Gammaflip}), the Theta function now representing the contribution (elliptic genus) of additional degrees of freedom on the boundary $\mathbb{T}^2$ with modulus $\sigma$. Similarly, by further dropping the KK modes labeled by $n_y$, after regularization one gets (up to exponentials of linear polynomials)
\be{
\mathcal{Z}^{\textrm{chi}(\text{D})}_{\textrm{1-loop}}(\Phi_{(0)})
 \to \det_{\mathcal{R}}\left[
\dfrac{
	1
}{\Gamma(1-r/2-\Phi_{(0)}/\sigma)
}
\right]~,\quad 
\mathcal{Z}^{\textrm{chi}(\text{R})}_{\textrm{1-loop}}(\Phi_{(0)})
 \to \det_{\mathcal{R}}\left[
	\Gamma(r/2+\Phi_{(0)}/\sigma)
	\right]~,
	}
where $\Gamma(u)$ is the ordinary Euler Gamma function. These results coincide with those in \cite{Hori:2013ika,Honda:2013uca}  for chiral multiplets on $\mathbb{D}^2$ with Dirichlet or Neumann boundary conditions respectively. The two determinants are now related by the identity
\be{
\det_{\mathcal{R}}\left[
\dfrac{
	1
}{\Gamma(1-r/2-\Phi_{(0)}/\sigma)
}
\right]= \det_{\mathcal{R}}\left[
	\Gamma(r/2+\Phi_{(0)}/\sigma)
	\right]\ \det_{\mathcal{R}}\left[
	\pi^{-1}\sin\pi( r/2+ \Phi_{(0)}/\sigma)
	\right]~,
}
the Sine function representing the contribution (flavored Witten index) of additional degrees of freedom on the boundary $\mathbb{S}^1$. This is consistent with the limit $\Theta(u;\sigma)\to \sin(\pi u/\sigma)$, up to a $u$-independent divergent factor. The limits for the vector multiplet can similarly be worked out and correctly reproduce the 1-loop determinants for vector multiplets in lower dimensions. 

\subsection{Anomalies and modularity}
The cubic polynomials appearing in the regularization of 1-loop determinants are known to encode the possible gauge, global and mixed-gauge anomalies \cite{Spiridonov:2012ww,Nieri:2015yia}. Therefore, in a physical anomaly-free theory, we can ignore the exponential factors associated with local symmetries, which must cancel out when building the block integral (\ref{eq:contours}). However, the constant terms can have a physical significance (at least in superconformal theories)  since they determine the asymptotic behavior of the partition function as a function of the moduli. In particular, in the case of the index background these terms have been successfully matched against Casimir energies and central charges \cite{DiPietro:2014bca,DiPietro:2016ond,Lorenzen:2014pna,Assel:2015nca,Genolini:2016sxe,Bobev:2015kza,Brunner:2016nyk,Ardehali:2016kza,Shaghoulian:2016gol,Closset:2019ucb}. It would be interesting to understand whether one can extract something new from the disk geometry, also in view of the almost perfect democracy between the torus and disk parameters $\tau$, $\sigma$ and the peculiar modular properties of the partition functions. In fact, it is worth noting that the very same cubic polynomials also appear in the modular transformation properties of the elliptic Gamma functions. These objects are not  sections of a line bundle over an elliptic curve, rather they are sections of a gerbe on the universal triptic curve \cite{felder2008}, and as such they enjoy $\text{SL}(3,\mathbb{Z})\rtimes\mathbb{Z}^3$ modular properties \cite{FELDER200044} rather than $\text{SL}(2,\mathbb{Z})\rtimes\mathbb{Z}^2$ as Theta functions, which do appear as 1-loop determinants of honest torus partition functions (\ref{eq:ellgenus}). Therefore, the $\mathbb{D}^2\times\mathbb{T}^2$ partition functions can be thought as defined on the torus of holonomies only in this generalized sense. In particular, the 1-loop determinants are not invariant under large gauge transformations, unless one imposes special relations between the torus and disk parameters and/or other global fugacities.
For instance, the failure of the double periodicity under $u\to u+\mathbb{Z}+\tau\mathbb{Z}$ is reflected in the law
\be{
\Gamma(u+k+n\tau;\tau,\sigma)=
\Theta(u;\sigma;\tau)_n\, \Gamma(u;\tau;\sigma)~,
}
where the $\Theta$-factorial is defined in appendix \ref{appell}. In view of (\ref{eq:Gammaflip}), we may also say that such a shift induces additional boundary contributions, but unfortunately we do not have a deep understanding for those.\footnote{However, it has been observed that these factors cancel out in compact space partition functions \cite{Nieri:2015yia}, hence supporting the boundary interpretation.}
We also observe that by using the modular property\footnote{
For the $\mathbb{S}^2\times\mathbb{T}^2$ geometry, the action of an $\text{SL}(3,\mathbb{Z})$ subgroup on $(\tau,\sigma)$ gives rise to equivalent complex structures \cite{Closset:2013sxa}, i.e. the same identifications (\ref{twist}).
} 
\be{
\Gamma\Big(u; \tau, \sigma\Big)\Gamma\Big(\frac{u}{\tau};\frac{1}{\tau},\frac{ \sigma}{\tau}\Big)\Gamma\Big(\frac{u}{\sigma};\frac{\tau}{\sigma},\frac{1}{\sigma}\Big)=\e^{-\frac{\i\pi}{3}B_{33}(u;1,\tau,\sigma)}~,
}
the 1-loop determinants may also be rewritten in terms of modified elliptic Gamma functions \cite{2003math......3205S}, which may be useful for studying the unrefined limit $\sigma\to 0$ and also gluings into compact geometries. Finally, it is conceivable that having a field theoretic construction of this class of special functions may be helpful for developing the field even further, especially in the context of automorphic forms or integrable systems.

%%%%%%%%%%%%%%%%%%%%
%%%%%%%%%%%%%%%%%%%%
%%%%%%%%%%%%%%%%%%%%

\section{Boundary supersymmetry}\label{sec:boundary}

In former sections, we have focused on bulk degrees of freedom only. However, the boundary has played several roles, from demanding the addition of certain terms to the actions to preserve supersymmetry, to requiring certain choices of boundary conditions for bulk fields. In flat space, upon imposing half-BPS boundary conditions, the restriction of bulk fields to the boundary gives rise to 3d $\CN=1$ multiplets. In this section, we describe how bulk degrees of freedom split up into boundary multiplets and summarize how the surviving curved space supersymmetry algebra acts on the latter.  

\subsection{3d $\mathcal N=1$  supersymmetry}\label{subsec:3dBil}

In Euclidean signature, the minimal three dimensional flat space supersymmetry algebra is generated by a Dirac supercharge\footnote{In Lorentzian signature, one can impose a Majorana condition on spinors. We will discuss later on what is the Euclidean analog of this condition. We also refer to \cite{VanProeyen:1999ni} for an exhaustive classification of spinors in arbitrary dimension.} $Q_A$ subject to 
\be{\label{eq:3dSUSYalg}
\{Q_A,Q_B\}=-2 \Gamma_{AB}^{\hat a} \text{P}_{\hat a}~,
}
where $\text{P}_{\hat a}$ is the (covariant) momentum operator,  $\hat{a}=\hat 1,\hat 2,\hat 3$  a three dimensional flat index and $\Gamma^{\hat a\, B}_A$  the Dirac matrices 
\be{
\Gamma^{\hat 1}  =-\i\sigma^2~,\qquad \Gamma^{\hat 2}  =\i\sigma^1~,\qquad
\Gamma^{\hat 3} = {-}\i\sigma^3~, \qquad \Gamma^{\hat a}\Gamma^{\hat b} = - \delta^{\hat a \hat b} + \varepsilon^{\hat a \hat b \hat c } \, \Gamma_{\hat c} ~ ,
}
with the convention $ \varepsilon^{\hat 1 \hat 2 \hat 3 } = +1$. 
Spinor indices are raised and lowered with the charge conjugation matrix $\ep^{AB}$. In order to study supersymmetry in a curved but conformally flat space, one can start by  solving  the conformal Killing spinor equation for a Dirac spinor $\xi$
\be{\label{eq: ckse}
\nabla_m \xi+\f13\,\Gamma_m \, \Gamma^n\,\nabla_n \xi=0~,
}
where we use $m,n$ for three dimensional spacetime indices. In the case of our interest, we have to solve (\ref{eq: ckse}) with the line element induced from the bulk. We consider again the K{\"a}hler metric (\ref{eq:Kmetric}), leading to the (flat) induced metric and volume form
\be{
\d \text{s}^2 =\d w \d\bar w + \d t^2~,\qquad \text{vol}_{\mathbb{T}^3}=\frac{\i}{2}\, \d w\wedge\d\bar w\wedge \d t~, 
}
where the adapted ``holomorphic" coordinates $x^m=(w,\bar w,t)$ describe the twisted $\mathbb{T}^3$ given by the global identifications
\be{\label{eq:twistedT3}
(w,t)\sim (w,t+2\pi)\sim (w+2\pi,t+2\pi\alpha)\sim (w+2\pi\tau,t+2\pi\beta)~.
}
Close to the boundary, the coordinate $t=\text{arg}(z)=-\text{arg}(\bar z)$ can be identified with a coordinate along the tangential direction $T$ (\ref{eq:tcoord}) complementary to the normal $N$ (\ref{eq:Nvec}). In the ``holomorphic" frame
\be{
\theta^{\hat 1}+\i\theta^{\hat 2} \equiv \d w~,\qquad \theta^{\hat 3}\equiv \d t~,
}
the general solution to (\ref{eq: ckse}) reads
\be{
\xi_A=\left(\begin{array}{c}a_1\\
 a_2\end{array}\right)_A+x^m {\Gamma_m}_A^{~~ B}\left(\begin{array}{c}b_1\\
 b_2\end{array}\right)_B~,
}
and it encodes four supercharges, one for each constant parameter  $a_i,b_i$ . In particular, $a_i$ correspond to   Poincar\'e supercharges $Q_A$, whereas $b_i$ correspond to superconformal  supercharges, which are however broken due to the identifications  (\ref{eq:twistedT3}).  The Killing spinor
\be{\label{eq:3dxi}
   \xi_A \equiv \f1{\sqrt{2}}\, \delta^+_A~,
}
and its conjugate thus generate the full 3d  $\mathcal N=1$ supersymmetry algebra (\ref{eq:3dSUSYalg}), in particular    
\be{\label{eq:3ddelta}
\{\delta,\delta\}=-2\i\delta_{k}~,
}
where we set $\delta\equiv \xi^A Q_A$ and $\delta_k$ is a (covariant) momentum along the vector
\be{
k^m \equiv \xi\Gamma^m\xi~.
}
Indeed, the Killing spinor $\xi_A$ and its conjugate $\xi^{\dagger\,A}$ can be used to form bilinears yielding a complete three dimensional frame, namely
\be{
\bar k^{\,m}  \equiv \f{\xi^\dagger\Gamma^m\xi^\dagger}{|\xi|^4}\,,\qquad y^m  \equiv  \f{ \xi^\dagger \Gamma^m\xi}{|\xi|^2}~,
}
which in adapted ``holomorphic" coordinates simply reads as
\be{
k=-\partial_{\bar w}~,\qquad \bar k=-4\partial_{w}~,\qquad y= {-}\im\partial_t~,
}
and similarly for the dual 1-forms, which we denote with the same symbols
\be{
k=-\frac{1}{2}\d w~,\qquad \bar k=-2 \d \bar w~,\qquad y= {-} \im \d t~.
}
By using Fierz identities, one finds that Killing spinor bilinears satisfy the relations\footnote{We use the $\varepsilon$ tensor $\varepsilon_{w\bar w t}=\i\sqrt{|g|}\times 1$.}
\be{
k^m{\bar k}^n-y^m y^n=g^{m n} {+} \varepsilon^{mn\ell}\,y_\ell\,,  \qquad  2\,y^{[m} \bar k^{n]} =  \varepsilon^{mn\ell}\bar k_\ell \,,   \qquad  2\,y^{[m}   k^{n]} = {-}    \varepsilon^{mn\ell}  k_\ell  ~,
}
and the vector $y^m$ can be used to define an almost contact metric structure $(y^m, {\Phi^m}_n)$\footnote{Note that we are using conventions where $y$ is purely imaginary.}
\begin{align}
& {\Phi^m}_n \equiv  \im { \varepsilon^m}_{n \ell} \, y^\ell \,, & y^m y_m = -1 \,, \qquad  {\Phi^m}_\ell {\Phi^\ell}_n = -{\delta^m}_n - y^m y_n~.
\end{align}
This is trivially integrable and it induces a complex structure on the transverse torus described by the holomorphic coordinate $w$, w.r.t. which $k^m$ is anti-holomorphic and $\bar k^m$ is holomorphic. In fact, the Killing spinor bilinears, and hence the (integrable) almost contact metric structure, directly descend from the bulk 
\begin{align}\label{eq: bulkframetobdryframe}
K^\mu|_\partial  = & \   - k^{\mu}~, & Y^\mu|_\partial =& \ Y_\perp|_\partial \, (N^\mu|_\partial -y^{\mu})~,\nn\\
%%%%
\bar K^\mu|_\partial  = &\   -\bar k^{\mu}~,  & \bar Y^\mu|_\partial  =& \ \bar Y_\perp|_\partial\, (N^\mu|_\partial + y^{\mu})~, 
%%%%
\end{align}
where the vanishing of the normal component of boundary vectors is to be understood.

%%%%%%%%%%%%%%%%%%%%
%%%%%%%%%%%%%%%%%%%%

\subsection{Supersymmetry multiplets and actions}
In this section, we recall the minimal three dimensional supersymmetry multiplets and construct $\delta$-exact supersymmetric actions.

\subsubsection{Real multiplet}

In Lorentzian signature, 3d $\mathcal N = 1$ matter  is described by the real multiplet   containing a real scalar $\varphi$, a Majorana  spinor $\chi$ and a real auxiliary field $f$. Their supersymmetry transformations are (see e.g. \cite{Gates:1983nr,Drukker:2017xrb,Drukker:2017dgn})
\be{
 \delta \varphi = \xi\chi\,,\qquad \delta \chi =  f\, \xi -\im \Gamma^m \xi \, \partial_m\varphi \,,\qquad  \delta f = - \im \xi\Gamma^m\partial_m\chi \,.
%%%%
}
In Euclidean signature,  the definition of a single real  multiplet is problematic due to the absence of Majorana spinors, even though attempts  to overcome the obstruction  have appeared in the literature   \cite{Nicolai:1978vc, vanNieuwenhuizen:1996tv}. Therefore, in our setup all the multiplets are actually complex. However, we can mimic a real setup by assuming that all the fields transform under a $\text{USp}(2N_\text{f})$ global symmetry as the latter  allows for describing fermionic fields by means of symplectic Majorana spinors  $\chi^i$, $i=1,\dots, 2N_\text{f}$. These can be defined in 3d Euclidean signature, and they satisfy the reality condition 
\be{
\p{\chi^{i\,A}}^\dagger=\chi_{i\,A} \equiv \omega_{ij}\epsilon_{AB}\chi^{j\, B}~,
}
where $\omega_{ij}$ is the standard invariant symplectic form. Slightly more generally, given  $2N_\text{f}\times 2 N_\text{c}$      real multiplets $\p{\varphi^{ia},\chi^{ia},f^{ia}}$ transforming in  the bifundamental representation of  $\text{USp}(2N_\text{f})\times \text{USp}(2N_\text{c})$, where $i=1,\dots , 2N_\text{f}$ is a flavour index and $a=1,\dots , 2N_\text{c}$ a gauge index, we have the supersymmetry transformations 
\be{\label{eq: euclsymprealmult}
 \delta \varphi^{ia} = \xi\chi^{ia} \,,\qquad \delta \chi^{ia}  =  f^{ia} \, \xi -\im \Gamma^m\xi \, \mathcal D_m\varphi^{ia}  \,,\qquad  \delta f^{ia}  =  -\im \xi\Gamma^m\mathcal D_m\chi^{ia}  \,,
%%%%
}
with the fields fulfilling the reality conditions 
 \be{ 
\p{\varphi^{ia}}^\dagger = \varphi_{ia}\,,\qquad \p{\chi^{ia}}^\dagger = \chi_{ia}\,,\qquad \p{ f^{ia}}^\dagger = - f_{ia}\,,
 }
where indices are raised and lowered by the corresponding symplectic form.\footnote{These reality conditions are reminiscent, but do not coincide with, those for symplectic Majorana spinors. The reason is that they involve the product of two antisymmetric invariant tensors (respectively for the groups $\text{USp}(2N_\text{c})$ and $\text{USp}(2N_\text{f})$), resulting in a  overall symmetric invariant tensor. Similar conditions are also considered e.g. in \cite{Benini:2017aed}. We thank the referee for calling our attention to this detail.} The covariant derivative $\mathcal D_m$ may  contain a background field $\textsl{v}_m$ in the adjoint representation of $\text{USp}(2N_\text{f})$ as well as a gauge field $\textsl{a}_m$ in the adjoint of $\text{USp}(2N_\text{c})$, namely
\be{
\mathcal D_m\varphi^{ia}\equiv\partial_m \varphi^{ia}-\im {{\textsl{v}_m}^i}_j\,\varphi^{ja}-\im {{\textsl{a}_m}^a}_{b}\,\varphi^{ib}~.
}
The supersymmetric variation  of the fermionic functional 
\be{\label{eq: euclrmdefterm}
 \mathscr{V}_{\rm RM}\equiv\f1{|\xi|^2}\delta\chi^{ia}  \chi_{ia} = -\f1{|\xi|^2}f^{ia}\, \xi^\dagger \chi_{ia}  -\f{\im}{|\xi|^2} \xi^\dagger\Gamma^m \chi_{ia}  \, \mathcal D_m\varphi^{ia}
}
produces by construction a $\delta$-exact Lagrangian with positive semi-definite bosonic term 
\be{\label{eq: euclrmactioncan}
\mathscr{L}_{\rm RM} \equiv \delta \mathscr{V}_{\rm RM} =\mathcal D_m\varphi^{ia}\mathcal D^m\varphi_{ia}-f^{ia}f_{ia}-\im\chi^{ia}\Gamma^m\mathcal D_m\chi_{ia} \,,
}
where we omitted the total derivative $\mathcal \partial_m\comm{-\im(\xi\chi_{ia} )(\chi^{ia}\Gamma^m\xi^\dagger)}$. 

 \textbf{Remark}. In order to avoid cluttering, in the following will often omit any gauge index when not explicitly needed.
 
%%%%%%%%%%%%%%%%%%%%

\subsubsection{Vector multiplet}

The 3d $\mathcal{N}=1$ vector multiplet consists of a gauge field $\textsl{a}_m$ and a Dirac spinor $\rho$ in the adjoint representation of the gauge group. As before, for $\text{USp}(2N_\text{c})$ we can impose a reality condition. The supersymmetry transformations read as follows
\be{\label{eq: 3dN1vecmultcfsusytransf}
 \delta \textsl{a}_m = -\im \xi \Gamma_m \rho ~,\qquad \delta \rho  = {\f12} \varepsilon^{mn\ell}\textsl{f}_{mn}\,\Gamma_\ell \xi  ~.
}
The latter can be used to write down a $\delta$-exact  Lagrangian containing a Yang-Mills term for the gauge field and a kinetic term for the gaugino (a $\text{Tr}$ is left implicit for non-Abelian theories)
\be{
\mathscr{V}_\text{VM}^\text{3d}\equiv \frac{1}{2|\xi|^2}\, \p{\delta\rho}^\dagger \rho~,\qquad \mathscr{L}^{\rm 3d}_{\rm VM} \equiv \delta  \mathscr{V}_\text{VM}^\text{3d}= \frac{1}{4} \textsl{f}_{mn}\textsl{f}^{\,mn} - { \frac{\i}{2}} \rho\, \Gamma^m\mathcal D_m \rho ~.
}
Furthermore, one can also consider a Chern-Simons term at level $k\in \mathbb Z$ given by
\be{
\mathscr{L}^{\rm 3d}_{\rm CS} \equiv \f{\i k}{4\pi}\left(\varepsilon^{mn\ell}(\textsl{a}_m\partial_n \textsl{a}_\ell -\f{2\im}3\textsl{a}_m \textsl{a}_n \textsl{a}_\ell)   - \im \rho\rho\right) ~,
}
which is $\delta$-closed.

%%%%%%%%%%%%%%%%%%

\subsubsection{Cohomological fields} 

By contracting all the spinors with the Killing spinor $\xi$ and its conjugate $\xi^\dagger$, we can introduce fermionic scalar variables as we did for the bulk multiplets.

\textbf{Real multiplet}. For the matter spinor fields $\chi^{i}$ we introduce 
\be{
c^{i} \equiv \xi\chi^{i}\,,\qquad b^{i}\equiv\frac{\xi^\dagger}{|\xi|^2}\chi^{i}\,,\qquad \chi^{i} \equiv \xi b^{i} - \frac{\xi^\dagger}{|\xi|^2} c^{i}\,.
}
These allow us to rewrite the supersymmetry transformations for the real multiplet as
\begin{align}\label{eq:3dcoh}
\delta\varphi^{i} =& \  c^{i}~, & \delta c^{i} =& \ -\im\mathcal L_k\varphi^{i}~,\nn\\
\delta b^{i}=& \ f^{i} -\im \mathcal L_y \varphi^{i}~, & \delta f^{i}=&\  -\im\mathcal L_k b^{i}+\im\mathcal L_y c^{i}~,
\end{align}
where $\mathcal{L}_k$, $\mathcal{L}_{\bar k}$, $\mathcal{L}_y$ denote the total covariant derivatives along $k$, $\bar k$ and $y$ respectively. Note  that the supersymmetry algebra in this notation is given by (\ref{eq:3ddelta}) with $\delta_k =\mathcal{L}_k$. The real multiplet Lagrangian (\ref{eq: euclrmactioncan})  in terms of cohomological fields reads as
\begin{align}\label{eq: rmactioncoh}
\mathscr{L}_{\rm RM} &\  =\mathcal L_{\bar k}\varphi^{ia}\mathcal L_k\varphi_{ia}-\mathcal L_y\varphi^{ia}\mathcal L_y\varphi_{ia} -f^{ia}f_{ia}+\nn\\
&\qquad\qquad\qquad - \im  b^{ia}\mathcal L_k b_{ia}+\im  c^{ia}\mathcal L_y b_{ia} +\im  b^{ia}\mathcal L_y  c_{ia} -\im  c^{ia}\mathcal L_{\bar k} c_{ia}\,.
\end{align}
We recall that this  action is $\delta$-exact w.r.t. the fermionic functional (\ref{eq: euclrmdefterm}), which in cohomological fields is simply 
\be{
 \mathscr{V}_{\rm RM}= \delta b^{ia} b_{ia}+ |\xi|^{-4}\delta c^{ia} c_{ia}= -f^{ia}\,b_{ia}  - \im b_{ia}\mathcal L_y\varphi^{ia} +  \im c_{ia}\mathcal L_{\bar k}\varphi^{ia} \,. 
}

\paragraph{Vector multiplet.} Similarly, for the vector multiplet we can introduce  the odd scalars
\be{
\bar\psi_\text{v} \equiv 2\im {\sqrt2}\,\frac{\xi^\dagger}{|\xi|^2} \rho\,, \qquad  \lambda_\text{v} \equiv {\sqrt2}\, \xi\rho \,, \qquad \rho \equiv \frac{\xi}{2\im {\sqrt2}\,}   \bar\psi_\text{v} - \frac{\xi^\dagger}{{\sqrt2}\,|\xi|^2} \lambda_\text{v} \,,
}
and then the supersymmetry transformations are 
\begin{align}
& \delta \textsl{a}_m = -\frac{1}{2 {\sqrt2}\, }  k_m  \bar\psi_\text{v} + \f\i{{\sqrt2}\,} y_m \, \lambda_\text{v}\,, \quad \delta \lambda_\text{v} =  {\sqrt2}\, k^ m y^n \,   \, \textsl{f}_{mn}  \,, \quad \delta  \bar\psi_\text{v} =    \im {\sqrt2}\, k^m \, \bar k^n \, \textsl{f}_{mn}  \,.
\end{align}
Note that the variation of the gauge field is missing a component, namely $\delta(k^m \textsl{a}_m) =0$. This is again due to the fact that $k^m \textsl{a}_m$ appears as a gauge parameter in $\delta^2$. For later purposes, it is  convenient to rewrite the cohomological vector multiplet in terms of scalar components\footnote{To be precise, such components are scalars up to shifts induced by gauge transformations.} obtained by contraction with the Killing spinor bilinears:
\be{
\textsl{a}_\|\equiv \i y^m \textsl{a}_m~,\qquad \phi_\textrm{v}\equiv k^m \textsl{a}_m~,\qquad \bar\phi_\textrm{v}\equiv \bar k^m \textsl{a}_m~,
}
so that the supersymmetry transformations are 
\begin{align}\label{eq:3dveccoh}
\delta \textsl{a}_\| = & \ \frac{1}{\sqrt{2}}\, \lambda_\text{v}\,~,& \delta \lambda_\text{v}  = & \  {\sqrt2}\,  k^ m y^n \,   \, \textsl{f}_{mn}~,\nn\\
\delta \bar\phi_\text{v} = & \ - \frac{1}{\sqrt{2}}\, \bar\psi_\text{v}~, & \delta  \bar\psi_\text{v} = & \    \im{\sqrt2}\, k^m \, \bar k^n \, \textsl{f}_{mn}~, & \delta \phi_\text{v} = & \ 0~.
\end{align}
Since we will not need to consider boundary vector multiplets, we omit writing their lagrangian in cohomological form.

%%%%%%%%%%%%%%%%%%%%

\subsection{Induced supersymmetry}

After having reviewed basic facts about three dimensional minimal  supersymmetry which are relevant for our boundary, we are ready to study in more detail how the bulk supersymmetry acts on the latter. In particular, since  the supercharge $\Q$ played a major role for localization in the bulk, the natural question is whether it can be interpreted as a boundary supercharge as well. In this case, it is possible to consider bulk-boundary coupled systems and perform (in principle) localization w.r.t. the very same supercharge. In flat space, the boundary would preserve half of the bulk supercharges at most. In curved space, the full supersymmetry algebra is already broken down to the subalgebra (\ref{deltaK}), and we recall that the selected localizing supercharge is the linear combination $\Q=\delta_\zeta+\delta_{\tilde\zeta}$ of supercharges preserved by the bulk. Therefore, the minimal setup we would like to consider is when $\Q$ is a preserved supercharge of the 3d $\mathcal{N}=1$ supersymmetry algebra we have just reviewed. Following our previous discussion, it is clear that the correct candidate is the supercharge $\delta$ defined in  (\ref{eq:3ddelta}) and associated with the spinor (\ref{eq:3dxi}). In this section, we show how this comes about, closely following the approach of \cite{Drukker:2017dgn,Drukker:2017xrb}.

We start by considering embedding functions $\eta^\alpha_A$, $\tilde\eta^{\dot\alpha}_A$ and the following linear combination of bulk supercharges
\be{
\mathcal{Q}_A\equiv \frac{1}{\sqrt{2}}\left(\eta^\alpha_A Q_\alpha+\tilde\eta^{\dot\alpha}_A \tilde Q_{\dot\alpha}\right)~,
}
which, by plugging in (\ref{eq:4dSUSYalggg}),  satisfy the algebra
\be{\label{eq:3dSUSY}
\{ \Q_A,\Q_B\}=\eta^\alpha_{(A}\tilde\eta^{\dot\alpha}_{B)}\{ Q_\alpha,\tilde{Q}_{\dot\alpha}\}=2\eta^\alpha_{(A}\tilde\eta^{\dot\alpha}_{B)}\sigma^\mu_{\alpha\dot\alpha} \text{P}_{\mu}~,
}
where $A,B$ are $\text{SU}(2)$ indices  to be identified with boundary spinor indices. In order for the algebra (\ref{eq:3dSUSY}) to match with the algebra (\ref{eq:3dSUSYalg}) and hence for it to describe the minimal supersymmetry of a boundary theory, the embedding functions must satisfy certain conditions. First of all, the matrix
\be{\label{eq:Gammaind}
\Gamma^\mu_{(AB)}\equiv -\eta^\alpha_{(A}\tilde\eta^{\dot\alpha}_{B)}\sigma^\mu_{\alpha\dot\alpha}
}
must act as a projector on the transverse space w.r.t. the normal $N$, that is
\be{
\Gamma^\perp_{(AB)} =0~.
}
Therefore, the projection on the normal must be captured by the anti-symmetric part
\be{
\ep^{AB}\eta^\alpha_{A}\tilde\eta^{\dot\alpha}_{B}\equiv \tilde\sigma^{\dot\alpha\alpha}_\perp ~.
}
Then, (\ref{eq:Gammaind}) can be interpreted as boundary Dirac matrices, namely they must satisfy
\be{\label{eq:3dDiracAlg}
\{\Gamma^\mu,\Gamma^\nu\}_A^{~~B}=-2(g^{\mu\nu}-N^\mu N^\nu)\delta_A^B~,
}
where indices are moved with $\epsilon^{AB}$. The orthogonality condition can be solved by setting $\Gamma^\mu_{(AB)}\equiv (\Gamma^{\mu\nu}N_\nu)_{(AB)}$ for some matrix-valued anti-symmetric 2-tensor $\Gamma^{\mu\nu}$, and we can almost immediately identify (up to signs)
\be{
\Gamma^\mu_{(AB)}\equiv -2\sigma^{\mu\nu}_{(AB)}N_\nu=2\sigma^{\perp \mu}_{AB}~.
}
Indeed, one can verify that (\ref{eq:3dDiracAlg}) is satisfied and the algebra  (\ref{eq:3dSUSY}) reproduces (\ref{eq:3dSUSYalg}). A concrete choice for the embedding functions can be given by noticing that $\Q_A$ must be R-symmetry neutral and that in three dimensions there is no distinction between dotted and undotted spinors, a convenient identification being provided by $\sigma^\perp_{\alpha\dot\alpha}$ and $\tilde\sigma_\perp^{\dot\alpha\alpha}$. For definiteness, we identify the $\text{SU}(2)$ indices with undotted indices, and then it is easy to show that the following choice meets all the requirements\footnote{Because of the four dimensional R-symmetry, this identification is up to local phase rotations.}
\be{
\eta^\alpha_A \equiv \sqrt{Y_\perp}\, \delta^\alpha_A~,\qquad \tilde\eta^{\dot\alpha}_A \equiv - \sqrt{\bar Y_\perp}\, \tilde\sigma_\perp^{\dot\alpha\beta}\epsilon_{\beta \alpha}\delta^\alpha_A~,
}
where we set $Y_\perp\equiv \iota_N Y$, $\bar Y_\perp\equiv \iota_N \bar Y$. Let us now define a spinor $\xi_A$ such that
\be{
\zeta^\alpha\equiv\xi^A\eta_A^\alpha =\xi^A\sqrt{Y_\perp}\,\delta^\alpha_A~,\qquad \tilde\zeta^{\dot\alpha}\equiv-\xi^A\tilde\eta^{\dot\alpha}_A=\xi^A\sqrt{\bar Y_\perp}\, \tilde\sigma_\perp^{\dot\alpha\beta}\epsilon_{\beta \alpha}\delta^\alpha_A~.
}
Then it follows that the supercharge $\xi^A \Q_A$ can be identified (up to a factor of $\sqrt{2}$) with the localizing supercharge $\Q$. Moreover, since we have the inverse relations $\eta^\alpha_A \eta^B_\alpha=- Y_\perp\delta_A^B$, $\tilde\eta^B_{\dot\alpha}\tilde\eta^{\dot\alpha}_A=   \bar Y_\perp\delta^B_A$,
we can write the spinor $\xi_A$ also as
\be{
\xi_A=-\ep_{AB}\frac{\eta^B_\alpha}{Y_\perp}\zeta^\alpha=\frac{1}{\sqrt{Y_\perp}}\,\delta^\alpha_A\zeta_\alpha~,\qquad \xi_A=-\ep_{AB}\frac{\tilde\eta^B_{\dot\alpha}}{\bar Y_\perp}\tilde\zeta^{\dot\alpha}=-\frac{1}{\sqrt{\bar Y_\perp}}\delta^\alpha_A\sigma^\perp_{\alpha\dot\alpha}\tilde\zeta^{\dot\alpha}~,
}
or more democratically
\be{\label{eq:4d-3d-spinors}
\xi_A=-\frac{\ep_{AB}}{2}\left(\frac{\eta^B_\alpha}{Y_\perp}\zeta^\alpha+\frac{\tilde\eta^B_{\dot\alpha}}{\bar Y_\perp}\tilde\zeta^{\dot\alpha}\right)~.
}
One can verify that the spinor bilinears given in section \ref{subsec:3dBil} are correctly reproduced, allowing us to finally identify $\xi_A$ defined here with the spinor defined in (\ref{eq:3dxi}),\footnote{Note that, because of the different frames, there is a relative Lorentz rotation to take into account. This does not affect the quantities with all the spinor indices contracted.} and hence $\xi^A \Q_A$ with the supercharge $\delta$ defined in (\ref{eq:3ddelta}).

\subsection{Mapping 4d multiplets to  3d multiplets}

We are now ready to discuss how bulk and boundary supersymmetry can be related. 

\textbf{Matter multiplets}. Let us start by considering 4d $\mathcal{N}=1$ chiral and anti-chiral multiplets, transforming in the fundamental and anti-fundamental representations of the gauge group $\text{G}$ respectively. If these representations are $N_\text{c}$ dimensional,  they  can be accommodated into a pair of 3d $\mathcal N=1$ real multiplets transforming in the fundamental of $\text{USp}(2N_\text{c})$ by  embedding   the $N_\text{c}\oplus \overline N_\text{c}$ representation of $\text{G}$ into the fundamental of $\text{USp}(2N_\text{c})$.\footnote{This is not necessary but we decided to mimic a real structure as discussed before.} For concreteness, we may consider $\text{G}=\text{(S)U}(N_\text{c})$. Using twisted variables, we can then write  $(\phi^a,C^a,B^a,F^a)$ for the chiral multiplet and  $(\tilde \phi^a,\tilde C^a,\tilde B^a,\tilde F^a)$ for the anti-chiral multiplet, where $a=1,\dots ,2N_\text{c}$ is a $\text{USp}(2N_\text{c})$ index, and eventually set the last $N_\text{c}$ components to zero to recover the original multiplets. In analogy with    \cite{Drukker:2017xrb,Drukker:2017dgn, Dedushenko:2018tgx}, we introduce 
\begin{align}\label{eq:3dcurvedtwistedfieldschi}
 \phi^{ia} \equiv & \   \phi^a \, \delta^i_1 -    \tilde \phi^a \, \delta^i_2  ~ , \nn\\
C^{ia} \equiv & \   C^a \, \delta^i_1 -    \tilde C^a \, \delta^i_2 ~,\nn\\
B^{ia} \equiv & \  \p{\bar Y_\perp}^{-1} B^a \, \delta^i_1 -      \p{  Y_\perp}^{-1} \tilde B^a\, \delta^i_2 ~ ,   \qquad \qquad \qquad   ~ , \nn\\
 F^{ia} \equiv & \  \delta^i_1 \comm{\p{\bar Y_\perp}^{-1} F^a-\im \mathcal L_\perp\phi^a}-\delta^i_2 \comm{   \p{  Y_\perp}^{-1} \tilde F^a + \im \mathcal L_\perp\tilde\phi^a } ~ ,
\end{align}
where the  index $i=1,2$ can be associated with a (broken) $\text{USp}(2)$ flavor symmetry. 
In particular, the new bosonic fields can be taken to satisfy the reality conditions
\begin{align}
& \p{\phi^{ia}}^\dagger = \phi_{ia} \equiv \epsilon_{ij}\,\omega_{ab}\,\phi^{\,jb}\,, & \p{F^{ia}}^\dagger = - F_{ia} \equiv -\epsilon_{ij}\,\omega_{ab}\,F^{\,jb} ,
\end{align}
which can be though of as imposing the real contour $(\phi,F)^\dagger=(\tilde \phi,-\tilde F)$ in terms of the original bulk fields. A similar reality condition is to be imposed on the fermions.

Having set up some notation, we can now rewrite in 3d language the SUSY complex  of the 4d chiral multiplet. Using the relations (\ref{eq: bulkframetobdryframe}), we can write the bulk transformations at the boundary as
\begin{align}
& \mathcal Q \phi=\sqrt2\, C\,, \qquad  \mathcal Q \tilde \phi=\sqrt2\, \tilde C\,,  \nn\\
%%%%
& \mathcal Q\comm{ \p{\bar Y_\perp}^{-1} B} = \sqrt2\comm{ \p{\bar Y_\perp}^{-1} F - \im\mathcal L_\perp \phi - \im \mathcal L_y \phi} \,,\nn\\
%%%%
&  \mathcal Q\comm{  \p{  Y_\perp}^{-1} \tilde B} = \sqrt2\comm{  \p{  Y_\perp}^{-1} \tilde F + \im \mathcal L_\perp\tilde\phi-\im\mathcal L_y\tilde\phi} \,, \nn\\
%%%%
& \mathcal Q \comm{ \p{\bar Y_\perp}^{-1} F - \im\mathcal L_\perp \phi} = -  \im \p{\bar Y_\perp}^{-1} \sqrt2 \mathcal L_k B +     \im  \sqrt2\mathcal L_{ y }C \,,  \nn\\
%%%%
& \mathcal Q \comm{  \p{ Y_\perp}^{-1} \tilde F + \im \mathcal L_\perp\tilde\phi} = - \im  \p{ Y_\perp}^{-1} \sqrt2 \mathcal L_k \tilde B   +  \im\sqrt2 \mathcal L_y \tilde C  \,,
\end{align}    
which are easily seen to coincide with the 3d $\mathcal{N}=1$ cohomological complex (\ref{eq:3dcoh}) by using the definitions (\ref{eq:3dcurvedtwistedfieldschi}) and the capital/lower case map
\be{
\Q\to \sqrt{2}\delta~,\qquad (\phi^{i},C^{i})\to (\varphi^{i},c^{i})~,\qquad (B^{i},F^{i})\to (b^{i},f^{i})~.
}
The covariant derivatives along the boundary directions contain the induced gauge and R-symmetry connections, namely
\be{
\mathcal D_m \equiv \nabla_m -\i \A_m-\im r A^{(R)}_m~,\qquad  {A_m^{(R)\;i}}_{~j} \equiv A_{m} {R^{i}}_j\,,
}
where ${R^i}_j$ is the diagonal Pauli matrix. Note that all the fields have now the same R-charge magnitude, with the $i=1,2$ components representing the $\pm 1$ eigenspaces. Also, the background $A^{(R)}$ explicitly breaks the global $\text{USp}(2)$ to  its $\text{U}(1)$ Cartan. Equivalently, the bulk R-symmetry is realized at the boundary as the Cartan of the putative $\text{USp}(2)$ global symmetry for which a Wilson line is turned on.

\textbf{Vector multiplets}. Working with scalar components only, at the  boundary the gauge field can be split as $\mathcal A_\mu = \p{\A_\perp  ,  \mathcal A_m} \equiv \p{ \A_\perp,   \A_\| , \bar\Phi , \Phi}$, while the gauginos as $(\lambda_\alpha,\tilde\lambda^{\dot\alpha}) \equiv \p{\Lambda_\perp , \Lambda_\|, \Psi,\bar\Psi}$, where we defined
\begin{align}\label{eq: 3dn1veccohfields}
 \A_\perp \equiv & \ \f12\p{ {\bar Y_\perp} Y^\mu  + {  Y_\perp} \bar  Y^\mu }\mathcal A_\mu \,, &  \A_\| \equiv& \ \f1{2\im} \p{ {\bar Y_\perp} Y^\mu  -  {  Y_\perp} \bar  Y^\mu } \mathcal A_\mu = \im y^m  \mathcal A_m \,, \nn\\
%%%%
 \bar\Phi  \equiv&\  - \bar K^\mu \mathcal A_\mu = \bar k^m \mathcal A_m \,, &    \Phi  \equiv& \  -   K^\mu \mathcal A_\mu =  k^m \mathcal A_m \,,  \nn\\
%%%%
 \Lambda_\perp \equiv& \  \f\i{{\sqrt2}\,}  {\bar Y_\perp} \p{\zeta\lambda} -  \f\i{{\sqrt2}\,}   {\bar Y_\perp} \p{ \tilde \zeta \tilde \lambda} \,, & \Lambda_\| \equiv &\     {\bar Y_\perp}\p{\zeta\lambda}  +  {  Y_\perp}\p{\tilde \zeta \tilde \lambda }  \,, \nn\\
%%%%
 \bar\Psi \equiv & \ 2\i\left(\f{\zeta^\dagger\lambda}{|\zeta|^2} + \f{\tilde\zeta^\dagger \tilde\lambda}{| \tilde \zeta|^2} \right)\,, & \Psi_\perp \equiv & -\f\Psi{{2\sqrt2}\,} = -\f{\im}{{\sqrt2}\,} \left(  \f{\tilde\zeta^\dagger \tilde\lambda}{| \tilde \zeta|^2} -\f{\zeta^\dagger\lambda}{|\zeta|^2}\right)\,. 
\end{align}
Moreover, normal derivatives of fields yield or redefine auxiliary fields
\be{
D_\A \equiv   {{-}\,} \p{ D - \mathcal D_\perp \A_\|}  ~,\qquad D_\Phi \equiv \  \mathcal D_\perp \Phi ~.
}
In this notation,  part of the 4d $\mathcal{N}=1$ vector multiplet  gives rise to  a  3d  $\mathcal N=1$ vector multiplet with supersymmetry transformations
\begin{align}\label{eq:bdryvecmult}
\Q \A_\| =& \  \Lambda_\| \,, & \Q \Lambda_\| = & \ 2\, k^m \, y^n \, \F_{mn} \,, &  \nn\\
%%%%
\Q \bar\Phi = & \ -\bar\Psi \,,   & \Q \bar\Psi =&\  2\im k^m \, \bar k^n \, \F_{mn}\,,   & \qquad   \Q \Phi =& \  0\,, 
\end{align}
which is indeed identical to the  cohomological complex (\ref{eq:3dveccoh}) upon the identifications  
\be{
 \Q\to \sqrt{2} \, \delta ~,\qquad ( \A_\|  , \Lambda_\|  ) \to \p{ \textsl{a}_\| ,  {\lambda_\text{v}} } ~ , \qquad ( \bar\Phi  , \bar\Psi  ) \to   \p{ \bar\phi_\text{v} ,  \bar\psi_\text{v} }~, \qquad \Phi \to \phi_\text{v}  ~.
}
The remaining fields have supersymmetry transformations
\begin{align}\label{eq:bdryvecrealmult}
 \Q \A_\perp & = \ {{\sqrt2}\,}  \Lambda_\perp \,, & \Q \Lambda_\perp = & \ -\im {{\sqrt2}\,}  \mathcal L_k \A_\perp +\im {{\sqrt2}\,}  D_\Phi\,, \nn\\
%%%%
 \Q \Psi_\perp & = \     {{\sqrt2}\,}  D_\A -  \im {{\sqrt2}\,}  \mathcal L_y \A_\perp \,,  &  \Q D_\A = & \ -\im{{\sqrt2}\,} \mathcal L_k \Psi_\perp  +\im{{\sqrt2}\,}  G_\Phi \Psi_\perp +\im  {{\sqrt2}\,}  \mathcal L_y \Lambda_\perp \,,\nn\\
\Q D_\Phi & =\    {{\sqrt2}\,}  G_\Phi \Lambda_\perp ~.
\end{align}
Up to the additional auxiliary field $D_\Phi$, which is a remnant of the  bulk gauge symmetry, (\ref{eq:bdryvecrealmult}) are the  supersymmetry transformations of an adjoint real multiplet. Indeed, they coincide with (\ref{eq:3dcoh}) if the map
 \be{
 \Q\to \sqrt{2} \, \delta ~,\qquad ( \A_\perp  ,  \Lambda_\perp  ) \to (\varphi ,  c )~,\qquad ( \Psi_\perp , D_\A )\to (  b ,   f ) ~ ,
}
is performed. In particular, we see that Dirichlet or Neumann boundary conditions on the bulk vector multiplet correspond to killing either the 3d $\mathcal{N}=1$ vector or real adjoint multiplets respectively at the boundary. This is an example of dual boundary conditions, which we study in more detail in the next section.

%%%%%%%%%%%%%%%%%%%%
%%%%%%%%%%%%%%%%%%%%
%%%%%%%%%%%%%%%%%%%%

\section{Dual boundary conditions from 3d multiplets}\label{sec:bc-flips}

We will now explore basic aspects of a rather interesting interplay between boundary conditions for bulk fields and couplings to boundary degrees of freedom. 
Using the $\mathbb{D}^2\times \mathbb{T}^2$ partition functions computed previously, together with the discussion of boundary supersymmetry of the previous section, we study how it is possible to change the boundary conditions through the inclusion of boundary fields. 
Our main goal is to give a microscopic derivation of the relation (\ref{eq:Gammaflip}). Very similar phenomena were discovered for 3d $\CN=2$ theories with 2d $\mathcal{N}=(0,2)$ boundary degrees of freedom in \cite{Dimofte:2017tpi}. We expect that many results of this reference should have an interesting uplift to four dimensions, vastly generalizing the discussion of this section. 

%%%%%%%%%%%%%%%%%%%%
 
\subsection{Chiral multiplet boundary conditions}

Let us start by writing down the action of a  4d $\mathcal N=1$ chiral multiplet  in  3d $\mathcal N=1$ language. For $F_\text{on-shell}=\tilde F_\text{on-shell}=0$, we find\footnote{We focus on the chiral multiplet alone, omitting the coupling to the vector multiplet.}  
\begin{align}
\mathscr{L}_{\rm chi}= & \ {\mathcal L}_{\bar K}\tilde \phi_a  {\mathcal L}_{  K}  \phi^a  + {\mathcal L}_{\bar Y} \phi^a  {\mathcal L}_{  Y} \tilde  \phi_a  + \im \tilde B_a {\mathcal L}_K B^a +\im \tilde B_a  {\mathcal L}_{\bar Y}C^a -\im  B^a {\mathcal L}_Y\tilde C_a + \im \tilde C_a {\mathcal L}_{\bar K}C^a \nn\\
= & \ \f12 {\mathcal L}_{\bar k} \phi_{ia}  {\mathcal L}_{  k}  \phi^{ia}  + \f12\mathcal L_\perp \phi^{ia}\mathcal L_\perp \phi_{ia} -  \f12\mathcal L_y \phi^{ia}\mathcal L_y \phi_{ia}+  \mathcal L_\perp \phi_{ia}{R^i}_j\mathcal L_y\phi^{ja} +\nn\\
   &\qquad\qquad\qquad - \f\im2\,  B_{ia} {\mathcal L}_k B^{ia} +\im\,  B_{ia} {R^i}_j {\mathcal L}_{\perp }C^{ja}    +\im B_{ia}\mathcal L_y C^{ia} - \f \im2\,  C_{ia} {\mathcal L}_{\bar k}C^{ia}\,.
   %%%%
\end{align}

We may ask whether this Lagrangian is $\mathcal Q$-exact with respect to the boundary supersymmetry. To answer this question, we can consider  the fermionic deformation term    
\begin{multline}
4 \mathscr V^{\rm (R)}_{\rm chi}  \equiv   (\mathcal Q\, B^{ia}) B_{ia}+|\xi|^{-4}(\mathcal Q \, C^{ia} )C_{ia} -2 \im \sqrt2\, B_{ia}{R^i}_j \mathcal L_\perp \phi^{ja}=\\
=   -\sqrt2 \p{F^{ia}+\im\mathcal L_y\phi^{ia}}B_{ia}+ \im\sqrt2\mathcal L_{\bar k}\phi_{ia} C_{ia} - 2\im \sqrt2\, B_{ia}{R^i}_j \mathcal L_\perp \phi^{ja}\,, 
\end{multline}
  leading   to the  $\mathcal Q$-exact Lagrangian
\begin{multline}\label{eq: n1chimult3dtf}
 \mathscr{L}^{\rm (R)}_{\rm chi}  \equiv    \mathcal Q \mathscr{V}^{\rm (R)}_{\rm chi} =   \f12 \mathcal L_{\bar k}\phi^{ia} \mathcal L_k\phi_{ia} - \f12 { \mathcal L_y\phi^{ia}} {\mathcal L_y\phi_{ia}}   - \im  \p{F_{ia}-\im\mathcal L_y\phi_{ia}}{R^i}_j \mathcal L_\perp \phi^{ja} +\\
 - \f12 {F^{ia} } {F_{ia} } - \f\im2 B^{ia}   \mathcal L_k B_{ia}+\im B^{ia} \mathcal L_y C_{ia} - \f\im2 C^{ia} \mathcal L_{\bar k} C_{ia} +  \im \, B_{ia}{R^i}_j \mathcal L_\perp  C^{ja} \,.
\end{multline}
We have that $F_\text{on-shell}^{i} = -\im {R^i}_j\,\mathcal L_\perp \phi^{j}$, thus the actions  $\mathscr S_{\rm chi}$ and  $ \mathscr  S^{\rm (R)}_{\rm chi}$ are identical  upon integrating out the auxiliary fields. Moreover, from (\ref{eq: n1chimult3dtf}) we see that the  equations of motion  of $B_{i}, F_{i}$  give bulk terms only. Instead, those of $\phi_{i}, C_{i} $ yield  boundary terms
\be{\label{eq: 4deomphic}
	\hat\delta(\phi,C) \mathscr  S^{\rm (R)}_{\rm chi}  =  \text{bulk e.o.m.} + \i  \int_\partial\sqrt{g}\, \d^3 x \left( {B_{ia}{R^i}_j   \hat\delta C^{ja}  - \p{F_{ia}-\im\mathcal L_y\phi_{ia}}{R^i}_j \hat\delta \phi^{ja}}\right)~, 
}
where $\hat\delta$ denots field variations producing the equations of motion, not to be confused with supersymmetry variations. The boundary terms in (\ref{eq: 4deomphic}) can be removed  in two ways compatible with supersymmetry:
\begin{itemize}
	\item Dirichlet  boundary conditions
	\begin{align}  
	(\text{D}): \ \ 
		 & \phi_{i} |_\partial =\text{const.}\,,  &   C_{i} |_\partial =\text{const.} \, . 
	\end{align} 
	\item Robin-like  boundary conditions
	\begin{align}  
	(\text{R}): \ \ 
		 & F_{i}|_\partial=\im \p{ \mathcal L_y\phi_{i}}_\partial  \,,  &  B_{i} |_\partial =0 \,.
	\end{align}
\end{itemize}
Especially, if we choose not to constrain the field variations   $\hat\delta\phi_{i}, \hat\delta C_{i} $, the action  (\ref{eq: n1chimult3dtf}) naturally encodes (R).

 %%%%%%%%%%%%%%%%%%%%

\subsection{From Robin to Dirichlet boundary conditions}\label{sec:dirichlet-robin}

We are now ready to explain the relation (\ref{eq:Gammaflip}) in greater detail. In the following, we show that a suitable modification of the Lagrangian leads to different constraints on boundary fields, turning Robin-like conditions into Dirichlet. In order to do this, let us start from the action given in (\ref{eq: n1chimult3dtf}), with $i=1,2$ for simplicity. We construct a new action  $\hat{\mathscr{S}}^{\text{(D)}}_{{\rm chi}}$   by adding to   $\mathscr{S}^{\text{(R)}}_{{\rm chi}} $ a positive and  $\Q$-exact  boundary action for the 3d multiplets $(b_{i},\Q  b_{i})$ along with a  bulk-boundary coupling
\begin{align}\label{eq:bb-cplg-R2D}
	 \hat{\mathscr{S}}^{\text{(D)}}_{{\rm chi}}  
	 & \equiv \mathscr{S}^{\text{(R)}}_{{\rm chi}}  + \i\int_\partial \sqrt{g}\, \d^3 x \Big( \sqrt{2}\p{f_{ia}-\im\mathcal L_y\varphi_{ia}}{R^i}_j\phi^{ja}- b_{ia}{R^i}_jC^{ja}\Big) + {\rm kinetic}\,.
\end{align}
Note that, since $\Q\sim\delta$ at the boundary, we just use $\Q$. The equations of motion of $b_{ia}, f_{ia}$ now provide Dirichlet boundary conditions
\begin{equation}
\phi_{i}|_\partial=0~,\qquad C_{i}|_\partial=0~.
\end{equation}
Since $\mathcal Q\phi_{i}\sim C_{i}$, these boundary conditions are trivially supersymmetric. On the other hand,  the equations of motion of $\phi_{i},C_{i}$ lead to 
\begin{equation}\label{eq: bulkboundarybbprime}
B_{i}|_\partial=b_{i}\,,\qquad F_{i}|_\partial=f_{i}~.
\end{equation}
Supersymmetry imposes the further conditions $\varphi_{i} = \phi_{i}|_\partial$ and $c_{i} =C_{i}|_\partial$. Therefore, Dirichlet boundary conditions on $\phi_{i}$ and $C_{i}$ imply $\varphi_{i} = c_{i} =0 $. As a result, the   degrees of freedom in the kinetic term in (\ref{eq:bb-cplg-R2D}) effectively reduce to   $b_{i},f_{i}$ and only these latter will contribute to the 1-loop determinant of the boundary action. At the end of the day, the action $\hat{\mathscr{S}}^{\text{(D)}}_{{\rm chi}}$ correctly encodes Dirichlet conditions. Consequently, we can borrow the flipping argument from \cite{Dimofte:2017tpi} and argue that  the partition functions computed with different boundary conditions should be related as 
\begin{equation}\label{eq: ZflipRtoD}
\mathcal Z^\text{(D)}[\phi_{i}, C_{i},B_{i},F_{i}; b_{i},\Q  b_{i}]=\mathcal Z^\text{(R)}[\phi_{i}, C_{i},B_{i},F_{i}]\,\mathcal Z_\partial[b_{i},\Q  b_{i}]\,,
\end{equation}
where $\mathcal Z_\partial[b_{i},\Q  b_{i}]$ is the partition function of the boundary theory on $\partial (\mathbb{D}^2\times\mathbb{T}^2)\simeq \mathbb{T}^3$. 

The bulk-boundary couplings in (\ref{eq:bb-cplg-R2D}) deserve some comment. First of all, these terms do not respect full 3d $\mathcal{N}=1$ supersymmetry since only half-multiplets appear. 
This unusual fact is consistent with our setup since the bulk preserves two of the four flat space supercharges, and the boundary is only required to preserve a further half of these, namely the linear combination $\Q$. Interactions are only required to transform in representations of the surviving supersymmetry subalgebra, and the half-multiplets appearing in (\ref{eq:bb-cplg-R2D}) are by construction representations of the subalgebra generated by $\mathcal{Q}$. However, note that since our boundary is flat, if taken alone it could preserve all supercharges: the breaking of supersymmetry is a consequence of coupling to a curved bulk. This differs from the case of a flat space with boundary \cite{Dimofte:2017tpi}, where half of the bulk supercharges would be preserved, leading to a full 3d $\mathcal{N}=1$ theory on the boundary in our case.

We can now turn to discussing the boundary partition function, which is given by the 1-loop determinant of the kinetic operator $-\i\mathcal{L}_k\sim  \Q^2$ computed over the vector space of $b_{i}$ modes, in the constant background $\A^{(0)}$. Calculations can be simplified by inverting the maps in (\ref{eq:3dcurvedtwistedfieldschi}), that is
\be{
	  b \equiv   {\bar Y_\perp} B^{1}|_\partial ~,\qquad  \tilde b  \equiv   -  { Y_\perp}  B^{2}|_\partial   ~.
}
In this language, the on-shell conditions (\ref{eq: bulkboundarybbprime})   simply read as $B|_\partial = b $ and $\tilde B|_\partial = \tilde b$. Therefore, the mode expansions of $b,\tilde b$ should coincide with that of $B,\tilde B$ restricted to the boundary, namely
\be{
b(w,\bar w,t)=\sum_{n_x,n_y,m\in\mathbb{Z}}b_{n_x,n_y,m}\e^{\i m t}F_{n_x,n_y}^{(r/2-1-m)}(w,\bar w)~,
}
and similarly for the conjugate field $\tilde b$. Note that on $\mathbb{T}^3$ all the modes are normalizable, hence $m\in \mathbb Z$. Up to an overall constant, the  boundary partition function is  then     
\begin{align}
	\mathcal Z_\partial[b_i,\Q  b_i]  & \equiv  {\det}_{b} ( -\im\mathcal L_{ k})
		= \det_\mathcal{R}\Big[\prod_{n_x,n_y,m\in\mathbb Z}  \frac{\i}{\text{Im}(\tau)}\Big(\tau\,n_x-n_y+\sigma(r/2-1-m)+\Phi_{(0)}\Big)\Big]~.
\end{align}
The condition $B|_\partial=b$ descending from  (\ref{eq: bulkboundarybbprime}) instructs us to regularize this infinite product by factoring  out the bulk contribution $\mathcal Z_\text{1-loop}^{{\rm chi} (\text{D})}(\Phi_{(0)})$, yielding  

\begin{align}\label{eq:detBLKreg}
\mathcal Z_\partial[b_i,\Q  b_i] &  = \det_\mathcal{R}\Big[ \prod_{\substack{n_x,n_y\in\mathbb Z\\ m\in\mathbb{Z}_{\geq 0}}}   -\frac{\i}{\text{Im}(\tau)}\Big(\tau\,n_x-n_y+\sigma(m+1-r/2)-\Phi_{(0)}\Big)\times \nn\\
	& \times \prod_{\substack{n_x,n_y\in\mathbb Z\\ m\in\mathbb{Z}_{\geq 0}}}  \frac{\i}{\text{Im}(\tau)}\Big(\tau\,n_x-n_y+\sigma(m+r/2)-\Phi_{(0)}\Big)\Big]  = \frac{\mathcal Z_\text{1-loop}^{{\rm chi}(\text{D})}(\Phi_{(0)})}{\mathcal Z_\text{1-loop}^{{\rm chi}(\text{R})}(\Phi_{(0)})}~,
\end{align}
where in the last step we have used the very same regularization employed in (\ref{eq:Z-1loop-B}), (\ref{eq:Z-1loop-phi}). This result explains the physics behind the relation observed in (\ref{eq:Gammaflip}).  

%%%%%%%%%%%%%%%%%%%%

\subsection{From Dirichlet to Robin boundary conditions}

We have just seen how adding matter fields at the boundary with a coupling to the bulk fields can be used to switch from Robin-like to Dirichlet boundary conditions. Here we discuss the reverse mechanism. Let us start from the alternative Lagrangian
\begin{multline}\label{eq: n1chimult3dtfD}
 \mathscr L^{\rm (D)}_{\rm chi}  \equiv  \f12 \mathcal L_{\bar k}\phi^{ia} \mathcal L_k\phi_{ia} - \f12 { \mathcal L_y\phi^{ia}} {\mathcal L_y\phi_{ia}} - \f12 {F^{ia} } {F_{ia} }  + \im {R^i}_j  \phi^{ja}  \, \mathcal L_\perp\p{F_{ia}-\im\mathcal L_y\phi_{ia}} +\\
 - \f\im2 B^{ia}   \mathcal L_k B_{ia}+\im B^{ia} \mathcal L_y C_{ia} - \f\im2 C^{ia} \mathcal L_{\bar k} C_{ia} + \im {R^i}_j  C^{ja} \mathcal L_\perp  B_{ia} ~,
\end{multline}
which encodes  (D) through the equations of motion of $B_{ia},F_{ia}$. We can obtain (R) by adding a positive and $\Q$-exact action for boundary multiplet $(\varphi_i,\Q \varphi_i)$ coupled to bulk 
\begin{align}
 \hat{\mathscr{S}}^{\text{(R)}}_{{\rm chi}}  
	 & \equiv \mathscr{S}^{\text{(D)}}_{{\rm chi}}   - \i\int_\partial\sqrt{g}\,\d^3 x\Big(\sqrt{2}B_{ia}{R^i}_j c^{\;ja} +  \p{F_{ia}-\im\mathcal L_y\phi_{ia}}{R^i}_j\varphi^{\; ja}\Big) + {\rm kinetic}\,.
\end{align}
The equations of motion of $\varphi_i, c_i$ set Robin-like  conditions $\Q  B_i |_\partial=B_i |_\partial =0$,  while those of $B_i,F_i$ set $\phi_i|_\partial=\varphi_i$ as well as $C_i|_\partial=c_i$. According to the flipping argument presented in the previous subsection,  the partition functions computed with different boundary conditions should then  satisfy  
\begin{equation}\label{eq:ZflipDtoR}
\mathcal Z^\text{(R)}[\phi_i, C_i,B_i,F_i; \varphi_i,\Q  \varphi_i]=\mathcal Z^\text{(D)}[\phi_i, C_i,B_i,F_i]\,\mathcal Z_\partial[\varphi_i,\Q  \varphi_i]~,
\end{equation}
where $\mathcal Z_\partial[\varphi_i,\Q  \varphi_i]$ is the partition function of the boundary theory, which is computed by the 1-loop determinant of the kinetic operators $-\mathcal L_{\bar k}\mathcal L_k$ and  $-\im \,\mathcal L_{\bar k}$ over the vector spaces of modes of the boundary fields $\varphi_i,c_i$. The maps in (\ref{eq:3dcurvedtwistedfieldschi}) suggest to consider the fields 
\be{
\varphi \equiv \varphi^{1}   \,,  \qquad  \tilde \varphi \equiv  - \varphi^{2}  \,,  \qquad   c \equiv c^{1}   \,,   \qquad  	 {\tilde c} \equiv  - c^{2}  \,.
}
The boundary partition function is then    
\begin{equation}
\mathcal Z_\partial[\varphi_i,c_i]\equiv 
\frac{	\det_{c}(-\i\,\mathcal L_{\bar k})}{\det_{\varphi}(-\mathcal L_{\bar k}\mathcal L_{ k})}
		=
		\frac{1}{\det_{\varphi} (-\i\,\mathcal L_{  k})}~,
\end{equation}
as expected from the cohomological structure of the  multiplets. In this notation, the on-shell conditions relating bulk and boundary fields  reduce to  $\phi|_\partial = \varphi$, $C|_\partial = c$ and similarly for the tilded fields. The mode expansion of boundary fields should therefore match with their bulk counterparts, for instance
\be{
\varphi(w,\bar w,t)=\sum_{n_x,n_y,m\in\mathbb{Z}}\varphi_{n_x,n_y,m} \e^{-\i m t}F_{n_x,n_y}^{( r/2+m)}(w,\bar w)~.
}
The relation $\phi|_\partial = \varphi$ teaches us to regularize the determinant by factoring out $\mathcal Z^\text{chi(R)}_\text{1-loop}$, and a computation analogous to (\ref{eq:detBLKreg}) directly leads to   (\ref{eq:ZflipDtoR}), namely
\be{\label{eq:detBLKreg2}
\mathcal Z_\partial[\varphi_i,c_i] =\frac{\mathcal{Z}^\text{chi(R)}_\text{1-loop}(\Phi_{(0)})}{\mathcal{Z}^\text{chi(D)}_\text{1-loop}(\Phi_{(0)})}~.
}

\textbf{Remark}. The partition functions $Z_\partial[\varphi_i,\Q \varphi_i] $ and $Z_\partial[b_i,\Q b_i] $ are formally the same as those of chiral and Fermi multiplets on a two-torus with modular parameter $\sigma$. This seems to be consistent with the previous observation made after (\ref{eq: ZflipRtoD}) that the peculiar boundary theories we have considered behave essentially as a 2d $\mathcal{N}=(0,2)$ theories due to the coupling with the bulk. 
It would be natural to expect $\tau$ to emerge as the modular parameter, instead of $\sigma$. 
The appearance of $\sigma$ can be traced to the asymmetry between the two moduli introduced in the choice of regularization made in (\ref{eq:detBLKreg}), which is dictated by coupling with the bulk.

\subsection{A conjecture for the vector: from Neumann to Dirichlet}\label{subsec:Dirichletvec}

In the previous two subsections, we have described a mechanism for switching from Robin-like to Dirichlet conditions for the 4d $\mathcal{N}=1$ chiral multiplets. From the viewpoint of the bulk, imposing one type or the other translates to the vanishing of either submultiplet $(B,\Q B)$ or $(\phi,\Q\phi)$ respectively. From the viewpoint of the boundary, these two choices can be related by the coupling additional degrees of freedom, which can restore one submultiplet at the expense of the other. Using the explicit results (\ref{eq:Z-1loop-B}), (\ref{eq:Z-1loop-phi}) for the 1-loop determinants  with either boundary conditions, we have been able to explain the reflection property (\ref{eq:Gammaflip}) through 1-loop computations on the boundary (\ref{eq:detBLKreg}), (\ref{eq:detBLKreg2}).

In (\ref{eq:bdryvecmult}), (\ref{eq:bdryvecrealmult}), we have managed to split the 4d $\mathcal{N}=1$ vector multiplet into two submultiplets of the boundary supersymmetry, one of which is set to zero by either Neumann or Dirichlet conditions for the gauge field. It is therefore natural to expect that a boundary condition changing mechanism should exist in this case too. Since throughout the main text we have assumed Neumann conditions in order to preserve gauge symmetry at the boundary, we expect that, upon considering suitable boundary couplings, it should be possible to obtain the results for Dirichlet conditions. Unfortunately, as discussed in section \ref{subsec:vecloop}, when dealing with the vector multiplet the proper gauge fixing is crucial, and since we have not worked out the full BRST complex, we are not able to present an exhaustive analysis. However, we can at least propose what the final result should be at the level of 1-loop partition functions: guided by the observation that, in the case of the chiral multiplet the 1-loop determinants of dual boundary conditions are related by a simple reflection property, we can assume that the same holds for the vector multiplet. Starting from the computation with Neumann conditions (\ref{eq:bdryvecmult}) and using  (\ref{Gammashift}), we are thus led to propose that
\be{
\mathcal{Z}^{\textrm{vec}(\text{D})}_{\textrm{1-loop}}(\Phi_{(0)})
 =\det_{\text{ad}}\left[
	\e^{\frac{\i\pi}{3} P_3(\sigma\mathfrak{m}/2 +\Phi_{(0)})} \,\Gamma(\sigma\mathfrak{m}/2+\Phi_{(0)};\tau,\sigma)
	\right]~,
}
where we have allowed any multiplet (including the vector) to have an effective shifted R-charge to take into account possible contributions of flux configurations with flux $\mathfrak{m}$, described around (\ref{eq:BPSflux}). In fact, while Neumann conditions have allowed us to discard such configurations due to $\F_{\perp\mu}|_{\partial}=0$, Dirichlet conditions are compatible with these non-trivial saddles. As mentioned in the main text, even though such configurations are not BPS, they cannot a priori be discarded as can represent the non-perturbative contribution to the path integral \cite{Benini:2015noa,Closset:2015rna}. Moreover, while Neumann conditions preserve gauge symmetry at the boundary and hence $\Phi_{(0)}$ is to be seen as a modulus of the theory to be integrated over, Dirichlet conditions allow only gauge transformations that coincide with the identity at the boundary, hence changing the interpretation of  $\Phi_{(0)}$ as a background chemical potential for a global symmetry acting at the boundary. 
On the other hand, now a summation\footnote{We can assume $\mathfrak{m}$ to be quantized precisely due to Cauchy boundary conditions. 
Configurations of the gauge field can be classified by relative cohomology classes, due to the presence of a boundary. The holonomy along the disk boundary captures the non-integral part of the flux, while the integral part is classified by quantized flux through the disk (corresponding to $\mathfrak{m}$).} is to be performed over the modulus $\mathfrak{m}$ to take into account all possible saddles. In support of this rather heuristic derivation, this picture seems to be consistent with a four dimensional lift of the discussion presented in \cite{Dimofte:2017tpi}.

%%%%%%%%%%%%%%%%%%%%
%%%%%%%%%%%%%%%%%%%%
%%%%%%%%%%%%%%%%%%%%

\section{Observables and examples}\label{sec:applications}

In this section, we discuss which observables are compatible with our background. We also discuss a relevant example, namely the $\text{SU}(N)$ theory coupled to (anti-)fundamental matter and check Seiberg duality.

\subsection{Superpotential}\label{sec:potential}

In order to begin with, let us  investigate  the effect of including  superpotential terms. In the bulk we consider a standard superpotential, described by  (anti-)holomorphic functions $W(\phi)$ and $\widetilde W(\tilde\phi)$ of R-charge $\pm2$ of the scalars in the (anti-)chiral multiplets. We directly use twisted fields to write the  F-term descending  from $W(\phi)$ as
\begin{equation}
\mathscr S_W\equiv \int \d^4x\,\sqrt g\,\mathscr L_W\,,\qquad \mathscr L_W\equiv W_I(\phi)\,F^I - W_{IJ}(\phi)B^I\,C^J~,
\end{equation}
where $W_I\equiv \partial_I W$, $W_{IJ}\equiv \partial_I\partial_J W$. Supersymmetry acts on the two pieces as 
\ale{
& \Q \p{W_I F^I}  = \sqrt2 W_{IJ} C^J F^I + \im \sqrt2 W_I\p{  \mathcal L_K B^I + \mathcal L_{ \bar Y}C^I -  \sqrt2\tilde\Lambda^-\phi^I} ~ , \nn\\
%%%%
& \Q \p{W_{IJ} B^I C^J} =  \sqrt2 W_{IJ}\p{F^I - \im \mathcal L_{\bar Y} \phi^I}C^J - \im \sqrt2 W_{IJ}B^I\mathcal L_K \phi^J ~ ,
}
where the Lie derivatives contain the gauge field $\mathcal A_\mu$, the R-symmetry connection $A_\mu$ and a flavour connection $A_\mu^{(F)}$. By taking into account that 
\ale{
-2\im \tilde\Lambda^- W_I \phi^I = \Q\p{  \iota_{\bar Y} \mathcal A ~  W_I \phi^I } - \sqrt2  \p{\iota_{\bar Y} \mathcal A }  W_{IJ} C^J \phi^I  - \sqrt2   \p{ \iota_{\bar Y} \mathcal A} W_I  C^I ~ ,
}
we find  
\ale{
\Q \mathscr{L}_W = & ~ \im \sqrt2 \nabla_\mu \p{     K^\mu W_I B^I  + \overline Y^\mu W_I C^I}  + \nn\\
%%%%
& -  \sqrt2       W_{I} {\Phi} B^{I} -  \sqrt2    B^{I}    W_{IJ}{\Phi}\phi^{J}   + \nn\\
%%%%
& -  \sqrt2 \Phi^{(R)}  B^I \comm{ \p{r_I - 2} W_I +r_J W_{IJ}\phi^J }    -  \sqrt2 \Phi^{(F)}  B^I \comm{ v_I  W_I + v_J W_{IJ}\phi^J }  + \nn\\
%%%%
& + \bar Y^\mu \Q\comm{    ~  W_{I} {\p{ \mathcal A_\mu}}\phi^{I}  +   r_I A_\mu ~  W_I \phi^I  +   v_I A^{(F)}_\mu ~  W_I \phi^I } ~ ,
%%%%
}
where we defined  $\Phi^{(R)} \equiv - \iota_K A$, $\Phi^{(F)} \equiv - \iota_K A^{(F)}$, while $r_I$, $v_I$ denote the R-symmetry and flavor charges respectively. The second line is a gauge transformation acting on $W_{I} B^{I}$
\be{
  \delta_G\p{W_{I}B^{I}} \equiv W_{I} {\Phi} B^{I} +    B^{I}    W_{IJ}{\Phi}\phi^{J} = 0 ~ .
}
Hence, the supersymmetric variation of the holomorphic superpotential  is a total derivative if the conditions  
\ale{
&  \delta_R \p{W_I \phi^I }\equiv  r_J W_{IJ}\phi^I\phi^J + r_I  W_I \phi^I=   2  W_I\phi^I  ~ ,  \nn\\
%%%%
& \delta_F \p{W_I \phi^I} \equiv v_J W_{IJ}\phi^I\phi^J + v_I  W_I \phi^I=   0 ~ ,     
}
are satisfied, simply meaning that $W(\phi)$ must have R-charge $+2$ and be a singlet under flavour symmetry. Consequently
\begin{multline}
 \Q \mathscr S_W  = \im \sqrt2 \int \d^4x\,\sqrt g\,   \nabla_\mu \p{   \bar Y^\mu W_I C^I}  =\\
 =  \im \sqrt2 \int_\partial \d^3 x\sqrt{g}\,  {\bar Y_\perp} W_I C^I 
  = \im   \Q\int_\partial \d^3 x\sqrt{g}\,  {\bar Y_\perp} W  ~ .
\end{multline}
Similarly, the anti-holomorphic superpotential term  reads
\be{
\mathscr S_{\widetilde W}\equiv\int \d^4x\,\sqrt g\,\mathscr L_{\widetilde W}\,,\qquad \mathscr L_{\widetilde W}\equiv\widetilde W_I(\tilde\phi)\,\tilde F^I- \widetilde W_{IJ}(\tilde \phi)\tilde B^I\,\tilde C^J~,
}
where $\widetilde W_I\equiv \partial_I \widetilde W$, $\widetilde W_{IJ}\equiv \partial_I\partial_J \widetilde W$. Repeating the same steps as above, one can confirm that $\widetilde W(\tilde \phi)$ must have R-charge $-2$ and neutral under gauge and flavour symmetries. Eventually
\begin{multline}
 \Q \mathscr S_{\widetilde W}  = - \im \sqrt2 \int \d^4x\,\sqrt g\,   \nabla_\mu \p{    Y^\mu \widetilde W_I \widetilde C^I}  = \\
 = - \im \sqrt2 \int_\partial \d^3 x\sqrt{g}\,  {  Y_\perp}  \widetilde W_I  \widetilde C^I = - \im  \Q\int_\partial \d^3 x\sqrt{g}\,  {  Y_\perp} \widetilde W  ~ .
\end{multline}
Therefore, we showed that the supersymmetric variations  of $\mathscr S_{  W}$ and $\mathscr S_{\widetilde W}$ do not vanish because of the boundary. This can be cured by introducing  the F-term
\begin{align}
\mathscr S_{\mathscr F} \equiv  - \im \int_\partial \d^3 x\sqrt{g}\,   \mathscr F ~,\qquad 
 \mathscr F  \equiv            {\bar Y_\perp}  W - {  Y_\perp} \, \widetilde  W    \,, \qquad \p{\mathscr F}^\dagger =  -  \mathscr F ~ . 
\end{align}
The supersymmetric variation of  $\mathscr S_{\mathscr F}$ exactly cancels the boundary terms generated by $\Q  \mathscr S_{  W}$ and $\Q \mathscr S_{\widetilde W}$, and the improved superpotential term 
 \begin{align}
   \hat{\mathscr S}_{  W  } \equiv  \mathscr S_{  W  } + \mathscr S_{  \widetilde W  }   +   \mathscr S_{\mathscr F}  
 \end{align}
is supersymmetric without imposing boundary conditions.

%%%%%%%%%%%%%%%%%%%%

\subsection{Fayet-Iliopoulos term}\label{sec: fiterm}

For a $\text{U}(1)$ factor in the gauge group, we can consider a Fayet-Iliopoulos (FI) term, which for the class of backgrounds we are considering in this paper reads \cite{Closset:2014uda}
\begin{equation}
\mathscr S_{\rm FI} \equiv \xi_{\rm FI}\,\int \d^4x\,\sqrt g\,\p{D-2\,\A_\mu V^\mu}\,.
\end{equation}
Notice that for a K{\"a}hler metric $V^\mu=\kappa K^\mu$, so gauge invariance  holds due to $\nabla_\mu V^\mu=0$ and $V^\perp=0$.  The  $\Q$-variation  of $\mathscr S_{\rm FI}$ vanishes up to boundary terms,  and we find that $\Q\mathscr S_{\rm FI}$ generates a boundary term that is exact w.r.t. the boundary supersymmetry, namely
\be{
\Q\mathscr S_{\rm FI}= \xi_{\rm FI}\,\int \d^4x\,\sqrt g\, \mathcal L_\perp\p{ {\bar Y_\perp} \zeta\lambda +  {  Y_\perp} \,\widetilde \zeta\widetilde \lambda}= \xi_{\rm FI}\,\int _\partial \d^3x\,\sqrt{g}\,  \Lambda_\| = \frac{\xi_{\rm FI}}{\sqrt{2}}\, \Q\int _\partial \d^3x\,\sqrt{g}\,   \A_\|   \,,
}
where we used the 3d cohomological fields defined in (\ref{eq: 3dn1veccohfields}).  
As a consequence, we can define an improved FI term
\begin{equation}
\hat{\mathscr S}_{\rm FI}\equiv \xi_{\rm FI}\,\int \d^4x\,\sqrt g\,\p{D-2\,\A_\mu V^\mu} -   \frac{\xi_{\rm FI}}{\sqrt{2}}\,\int _\partial \d^3x\,\sqrt{g}\,   \A_\|  \,,
\end{equation}
whose supersymmetry variation vanishes without imposing boundary conditions. Note that for our choice of background this term is zero on the localization locus we have considered, hence it cannot give classical contributions. However, it may contribute for more general choices and/or if one allows for flux configurations. Also, this term cannot be made fully invariant w.r.t.  arbitrary large gauge transformations, unless one imposes some rational condition on $\xi_\text{FI}$ and $\textrm{Im}(\tau)/\textrm{Re}(\tau)$. A similar  observation can be found in \cite{Aharony:2013dha,Closset:2014uda}.

\subsection{Surface defects}
In A-twisted 2d $\mathcal{N}=(2,2)$ theories on the sphere \cite{Closset:2014pda,Closset:2015rna}, correlation functions of local operators at the poles can be computed through localization. In 3d $\mathcal{N}=2$ theories, the natural lift is provided by Wilson loops \cite{Benini:2015noa,Kapustin:2009kz}. In our setup, Wilson loops cannot be defined because the Killing vector is complex and it generates the whole torus (together with its complex conjugate). Therefore, it is more natural to look for surface operators wrapping the torus. In terms of the twisted variables (\ref{eq:twistCHI}), the 4d $\mathcal{N}=1$ SUSY transformations (\ref{eq:twistSUSYCHI1}) of the chiral multiplet can naturally be interpreted in terms of 2d $\mathcal{N}=(0,2)$ SUSY transformations on the torus, with $(\phi,C)$ and $(B,F)$ representing chiral and Fermi multiplets respectively. Similarly, the 4d $\mathcal{N}=1$ SUSY transformations (\ref{eq:twistedSUSYVEC}) and the components $(\A_w,\A_{\bar w},\Lambda_{\bar w},\tilde\Lambda_{\bar w},D)$ can be interpreted in terms of a 2d $\mathcal{N}=(0,2)$ vector multiplet on the torus. Therefore, it is easy to couple to the bulk a 2d defect theory on the torus while preserving the $\Q$ supercharge. For Lagrangian theories, the partition functions of the defect theories can be again computed by localization, and the 1-loop determinants of chiral, Fermi and vector multiplets read \cite{Benini:2013nda,Benini:2013xpa,Gadde:2013ftv} (up to zero-point energies)
\be{\label{eq:ellgenus}
\mathcal{Z}_\text{c}(u)\equiv   \frac{1}{\Theta(u;\tau)}~,\quad 
\mathcal{Z}_\text{F}(u)\equiv   \Theta(u;\tau)~,\quad 
\mathcal{Z}_\text{v}(w)\equiv   \det_\text{ad}\left[\Theta(w;\tau)\right]~,
}
where $u$ is a $\text{U}(1)$ chemical potential and $w$ belongs to a Cartan subalgebra of the 2d gauge group. The coupling to the bulk can be accomplished by gauging a subgroup of the 2d flavor group with a 4d vector multiplet, namely by inserting the 2d partition function under the integral of the bulk partition function. More generally, provided that the partition function can be computed by other means, one can also consider non-Lagrangian defect theories. A very interesting class of 2d $\mathcal{N}=(0,2)$ theories was constructed in \cite{Gadde:2013sca} from a twisted compactification of the 6d $\mathcal{N}=(0,2)$ theory on a 4-manifold, whose torus partition functions are computed by certain affine characters. These objects represent the natural lift of ordinary characters to our setup,\footnote{See also \cite{Hayling:2018fgy} for another interesting approach.} which in 3d $\mathcal{N}=2$ theories can be used to compute the expectation values of Wilson loops.

\subsection{SQCD and Seiberg duality}
We can now apply the results of this paper to an interesting example, namely the $\text{SU}(N)$ theory with (anti-)fundamental matter. We will denote by $\mu\equiv m+\sigma r/2$ and $\bar\mu\equiv \bar m-\sigma r/2$ the combined flavor and R-symmetry chemical potentials, while we will denote the fundamental gauge chemical potential by $u$. The absence of gauge anomalies instruct us to consider $N_\text{f}$ anti-fundamental chirals with Robin-like  conditions and $N_\text{f}$ fundamental chirals with Dirichlet conditions with the correct R-charge assignment $r=1-N/N_\text{f}$ and flavor chemical potentials constrained by $\sum_{f}(\bar m_f-m_f)=N(1+\tau)$. We also  restrict to $N_\text{f}\geq N$. The 1-loop integrand of the localized partition function 
\be{
Z_\gamma[\text{SU}(N)]\equiv \oint_\gamma\frac{\d^N u}{(2\pi\i)^N N!} \, \delta(\sum_{a}u_a)\,\Upsilon_N(u,\mu,\bar \mu)~
}
reads  (up to $u$-independent normalization)
\be{
\Upsilon_N(u,\mu,\bar \mu)\equiv \prod_{1\leq a\neq b\leq N}\frac{1}{\Gamma(u_b-u_a;\tau,\sigma)}\prod_{a=1}^N\prod_{f=1}^{N_\text{f}}\frac{\Gamma(\mu_f-u_a;\tau,\sigma)}{\Gamma(\sigma+\bar \mu_f-u_a;\tau,\sigma)}~.
}
For the $\text{U}(N)$ case, the choice of contour would correspond here to a selection of $N$ fundamental chirals out of $N_\text{f}$, and as such there are $N_\text{f}!/(N_\text{f}-N)!N!$ possibilities. In fact, since the unconstrained $\text{U}(N)$ integral is easier to handle, we assume that the $\text{SU}(N)$ partition function can eventually be recovered from the former by either imposing the traceless condition by hand \cite{Yoshida:2014qwa} or through a Fourier-like transform on the FI parameter $\xi$ that can be introduced. Therefore, we consider the independent poles at\footnote{This pole prescription is essentially the 4d lift of the contours which are usually considered in 3d \cite{Pasquetti:2011fj,Beem:2012mb}.} 
\be{
u_a\in \{\mu_f+k_f\sigma~,f\in \gamma,~k_f\in\mathbb{Z}_{\geq 0}\}~,
}
where $\gamma$ is a label set of $N$ elements out of $N_\text{f}$. Taking into account the permutation symmetry, we thus obtain the residue series
\begin{multline}\label{eq:UN}
Z_\gamma[\text{U}(N)]=\e^{2\pi\i \xi\sum_{f\in\gamma}\mu_f}\Upsilon_N(\mu_{f\in\gamma},\mu,\bar\mu)\sum_{\{k_{f}\}}\e^{2\pi\i \xi\sigma\sum_{f\in\gamma}k_{f}}\times\\
\times\prod_{f'\in\gamma}\frac{1}{k_{f'}!}\prod_{f\in\gamma}\frac{\Theta(\mu_f-\mu_{f'};\tau,\sigma)_{-k_{f'}}}{ \Theta(\mu_f-\mu_{f'};\tau,\sigma)_{k_f-k_{f'}}}
\frac{\prod_{f\not\in\gamma}\Theta(\mu_f-\mu_{f'};\tau,\sigma)_{-k_{f'}}}{\prod_{f=1}^{N_\text{f}}\Theta(\sigma+\bar\mu_f-\mu_{f'};\tau,\sigma)_{-k_{f'}}}~,
\end{multline}
where the divergent factor  in $\Upsilon(\mu_{f\in\gamma})$ is to be understood as its residue at the pole. This expression can be identified with the elliptic vortex partition function of the theory in the vacuum where $N$ anti-fundamentals acquire an expectation value. In order to see this more explicitly, let us focus on the fixed vortex number $k=\sum_{f\in\gamma}k_f$ and compare this with the more conventional integral  
\begin{multline}\label{eq:Zk}
Z_k\equiv \frac{1}{k!}\oint_\text{J.K.} \prod_{i=1}^k\frac{\d \phi_i}{2\pi\i}\, \prod'_{1\leq i, j\leq k}\frac{\Theta(\phi_i-\phi_j;\tau)}{\Theta(-\sigma+\phi_i-\phi_j;\tau)}\times\\
\times\prod_{j=1}^k\frac{\prod_{f=1}^{N_\text{f}}\Theta(\phi_j+\bar a_f;\tau)}{\prod_{f\in\gamma}\Theta(-\phi_j-a_f;\tau)\prod_{f\not\in\gamma}\Theta(\phi_j+a_f-\sigma;\tau)}~,
\end{multline}
where the prime denotes that the zero factor ($i=j$) in the numerator is to be replaced by $\text{Res}_{u=0}\Theta(u,\tau)$. Notice that the 4d anomaly cancellation condition guarantees that the form of the 2d integrand is preserved under the identification $\phi_j\sim\phi_j+\mathbb{Z}+\tau\mathbb{Z}$ and modular transformation $\tau\to-1/\tau$. Recalling (\ref{eq:ellgenus}), we see that the $k$-vortex sector corresponds to the elliptic genus of a 2d $\mathcal{N}=(0,2)$ $\text{U}(k)$ theory coupled to $N$ anti-fundamental chirals, \mbox{$N_\text{f}-N$} fundamental chirals, $N_\text{f}$ fundamental and one adjoint Fermi multiplets, corresponding to the $k$-vortex theory of the 4d $\mathcal{N}=1$ theory we are considering.\footnote{See e.g. \cite{Poggi:2017kut,Chen:2014rca} for a recent and more detailed discussion related to our setup.} More explicitly, we distribute each of the $k$ integration variables around the $N$ chirals according to the unordered partition $\{k_f\}$ and integrate them one by one. What we eventually get is a net contribution from a tail of poles arising from the charge minus chirals starting at $\phi_j=-a_f$ for some $f\in\gamma$, namely
\be{
\phi_j\in\{-a_f-\sigma\ell~,\ell=0,\ldots,k_f-1~,f\in\gamma\}~.
}
The evaluation of the residues at these poles yields
\be{
Z_k=\!\!\!\sum_{\substack{\{k_f\}\\ k=\sum_{f\in\gamma}k_f}}\prod_{f'\in\gamma}\frac{1}{k_{f'}!\prod_{f\in\gamma}\Theta(a_{f'}-a_f-\sigma k_f;\tau;\sigma)_{k_{f'}}}\frac{\prod_{f\not\in\gamma}\Theta(-a_{f'}+ a_f;\tau;\sigma)_{-k_{f'}}}{\prod_{f=1}^{N_\text{f}}\Theta(\sigma-a_{f'}+\bar a_f;\tau;\sigma)_{-k_{f'}}}~.
}
We can now use the identity
\be{
\Theta(u;\tau;\sigma)_{k'-k}=\frac{\Theta(u-\sigma k;\tau;\sigma)_{-k'}}{\Theta(u-\sigma k;\tau;\sigma)_{-k}}~
}
to write
\be{
Z_k=\!\!\!\sum_{\substack{\{k_f\}\\ k=\sum_{f\in\gamma}k_f}}\prod_{f'\in\gamma}\frac{\prod_{f\in\gamma}\Theta(a_{f}-a_{f'};\tau;\sigma)_{-k_{f'}}}{k_{f'}!\prod_{f\in\gamma}\Theta(a_{f}-a_{f'};\tau;\sigma)_{k_{f}-k_{f'}}}\frac{\prod_{f\not\in\gamma}\Theta(a_f-a_{f'};\tau;\sigma)_{-k_{f'}}}{\prod_{f=1}^{N_\text{f}}\Theta(\sigma+\bar a_f-a_{f'};\tau;\sigma)_{-k_{f'}}}~,
}
coinciding  with the $k$-vortex part of  (\ref{eq:UN}) upon straightforward identifications of the parameters. This exercise allows us to test Seiberg duality. In fact, the integrand of the elliptic genus (\ref{eq:Zk}) is a meromorphic elliptic function of $\phi_j\sim \phi_j+\mathbb{Z}+\mathbb{Z}\tau$, and by deforming the integration contour to pick up the poles arising from the tail of charge plus chirals, the $\text{U}(N)$ vortex partition function (\ref{eq:UN}) has an alternative representation in which the role of $f\in\gamma$ and $f\not\in\gamma$ are swapped, namely it is equivalent to a dual $\text{U}(N_\text{f}-N)$ vortex partition function. What is left to be checked is the correct transformation law of the 1-loop determinant in the $\gamma$ vacuum, which is indeed invariant up to the appearance of the singlet contributions
\be{
\Upsilon_N(u=\mu_{f\in\gamma},\mu,\bar \mu)=\Upsilon_{N_\text{f}-N}(u=\hat\mu_{f\not\in\gamma},\hat\mu,\hat{\bar \mu})\prod_{f,f'=1}^{N_\text{f}}\frac{1}{\Gamma(\sigma+\bar\mu_f-\mu_{f'};\tau,\sigma)}\,
~,
}
with the dual variables defined by $\hat\mu\equiv -\mu+\sigma/2$, $\hat{\bar\mu}\equiv -\bar \mu-\sigma/2+1+\tau$.

%%%%%%%%%%%%%%%%%%%%
%%%%%%%%%%%%%%%%%%%%
%%%%%%%%%%%%%%%%%%%%

\section{Conclusions and future directions}\label{sec:conclusions}

In this paper, we have studied 4d $\mathcal{N}=1$ theories on $\mathbb{D}^2\times \mathbb{T}^2$ and certain BPS boundary conditions, and we have used supersymmetric localization techniques to perform the exact evaluation of the partition functions. We have met several subtle points during our analysis, part of which are also shared with the compact backgrounds (such as the presence of fermionic zero modes in A-twisted theories) and part of which are instead proper of the non-compactness of the background (such as the proper implementation of the boundary conditions in 1-loop computations) or due to the complex nature of the Killing vector arising from the localizing supercharge (such as the proper treatment of the cohomological complex for the vector multiplet). 
Despite these issues, the results  that we have presented in this work passed some nontrivial tests, such as the derivation of a consistent picture, the correct 3d and 2d limits, and the recovery of known results and dualities. 
One prominent question that remains unclear, is to find a microscopic understanding of the relation between boundary conditions and integration contours for the localized path integral, a closely related one is the map between IR/UV boundary conditions. 
One overarching motivation for trying to give more rigorous --or more satisfactory-- answers to this and other questions, is that localization techniques would eventually allow one to compute partition functions and indices on non-compact manifolds with diverse topologies, which provide important tools in a wide range of current research areas. For an illustration of this point, we conclude with a collection of few potential applications of our results and suitable extensions thereof.

\begin{enumerate}
\setcounter{enumi}{0}

\item 
It would be very interesting to extend our results to more general boundary conditions. 
In the case of the chiral multiplet, we found that there is a nice mechanism for switching between two choices which naturally arise from our localization setup. Classifying boundary conditions would conceivably lead to a larger group of similar dualities. 
Analogous programs have been carried out in 3d $\CN=2$ \cite{Dimofte:2017tpi} and $\CN=4$ theories \cite{Gaiotto:2008sa,Gaiotto:2008ak,Chung:2016pgt,Okazaki:2019bok,Okazaki:2019ony}, which uncovered rich duality actions on the space of boundary conditions.

\item
4d holomorphic blocks should exhibit interesting global behavior in parameter space, such as Stokes phenomena. In fact, viewing our setup as a double uplift of 2d $\mathcal{N}=(2,2)$ theories on a disk, it should be possible to establish a precise relation between (sums of products of) 4d holomorphic blocks and the topological-anti-topological amplitudes first considered by Cecotti and Vafa \cite{Cecotti:1991me}. Indeed, in \cite{Cecotti:2013mba} a four dimensional version of the $tt^*$ geometry was proposed, involving precisely 4d $\mathcal{N}=1$ theories on $\mathbb{D}^2\times \mathbb{T}^2$. Similar considerations for 3d holomorphic blocks were explored in \cite{Beem:2012mb}. It would be interesting to study to what extent suitable combinations of 4d holomorphic blocks can be used to construct solutions to $tt^*$ equations, and explore applications to moduli spaces  of hyper-holomorphic connections.

\item
3d holomorphic blocks can be characterized as solutions to a set of difference equations, physically interpreted as Ward identities for the algebra of half-BPS line operators \cite{Dimofte:2011ju,Dimofte:2011py,Dimofte:2011jd,Beem:2012mb}. In our case, the 1-loop determinants of the chiral multiplets also satisfy simple difference equations (we refer to \cite{Nieri:2015yia} for an interacting example), and it is tempting to identify them as basic Ward identities for the algebra of surface defects (constructions of such identities may presumably be obtained with the help of the algebraic interpretation of 4d holomorphic blocks \cite{Nieri:2015dts,Iqbal:2015fvd,Lodin:2017lrc,Kimura:2016dys}). This would allow the 4d holomorphic blocks, possibly enriched with defects, to be also characterized as solutions to elliptic difference equations. This perspective has successfully been applied in the context of class $\mathcal{S}$ theories \cite{Gaiotto:2009we} and index computations \cite{Bullimore:2014awa,Bullimore:2014nla,Gaiotto:2012xa,Razamat:2013qfa,Gaiotto:2015usa}.

\item
It may be possible to preserve more supercharges with suitable restrictions on the background geometry or by modifying the supergravity background fields. This would lead to the possibility of computing interesting protected indices of 4d $\mathcal{N}=2$ theories. One of these would be the AMNP 3d index \cite{Alexandrov:2014wca}. 
The AMNP index enjoys several interesting properties, including a relation to Darboux coordinates for twistor constructions of hyper-K\"ahler metrics, and a relation to solutions of TBA equations arising in the study of 4d $\mathcal{N}=2$ wall-crossing.
To employ our approach for its computation, one should take a limit of the background with a flat infinite disk and simultaneously stretch the complex structure of the torus, so as to obtain the required geometry of $\mathbb{R}^2\times \mathbb{R}\times \mathbb{S}^1$ (in fact, torus compactification and stretching were already employed in \cite{Alexandrov:2014wca} as a regulator).
Perhaps the main technical step to be taken in this direction would be to understand how to include instanton and monopole corrections.

\end{enumerate}

\acknowledgments

It is a pleasure to thank Francesco Benini, Stefano Cremonesi, Guido Festuccia, Dario Martelli, Joe Minahan, Yiwen Pan and Maxim Zabzine for discussions. 
We are especially grateful to Jian Qiu for numerous illuminating discussions on technical questions studied in this work, and to Achilleas Passias for collaboration during part of this project.

The work of PL is supported by a grant from the Swiss National Science foundation. He also acknowledges the support of the NCCR SwissMAP that is also funded by the Swiss National Science foundation. 
PL was also supported by the grants ``Geometry and Physics'' and ``Exact Results in Gauge and String Theories'' from the Knut and Alice Wallenberg foundation during part of this work.
The work of FN is supported by the German Research Foundation (DFG) via the Emmy Noether program ``Exact results in Gauge theories''. FN was also supported  by Vetenskapsr\r{a}det under grant \#2014-5517, by the STINT grant and by the grant ``Geometry and Physics'' from the Knut and Alice Wallenberg foundation during part of this work.
The work of AP is supported by the ERC STG Grant 639220.

%%%%%%%%%%  Bibliography  %%%%%%%%%%%%

\begin{appendix}

\section{Spinor conventions and identities}\label{app:conventions}

We mainly follow \cite{Wess:320631} adapted to the Euclidean signature. The $\sigma$-matrices with flat indices $a,b$ are defined by
\be{
\sigma^a_{\alpha\dot\alpha}=(\vec\sigma,-\im\,\id_2)\,,\qquad \tilde \sigma^{a\,\dot\alpha\alpha}=(-\vec\sigma,-\im\,\id_2)\,,\,\qquad \sigma^{ab}=\f12\sigma^{[a}\tilde\sigma^{b]}\,,\qquad  \tilde\sigma^{ab}=\f12\tilde\sigma^{[a}\sigma^{b]}\,,
}
where $\vec\sigma$ are the usual  Pauli matrices. The following identities hold
\begin{align}\label{eq: coolid}
\sigma_a\tilde\sigma_b=& \ -\delta_{ab}+2\,\sigma_{ab},\quad &\tilde\sigma_a\sigma_b=& \ -\delta_{ab}+2\,\tilde\sigma_{ab}\,,\nn\\
 \epsilon_{abcd}\,\sigma^{cd}= &\ 2\,\sigma_{ab}\,,\quad &\epsilon_{abcd}\,\tilde\sigma^{cd}=& \ -2\,\tilde\sigma_{ab}\,,\nn\\
(\sigma^a_{\alpha\dot\alpha})^*= & \ -\tilde\sigma^{a\,\dot\alpha \alpha}\,, \quad& \p{{{ \sigma^{ab}}_\alpha}^\beta}^*=& \ -  {{{{ \sigma^{ab}}_\beta}^\alpha}}\,, \qquad \p{{{  \tilde\sigma^{ab \,\dot\alpha}}_{\dot\beta}}}^*=-   {{{\tilde \sigma^{ab \,\dot\beta}}_{\dot\alpha}}}\,,
\end{align}
where $^*$ denotes complex conjugation. The transition to curved indices is achieved by defining the real Euclidean frame $\theta^a$, $a=1,2,3,4$, and using the vielbeins, $\theta^a\equiv e^a_{\;~\mu}\d x^\mu$.
In the main text, we have used the following Fierz identities for commuting spinors
\ale{
\p{\chi_1\chi_2}\p{\chi_3\chi_4}&=-\p{\chi_1\chi_3}\p{\chi_4\chi_2}-\p{\chi_1\chi_4}\p{\chi_2\chi_3}~,\nn\\
\p{\tilde\chi_1\tilde\chi_2}\p{\tilde\chi_3\tilde\chi_4}&=-\p{\tilde\chi_1\tilde\chi_3}\p{\tilde\chi_4\tilde\chi_2}-\p{\tilde\chi_1\tilde\chi_4}\p{\tilde\chi_2\tilde\chi_3}~,\nn\\
\p{\chi_1\chi_2}\p{\tilde\chi_1\tilde\chi_2}&=-\frac{1}{2}\p{\chi_1\sigma_a\tilde\chi_2}\p{\chi_2\sigma^a\tilde\chi_1}=-\frac{1}{2}\p{\tilde\chi_1\tilde\sigma_a\chi_2}\p{\tilde\chi_2\tilde\sigma^a\chi_1}~,\nn\\
\p{\chi_1\sigma^\mu\tilde\chi_2}\p{\chi_3\sigma_{\mu\nu}\chi_4}&=\frac{1}{2}\p{\chi_1\chi_3}\p{\chi_4\sigma_\nu\tilde\chi_2}+\frac{1}{2}\p{\chi_1\chi_4}\p{\tilde\chi_2\tilde\sigma_\nu\chi_3}~,\nn\\
\p{\tilde\chi_1\tilde\sigma^\mu \chi_2}\p{\tilde\chi_3\tilde\sigma_{\mu\nu}\tilde\chi_4}&=\frac{1}{2}\p{\tilde\chi_1\tilde\chi_3}\p{\tilde\chi_4\tilde\sigma_\nu \chi_2}+\frac{1}{2}\p{\tilde\chi_1\tilde\chi_4}\p{\chi_2\sigma_\nu\tilde\chi_3}~.
}
Conjugation on  spinors fulfils
\begin{equation}
(\zeta_\alpha)^*=(\zeta^\dagger)^\alpha,\qquad (\zeta^\alpha)^*=-(\zeta^\dagger)_\alpha,\qquad  (\tilde\zeta^{\dot\alpha})^*=(\tilde\zeta^\dagger)_{\dot\alpha},\qquad (\tilde\zeta_{\dot\alpha})^*=-(\tilde\zeta^\dagger)^{\dot\alpha}~.
\end{equation}
Assuming that $^{**}=\text{id}$, we also have
\begin{equation}
\zeta_\alpha=(\zeta_\alpha)^{**}=((\zeta^\dagger)^\alpha)^*=-(\zeta^{\dagger\dagger})_\alpha,\qquad \tilde \zeta^{\dot\alpha}=(\tilde\zeta^{\dot\alpha})^{**}=((\tilde\zeta^\dagger)_{\dot\alpha})^*=-(\tilde\zeta^{\dagger\dagger})^{\dot\alpha}~.
\end{equation}
Moreover
\begin{equation}
(|\zeta|^2)^*=((\zeta^\dagger)^\alpha\zeta_\alpha)^*=\zeta_\alpha(\zeta^\dagger)^\alpha=|\zeta|^2,\qquad (|\tilde\zeta|^2)^*=((\tilde\zeta^\dagger)_{\dot\alpha}\tilde\zeta^{\dot\alpha})^*=\tilde\zeta^{\dot\alpha}(\tilde\zeta^\dagger)_{\dot\alpha}=|\tilde\zeta|^2\,.
\end{equation}
In particular, the previous formulae yield 
\begin{align}
(K^\mu)^*= & \ (\zeta^\alpha\sigma^\mu_{\alpha\dot\alpha}\tilde\zeta^{\dot\alpha})^*=(\zeta^\dagger)_\alpha\tilde\sigma^{\mu\,\dot\alpha\alpha}(\tilde\zeta^\dagger)_{\dot\alpha}=|\zeta|^2|\tilde\zeta|^2 \bar K^\mu~,\nn\\
(Y^\mu)^*=& \ |\tilde\zeta|^{-2}(\zeta^\alpha\sigma^\mu_{\alpha\dot\alpha}(\tilde\zeta^{\dagger})^{\dot\alpha})^*=-|\tilde\zeta|^{-2} (\zeta^\dagger)_\alpha\tilde\sigma^{\mu\,\dot\alpha\alpha}\tilde\zeta_{\dot\alpha}=|\tilde\zeta|^{-2}| \zeta|^{2}\bar Y^\mu \,.
\end{align}

\section{Special functions and regularization of determinants}\label{appell}\label{sec:1-loop-details}
In this appendix, we collect useful definitions and properties of some special functions used in the main text. Our main reference is \cite{Narukawa:2003}. 

\textbf{Definitions}. We start by defining the (infinite) $q$-factorial 
\be{\label{eq:qPocch}
(x;q)_\infty\equiv\prod_{k\geq 0}(1-q^k x)~,\quad |q|<1~.
}
Using the representation
\be{
(x;q)_\infty=\e^{-{\rm Li}_2(x;q)}~,\quad {\rm Li}_2(x;q)\equiv \sum_{k\geq 1}\frac{x^k}{k(1-q^k)}~,
}
it can be extended to the domain $|q|>1$ by means of 
\be{
(q x;q)_\infty\to \frac{1}{(x;q^{-1})_\infty}~.
}
The short Jacobi Theta function is defined by
\be{\label{Theta}
\Theta(x;q)\equiv (x;q)_\infty(q x^{-1};q)_\infty~.
}
In order to avoid cluttering, in the main text we will often use the alternative notation 
\be{
\Theta(x;q)\equiv \Theta(u;\tau)\qquad \text{for}\qquad q\equiv \e^{2\pi\i\tau}~,\quad x\equiv \e^{2\pi\i u}~.
}
A useful property is
\be{\label{Thetashift}
\frac{\Theta(q^m x;q)}{\Theta(x;q)}=(-x q^{(m-1)/2})^{-m},\quad 
\frac{\Theta(q^{-m}x;q)}{\Theta(x;q)}=(-x^{-1}q^{(m+1)/2})^{-m}~.
}

The double (infinite) $q$-factorial is defined by 
\be{
(x;p,q)_\infty\equiv \prod_{k\geq 0}(1-p^j q^k x)~,\quad |p|,|q|<1~.
}
Using the representation
\be{
(x;p,q)_\infty=\e^{-{\rm Li}_3(x;p,q)}~,\quad {\rm Li}_3(x;p,q)\equiv\sum_{k\geq 1}\frac{x^k}{k(1-p^k)(1-q^k)}~,
}
it can be extended to other domains by means of
\be{\label{pqcont}
(q x;p,q)_\infty\to \frac{1}{(x;p,q^{-1})_\infty}~.
}
The elliptic Gamma function is defined by
\be{\label{ellgamma}
\Gamma(x;p,q)\equiv \frac{(p q x^{-1};p,q)_\infty}{(x;p,q)_\infty}~.
}
It has has zeros and poles at 
\be{
\text{zeros}:~ x=p^{m+1}q^{n+1}~,\quad \text{poles}:~ x=p^{-m}q^{-n}~,\quad m,n\in \mathbb{Z}_{\geq 0}~.
}
In order to avoid cluttering, in the main text we will often use the alternative notation 
\be{
\Gamma(x;p,q)\equiv \Gamma(u;\tau,\sigma)\qquad \text{for}\qquad q\equiv \e^{2\pi\i\tau}~,\quad p\equiv \e^{2\pi\i\sigma}~,\quad x\equiv \e^{2\pi\i u}~.
}
The $\Theta$-factorial is defined by
\begin{align}
\label{thetafac} \Theta(x;p;q)_n&\equiv\frac{\Gamma(q^n x;p,q)}{\Gamma(x;p,q)}=\left\{\begin{array}{ll}\prod_{k=0}^{n-1}\Theta(x q^k;p)&\quad \textrm{ if } n\geq 0\\[5pt]
\prod_{k=0}^{|n|-1}\Theta(q^{-1}x q^{-k};p)^{-1}&\quad \textrm{ if } n<0\end{array}\right.~,\\
\label{thetafacneg}\Theta(x;p,q)_{-n}&\equiv\Theta(q^{-n}x;p,q)_{n}^{-1}~.
\end{align}
Useful properties of the elliptic Gamma function are ($m,n\in\mathbb{Z}_{\geq 0}$):
\begin{itemize}
\item[-]Reflection
\be{
\Gamma(x;p,q)\Gamma(p q x^{-1};p,q)=1~.
}
\item[-]Shift
\begin{align}\label{Gammashift}
\frac{\Gamma(p^m q^n x;p,q)}{\Gamma(x;p,q)}= & \ (-x p^{(m-1)/2}q^{(n-1)/2})^{-m n}\Theta(x;p;q)_n\Theta(x;q;p)_m~,\nn\\
\frac{\Gamma(p^m q^{-n}  x;p,q)}{\Gamma(x;p,q)}= & \ (-x p^{(m-1)/2}q^{-(n+1)/2})^{m n}\frac{\Theta(x;q;p)_m}{\Theta(p q x^{-1};p;q)_n}~.
\end{align}
\item[-]Residues
\be{
\text{Res}_{x=y p^m q^n}\frac{\Gamma(y x^{-1};p,q)}{x}=\text{Res}_{x=1}\Gamma(x;p,q)~\frac{(-p q~q^{(n-1)/2}p^{(m-1)/2})^{m n}}{\Theta(p q;p;q)_n\Theta(p q;q;p)_m}~.
}
\end{itemize}

\textbf{Regularization of 1-loop determinants}. In the computation of functional 1-loop determinants, we have to deal with divergent expressions involving infinite products. In fact, we have to regularize two towers of KK torus modes and one tower of disk modes. We can do that by a two-step Hurwitz $\zeta$-function regularization. In particular, we use
\be{
\prod_{n,k\geq 0}\frac{1}{n\tau+k+X}\simeq \Gamma_2(X|1,\tau)~,
}
where $\Gamma_2(X|1,\tau)$ is the double Gamma function, and \cite{FRIEDMAN2004362}
\be{
\Gamma_2(X|1,\tau)\Gamma_2(1-X|1,-\tau)=\frac{\e^{-\frac{\i\pi}{2} B_{22}(X|1,\tau)}}{(\e^{2\pi\i X};\e^{2\pi\i\tau})_\infty}~,
}
where $B_{22}$ is the quadratic Bernoulli polynomial
\be{
B_{22}(X|1,\tau)\equiv \frac{X^2}{\tau}-\frac{1+\tau}{\tau}\left(X-\frac{1+\tau^2+3\tau}{6(1+\tau)}\right)~.
}
This prescription regularizes the torus modes, giving the standard result
\be{\label{eq:ThetaReg}
\prod_{n,k\in \mathbb{Z}}\frac{1}{n\tau-k+X}\simeq
\frac{\e^{-\i\pi B_{22}(X|1,\tau)}}{\Theta(\e^{2\pi\i X};\e^{2\pi\i\tau})}~.
}
Next, we consider the disk modes, yielding the result
\be{\label{Gammaellreg}
\prod_{j\geq 0}\prod_{n,k\in \mathbb{Z}}\frac{1}{n\tau-k+j\sigma+X}\simeq \e^{\frac{\i\pi}{3} P_3(X)}\Gamma(\e^{2\pi\i X};\e^{2\pi\i\tau},\e^{2\pi\i\sigma})~,
}
where $P_3$ is the cubic Bernoulli polynomial $B_{33}$ up to a constant
\begin{align}
P_3(X)\equiv&\ B_{33}(X|1,\tau,\sigma)-\frac{1-\tau ^2+\tau ^4}{24\sigma(  \tau +   \tau ^2)}~,\\
B_{33}(X|1,\tau,\sigma)\equiv &\ \frac{X^3}{\tau\sigma}-\frac{3(1+\tau+\sigma)X^2}{2\tau\sigma}+\nn\\
& \ + \frac{1+\tau^2\sigma^2+3(\tau+\sigma+\tau\sigma)}{2\tau\sigma}X
 -\frac{(1+\tau+\sigma)(\tau+\sigma+\tau\sigma)}{4\tau\sigma}~.
\end{align}
In the last step we used that
\be{
\zeta(s,X)\equiv \sum_{k\geq 0}(k+X)^{-s}
}
represents an order $|s+1|$ polynomial for $s<0$.

\section{Some computation with twisted fields}\label{app:vecactions}\label{app:chiralactions}

In this appendix, we spell out some detail about manipulations used to compute supersymmetry variations and t$\Q$-exact actions. 

\textbf{Vector multiplet}. Let us compute
\begin{align}
\Q\lambda&=\F_{\mu\nu}\p{\zeta\frac{\i}{2}J^{\mu\nu}-\frac{\zeta^\dagger}{|\zeta|^2}P^{\mu\nu}}+\i D\zeta~,\nn\\
(\Q\lambda)^\vee&=\frac{1}{2}F^\vee_{\mu\nu}\p{(K^\nu \bar K^\mu+Y^\nu \bar Y^\mu)\zeta^\dagger+2|\zeta|^2 \bar K^\mu \bar Y^\nu\zeta}-\i D^\vee\zeta^\dagger~.
\end{align}
Similarly
\begin{align}
\Q\tilde\lambda&=\F_{\mu\nu}\p{\tilde\zeta\frac{\i}{2}\tilde J^{\mu\nu}-\frac{\tilde\zeta^\dagger}{|\tilde\zeta|^2}\tilde P^{\mu\nu}}-\i D\tilde\zeta~,\nn\\
(\Q\tilde\lambda)^\vee&=\frac{1}{2}\F_{\mu\nu}^\vee\p{(K^\nu \bar K^\mu+Y^\mu \bar Y^\nu)\tilde\zeta^\dagger-2|\tilde\zeta|^2\bar K^\mu Y^\nu\tilde\zeta}+\i D^\vee\tilde\zeta^\dagger~.
\end{align}
Hence the localizing functional used in the main text reads
\be{
\mathscr{V}^{\rm loc}_{\rm vec}\equiv\frac{(\Q\lambda)^\vee\lambda}{4|\zeta|^2}+
\frac{(\Q\tilde\lambda)^\vee\tilde\lambda}{4|\tilde\zeta|^2}=\frac{\i}{16}\p{-\Psi Y^\mu\bar Y^\nu +\Xi_\alpha \bar K^\alpha \bar K^\mu K^\nu -4\Xi^\nu\bar K^\mu}\F^\vee_{\mu\nu}+\frac{1}{8}\Psi D^\vee~.
}
Given
\be{
D\equiv \frac{\Delta}{4}+\frac{\i}{2}Y^\mu \bar Y^\nu \F_{\mu\nu}~,\quad D^\vee\equiv \frac{\Delta^\vee}{4}+\frac{\i}{2}Y^\mu \bar Y^\nu \F_{\mu\nu}^\vee~,
}
one can also rewrite the above functional in terms of $\Delta^\vee$. 

\textbf{Chiral multiplet}. In the main text, we have considered the action constructed by acting with $\Q$ on the following functionals
\be{
{\mathscr V}_\text{chi}\equiv\f1{2|\zeta|^2}\left((\delta_\zeta \psi)^\vee\psi+(\delta_\zeta\tilde\psi)^\vee\tilde\psi\right)~,\quad \tilde{\mathscr V}_\text{chi}\equiv\f1{2|\tilde\zeta|^2}\left((\delta_{\tilde\zeta} \tilde\psi)^\vee\tilde\psi+(\delta_{\tilde\zeta}\psi)^\vee\psi\right)~.
}
Let us rewrite them in terms of the twisted fields. We use
\begin{align}
(\delta_\zeta\psi)^\vee\psi&=\sqrt{2}(F\zeta)^{\vee \, \alpha}\left(\zeta B-\frac{\zeta^\dagger}{|\zeta|^2}C\right)_\alpha~,\nn\\
(\delta_{\tilde\zeta}\psi)^\vee\psi&=\i\sqrt{2}\left(\zeta\L_{\bar Y}\phi+\frac{\zeta^\dagger}{|\zeta|^2}\L_K\phi\right)^{\vee\,\alpha}\left(\zeta B-\frac{\zeta^\dagger}{|\zeta|^2}C\right)_\alpha~,\nn\\
(\delta_\zeta \tilde\psi)^\vee\tilde\psi&=\i\sqrt{2}\left(-\tilde\zeta\L_Y\tilde\phi+\frac{\tilde\zeta^\dagger}{|\tilde\zeta|^2}\L_K\tilde\phi\right)^\vee_{\dot\alpha}\left(\tilde\zeta \tilde B-\frac{\tilde\zeta^\dagger}{|\tilde\zeta|^2}\tilde C\right)^{\dot\alpha}~,\nn\\
(\delta_{\tilde\zeta}\tilde\psi)^\vee\tilde\psi&=\sqrt{2}(\tilde F\tilde\zeta)^\vee_{\dot\alpha}\left(\tilde\zeta \tilde B-\frac{\tilde\zeta^\dagger}{|\tilde\zeta|^2}\tilde C\right)^{\dot\alpha}~.
\end{align}
In order to obtain a positive semi-definite bosonic part on the nose after varying with $\delta_\zeta$ or $\delta_{\tilde\zeta}$, we should identify ${}^\vee$ with ${}^\dagger$ on the spinor indices and complex conjugation on $\mathbb{C}$ numbers and vectors/forms. Since $\zeta^{\dagger\dagger}=-\zeta$, $\tilde\zeta^{\dagger\dagger}=-\tilde\zeta$, we obtain
\begin{align}
\frac{1}{|\zeta|^2}(\delta_\zeta\psi)^\vee\psi&=\sqrt{2}F^\vee B ~,\nn\\
\frac{1}{|\tilde\zeta|^2}(\delta_{\tilde\zeta}\psi)^\vee\psi&=\i\sqrt{2}\left(\frac{|\zeta|^2}{|\tilde\zeta|^2}(\L_{\bar Y}\phi)^\vee B-\frac{1}{|\zeta|^2|\tilde\zeta|^2}(\L_K\phi)^\vee C\right)~,\nn\\
\frac{1}{|\zeta|^2}(\delta_\zeta \tilde\psi)^\vee\tilde\psi&=-\i\sqrt{2}\left(\frac{|\tilde\zeta|^2}{|\zeta|^2}(\L_Y\tilde\phi)^\vee\tilde B+\frac{1}{|\zeta|^2|\tilde\zeta|^2}(\L_K\tilde\phi)^\vee \tilde C\right)~,\nn\\
\frac{1}{|\tilde\zeta|^2}(\delta_{\tilde\zeta}\tilde\psi)^\vee\tilde\psi&=\sqrt{2}\tilde F^\vee  \tilde B~.
\end{align}
Now notice that
\begin{align}
\L_{K}\phi&=L_{K}\phi-\i q_R K^\mu A_\mu\phi-\i  K^\mu \A_\mu .\phi~,\nn\\
\frac{1}{|\zeta|^2|\tilde\zeta|^2}\L_{K^\dagger}\tilde\phi&=L_{\bar K}\tilde\phi-\i \tilde{q_R}\bar K^\mu A_\mu\tilde\phi-\i \bar K^\mu \A_\mu .\tilde\phi~.
\end{align}
This means that for a real background  we can simply identify
\be{
\frac{1}{|\zeta|^2|\tilde\zeta|^2}(\L_{K}\phi)^\vee=\L_{\bar K}\tilde \phi~,
}
where we set $\phi^\vee=\tilde\phi$ and considered that $\A.\phi=-\A.\tilde\phi$, $\tilde q_R=-q_R$. Similarly
\begin{align}
\frac{1}{|\zeta|^2|\tilde\zeta|^2}(\L_K \tilde\phi)^\vee&=\L_{\bar K}\phi~,\quad \frac{|\zeta|^2}{|\tilde\zeta|^2}(\L_{\bar Y}\phi)^\vee=\L_{Y}\tilde\phi~,\quad \frac{|\tilde\zeta|^2}{|\zeta|^2}(\L_Y\tilde\phi)^\vee=\L_{\bar Y}\tilde\phi~,
\end{align}
where we set $\tilde\phi^\vee=\phi$. Then the sum of the four pieces becomes
\be{
2{\mathscr V}_{\text{chi}}+2\tilde{\mathscr V}_{\text{chi}}\equiv \sqrt{2}F^\vee B+\sqrt{2}\i (\L_Y \tilde\phi )B-\sqrt{2}\i(\L_{\bar K}\tilde\phi)C-\tilde B\sqrt{2}\i \L_{\bar Y}\phi-\tilde C\sqrt{2}\i \L_{\bar K}\phi+\sqrt{2}\tilde B \tilde F^\vee~.
}
Eventually, we may also want to set $(F,\tilde F)^\vee=(-\tilde F,-F)$ and identify ${}^\vee$ with ${}^\dagger$. Notice that if we act on ${\mathscr V}_{\text{chi}}$, $\tilde{\mathscr V}_{\text{chi}}$ with $\Q$, there are pieces which are manifestly positive semi-definite, but there are also mixed terms given by
\begin{align}
\frac{1}{|\zeta|^2}(\delta_\zeta \psi)^\vee \delta_{\tilde\zeta}\psi=& \ -2\i F^\vee\L_{\bar Y}\phi~,\quad & 
\frac{1}{|\tilde\zeta|^2}(\delta_{\tilde \zeta} \psi)^\vee \delta_{\zeta}\psi=& \ 2\i(\L_Y\tilde\phi)F~,\nn\\
\frac{1}{|\zeta|^2}(\delta_\zeta \tilde\psi)^\vee \delta_{\tilde\zeta}\tilde \psi=& \ -2\i\tilde F\L_{\bar Y}\phi~,\quad & \frac{1}{|\tilde\zeta|^2}(\delta_{\tilde\zeta} \tilde\psi)^\vee \delta_{\zeta}\tilde\psi=& \ 2\i\tilde F^\vee\L_Y\tilde\phi~.
\end{align}
However, they do cancel out in the summation if we indeed consider $F^\vee=-\tilde F$, $\tilde F^\vee=-F$. 

In terms of the twisted variables, there is another natural action we may consider, namely
\be{
2{\mathscr V}_{\text{twisted}}\equiv (\Q B)^\vee B+(\Q \tilde B)^\vee \tilde B+(\Q C)^\vee C+(\Q \tilde C)^\vee \tilde C~,
}
with the usual involution ${}^\vee$ acting as ${}^\dagger$. The two definitions only agree in special cases. In terms of the $\Q$-variations of the twisted fields we can write
\begin{multline}
2{\mathscr V}_{\text{chi}}+2\tilde{\mathscr V}_{\text{chi}}=\sqrt{2}(F^\vee+\i\L_Y\tilde \phi)B+\sqrt{2}\tilde B(\tilde F^\vee-\i\L_{\bar Y}\phi)-\sqrt{2}\i(\L_{\bar K}\tilde\phi)C-\sqrt{2}\i \tilde C\L_{\bar K}\phi=\\
=\sqrt{2}\left(F-\i\frac{|\zeta|^2}{|\tilde\zeta|^2}\L_{\bar Y} \phi\right)^\vee B+\sqrt{2}\left(\tilde F+\i\frac{|\tilde\zeta|^2}{|\zeta|^2}\L_{Y}\tilde\phi\right)^\vee \tilde B+\\
+\frac{1}{|\zeta|^2|\tilde \zeta|^2}(\Q C)^\vee C+\frac{1}{|\zeta|^2|\tilde \zeta|^2}(\Q \tilde C)^\vee \tilde C~.
\end{multline}
Therefore, we have the general relation
\begin{align}
2{\mathscr V}_{\text{chi}}+2\tilde{\mathscr V}_{\text{chi}}&=(\Q B)^\vee B+(\Q\tilde B)^\vee\tilde B+\frac{(\Q C)^\vee C}{|\zeta|^2|\tilde \zeta|^2}+\frac{(\Q \tilde C)^\vee \tilde C}{|\zeta|^2|\tilde \zeta|^2}+\nn\\
&\qquad+\left(1-\frac{|\zeta|^2}{|\tilde\zeta|^2}\right)\left(\sqrt{2}\i\L_{\bar Y}\phi\right)^\vee B-\left(1-\frac{|\tilde\zeta|^2}{|\zeta|^2}\right)\left(\sqrt{2}\i\L_{Y}\tilde\phi\right)^\vee \tilde B~,
\end{align}
which coincides with $\mathscr{V}_\text{twisted}$ for $|\zeta|=|\tilde\zeta|=1$. We may also observe that the whole point in setting $F^\vee=F^\dagger=-\tilde F$ is to obtain a positive semi-definite Lagrangian. In twisted variables, we see that there is another ``exotic" involution which can do the job, namely $F^\vee=F^\dagger |\zeta|^2/|\tilde \zeta|^2$, $\tilde F^\vee=\tilde F^\dagger |\tilde \zeta|^2/| \zeta|^2$.

\section{More details on the chiral multiplet}\label{app:spherical}\label{chiralKbKf}

In this section, we give more details on the computation of the 1-loop determinant for the chiral multiplet. We also use another basis for the mode expansion which confirms the results obtained in the main text through (anti-)holomorphic modes. 
The fermionic part of the localizing Lagrangian (\ref{V-loc-chi}) reads  
\be{
\Q\mathscr{V}_{\rm chi}^{\rm loc}\Big|_{\rm F}=\i\tilde B\mathcal{L}_K B+\i\tilde C\mathcal{L}_{\bar K}C+\i\tilde B\mathcal{L}_{\bar Y}C-\i\tilde C\mathcal{L}_{Y}B+\mathscr{L}^{\partial}_{\rm chi}\Big|_{\rm F}~,
}
where we omitted the term ${\mathscr V}_\lambda + \tilde{{\mathscr V}}_\lambda$ for simplicity. Upon integration by parts, this equation defines the boundary Lagrangian
\be{
\mathscr{L}^{\partial}_{\rm chi}\Big|_{\rm F}\equiv -\frac{\i}{2}\mathcal{L}_{K}(\tilde B B)-\frac{\i}{2}\mathcal{L}_{\bar K}(\tilde C C)+\i\mathcal{L}_{Y}(\tilde C B)~.
}

In order to compute the 1-loop determinant of Gaussian fluctuations around trivial field configurations, let us introduce the new auxiliary fields (with trivial Jacobian determinant)
\be{
X_F\equiv F-\i\mathcal{L}_{\bar Y}\phi~,\quad \tilde X_F\equiv\tilde F+\i\mathcal{L}_Y\tilde \phi~,
}
which allows us to recast the supersymmetry transformations into the cohomological form
\be{
\Q\varphi_{e,o}=\varphi'_{o,e}~,\quad \Q\varphi'_{o,e}=2\i\mathcal{L}_K\varphi_{e,o}~,
}
with the identifications  $\varphi_{e}=(\phi,\tilde\phi)$, $\varphi_o=(B,\tilde B)$, $\varphi'_o=\sqrt{2}(C,\tilde C)$, $\varphi'_{e}=\sqrt{2}(X,\tilde X)$. On the real  contour we have $\tilde\phi=\phi^\dagger$, $\tilde F=-F^\dagger$, namely $\tilde X_F=-X_F^\dagger+\i(1+\frac{\Omega^2}{|s|^2})\mathcal{L}_Y\tilde\phi$. Then we can write
\be{
-\tilde F F=(X_F^\dagger-\i\frac{\Omega^2}{|s|^2}\mathcal{L}_Y\tilde\phi)(X_F+\i\mathcal{L}_{\bar Y}\phi)=X_F^\dagger(X_F+\i\mathcal{L}_{\bar Y}\phi)+\tilde\phi\frac{\i\Omega^2}{|s|^2}\mathcal{L}_Y(X_F+\i\mathcal{L}_{\bar Y}\phi)+\mathscr{L}^{\partial}_{X_F}~,
}
where we defined the boundary term
\be{
\mathscr{L}^{\partial}_{X_F}\equiv\mathcal{L}_{Y}\Big(-\i\frac{\Omega^2}{|s|^2}\tilde\phi(X_F+\i\mathcal{L}_{\bar Y}\phi)\Big)~.
}
Now we can recast the localizing Lagrangian in the form
\be{
\Q\mathscr{V}^{\rm loc}_{\rm chi}=\left(\begin{array}{c} \phi \\ X_F \end{array}\right)^\dagger 
\text{K}_\text{B} \left(\begin{array}{c} \phi \\ X_F \end{array}\right)+\left(\begin{array}{c} C \\ B \end{array}\right)^\dagger  
\text{K}_\text{F} 
\left(\begin{array}{c} C \\ B \end{array}\right) + \mathscr{L}^{\partial}_{\rm bos}+\mathscr{L}^{\partial}_{\rm fer}~,
}
where we defined the kinetic operators
\begin{align}
\text{K}_\text{B}&\equiv \text{M}_{\rm L}\left(\begin{array}{cc} 
	\Delta^{(r)}~&~ 0^{(r-2)} \\
	0^{(r)}  ~&~  1^{(r-2)}
\end{array}\right)\text{M}_{\rm R}~,\quad& \text{K}_\text{F} \equiv   &
\left(\begin{array}{cc} 
	\i{\cal L}^{(r)}_{\bar K} & -\i{\cal L}^{(r-2)}_{Y}\\
	\i{\cal L}^{(r)}_{\bar Y} &  \i{\cal L}^{(r-2)}_{K}
\end{array}\right)~,\nn\\
\text{M}_{\rm L}&\equiv \left(\begin{array}{cc} 
	1^{(r)}~&~ \i\frac{\Omega^2}{|s|^2}\mathcal{L}^{(r-2)}_Y \\
	0^{(r)}  ~&~  1^{(r-2)}
\end{array}\right)~,\quad& \text{M}_{\rm R}\equiv& \left(\begin{array}{cc} 
	1^{(r)}~&~ 0^{(r-2)} \\
	\i\mathcal{L}^{(r)}_{\bar Y}  ~&~  1^{(r-2)}
\end{array}\right)~,
\end{align}
with $\Delta^{(r)}\equiv -\mathcal{L}^{(r)}_K\mathcal{L}^{(r)}_{\bar K}-\mathcal{L}^{(r)}_Y\mathcal{L}^{(r)}_{\bar Y}$ and we added a superscript to remind on which space the operators act on. The boundary terms are
\begin{align}
\mathscr{L}^{\partial}_{\rm B}&\equiv \frac{1}{2}\mathcal{L}_{K}(\tilde\phi\mathcal{L}_{\bar K}\phi)+\frac{1}{2}\mathcal{L}_{\bar K}(\tilde\phi\mathcal{L}_{ K}\phi)+\mathcal{L}_{Y}(\tilde\phi\mathcal{L}_{\bar Y}\phi)+\mathcal{L}_{Y}\Big(\tilde\phi(-\i\frac{\Omega^2}{|s|^2}X_F+(1+\frac{\Omega^2}{|s|^2})\mathcal{L}_{\bar Y}\phi)\Big)~,\nn\\
\mathscr{L}^{\partial}_{\rm F}&\equiv-\frac{\i}{2}\mathcal{L}_{K}(\tilde B B)-\frac{\i}{2}\mathcal{L}_{\bar K}(\tilde C C)+\i\mathcal{L}_{Y}(\tilde C B)~.
\end{align}
Now we can notice that there exists the operator 
\be{
\text{T}=\left(\begin{array}{cc} 
	\i\mathcal{L}_K^{(r)}~&~ 0^{(r-2)} \\
	  -\i\mathcal{L}_{\bar Y}^{(r)}~&~  1^{(r-2)}
\end{array}\right)~
}
such that 
\be{
\frac{\det \text{K}_{\rm F}}{\det \text{K}_\text{B}}=\frac{\det (\text{K}_{\rm F}\circ \text{T})}{\det (\text{K}_\text{B}\circ \text{T})}=\frac{\det (\text{K}_{\rm F}\circ \text{T})}{\det(\text{M}_{\rm L})\det (\Delta^{(r)})\det(\text{M}_{\rm R})\det( \text{T})}=\frac{\det \i\mathcal{L}_K^{(r-2)}}{\det\i\mathcal{L}_K^{(r)}}~,
}
where we used that $\text{K}_\text{F}\circ \text{T}$ is upper triangular thanks to $[\mathcal{L}_K,\mathcal{L}_{\bar Y}]=0$,  with determinant  $\det(\Delta^{(r)})\det(\i\mathcal{L}^{(r-2)}_K)$. The determinants on the r.h.s. are to be computed on the space of scalar fields of R-charge $r$ or $r-2$, and the modes to be kept are determined by the vanishing of the boundary actions. The only obvious ways to impose their vanishing on the boundary (consistently with supersymmetry) is either through Dirichlet boundary conditions on $\phi$, and by supersymmetry  Dirichlet conditions also on $C$, or $\mathcal{L}_{\bar Y}\phi |_\partial=0$ and Dirichlet conditions on $B$, and by supersymmetry  Dirichlet conditions also on $X_F$.

\end{appendix}

%{
%\bibliographystyle{utphys}
%\bibliography{refs}
%}

\providecommand{\href}[2]{#2}\begingroup\raggedright\endgroup

\end{document}